\documentclass[12pt]{amsart}

\usepackage{amssymb}
\usepackage{amsmath}
\usepackage{amsfonts}
\usepackage{mathrsfs}
\usepackage{graphicx}
\usepackage{placeins}
\usepackage{caption}
\usepackage{subcaption}
\usepackage{color}
\usepackage[onehalfspacing]{setspace}
\usepackage{natbib}
\usepackage{enumerate}
\usepackage{dcolumn}
\usepackage[charter,cal=cmcal]{mathdesign}
\usepackage[foot]{amsaddr}
\usepackage[colorlinks=true,citecolor=blue,urlcolor=blue,pdfpagemode=UseNone,pdfstartview=FitH]{hyperref}
\usepackage{apptools}
\usepackage{bibunits}
\makeatletter
\def\section{\@startsection{section}{1}
	\z@{0.8\linespacing\@plus\linespacing}{.5\linespacing}{\Large}}

\def\subsection{\@startsection{subsection}{2}
	\z@{.6\linespacing\@plus.4\linespacing}{.4\linespacing}{\large}}

\def\subsubsection{\@startsection{subsubsection}{3}
	\z@{.5\linespacing\@plus.4\linespacing}{-.5em}{\normalfont\bfseries}}
\makeatother

\numberwithin{equation}{section}

\newtheorem{theorem}{Theorem}[section]
\newtheorem{lemma}{Lemma}[section]

\theoremstyle{definition}
\newtheorem{definition}{Definition}[section]

\theoremstyle{definition}
\newtheorem{assumption}{Assumption}[section]

\theoremstyle{definition}

\DeclareTextFontCommand{\textbfit}{%
	\fontseries\bfdefault 
	\itshape
}

\title{}

\begin{document}
	\vspace*{5ex minus 1ex}
	\begin{center}
		\Large \textsc{Estimating Unobserved
				Individual Heterogeneity\\
				Using Pairwise Comparisons}
		\bigskip
	\end{center}

\begin{center}
Elena Krasnokutskaya, Kyungchul Song, and Xun Tang\\
\textit{Johns Hopkins University, University of British Columbia, Rice University}
\bigskip
\end{center}
\date{%
\today%
.}

\begin{abstract}
{\small 
	We propose a new method for studying environments with  unobserved individual heterogeneity. 
	Based on model-implied pairwise inequalities, the method classifies individuals in the sample into groups defined by discrete unobserved heterogeneity with unknown support.  We establish conditions under which the groups are identified and consistently estimated through our method. We show that the method performs well in finite samples through Monte Carlo simulation. 
	We then apply the method to estimate a model of lowest-price procurement auctions with unobserved bidder heterogeneity, using data from the California highway procurement market.}\medskip

{\footnotesize \ }


\end{abstract}

\thanks{We are grateful to the Associate Editor and three referees for valuable comments. All errors are ours. Corresponding address: Elena Krasnokutskaya, Department of Economics, Johns Hopkins University, 3100 Wyman Park Drive, Baltimore, MD 21218, USA. Email address: ekrasnok@jhu.edu}
\maketitle

\begin{bibunit}[econometrica]			
\section{Introduction}
The empirical analysis of many economic settings requires accounting for unobserved individual heterogeneity (UIH) which reflects agent-specific factors that influence agents' decisions but are not recorded in the data. Failing to account for UIH generally leads to biased estimates and affects the validity of counterfactual prediction.

In this paper, we consider a generic economic model where UIH induces a group structure among agents according to their types. We provide conditions for identification of the group structure, and propose a method to recover the group structure from data.

Our main idea is based on the insight that UIH often implies \textit{pairwise} inequality restrictions on endogenous observable quantities. For instance, in multi-attribute auctions, bidders with higher (unobserved) quality levels have a greater chance of winning the auction, controlling for the bid and the set of competitors. In a labor market setting, agents with higher (unobserved) productivity receive higher wages than less productive ones. In Section 2.2 we provide further examples of economic applications in which pairwise inequality restrictions arise naturally from the behavior of agents in equilibrium.

We develop a statistical method to recover the group structure (that is, to classify individuals into groups defined by UIH) using a pairwise comparison approach. Our method treats UIH as individual-specific discrete parameters which may affect the distribution of other observed or unobserved variables. We assume that the support of UIH is a finite, ordered set that is not known to the econometrician. Such flexibility is important in structural models where individuals interact strategically and the UIH of all agents jointly affects the equilibrium outcome.

One may attempt to order individuals on a pair-by-pair basis using pairwise inequality tests. However, it may not produce an ordering that is transitive in finite samples. Our method recovers the whole group structure by sequentially sub-dividing the set of agents on the basis of the $p$-values of tests of pairwise inequality restrictions. The method recovers the group structure for each assumed number of groups, and then selects the number of groups (and the associated group structure) using a penalization scheme. We show that our estimator of the group structure is consistent under mild regularity conditions and performs well in small samples. 

In many settings, classifying individuals into groups defined by UIH offers key economic insights. For example, our method can be used to identify colluding bidders in auctions, and firms' cost asymmetries or product quality differences. In addition, recovering the group structure also often serves as the first step for estimating structural models with strategic interactions, such as dynamic industry models or auctions with asymmetric bidders.\footnote{Estimation of \textit{discrete} unobserved individual heterogeneity does not affect subsequent estimation of other structural parameters in terms of pointwise asymptotics. However, establishing uniform asymptotics remains an open question. This problem is analogous to that of post-model-selection inference. For discussion on the issues, see \citet{Potcher1991}, \citet{LeebPotcher2005}, and \citet{AndrewsGuggenberger2009} and the references therein. Uniform asymptotics in our setup is complex because of the need to consider every possible direction of local perturbation from the actual group structure in data-generating process. A full theoretical investigation of the issue in our context merits a separate paper.} 

This approach offers a feasible way to identify and estimate games with UIH. Specifically, a traditional approach in a setting with UIH would be to treat UIH as ``fixed effects" and estimate them jointly with other structural parameters. This approach poses practical challenges in a setting with strategic interdependence among agents, especially when  equilibrium outcomes admit no closed-form expressions.  First,  it is generally not obvious what variation in the data may identify model components including the fixed effects in such settings. One of the contributions of our paper is to point out the variation which identifies the group structure associated with ``fixed effects.'' 

Further, we propose to separate the recovery of the group structure from the estimation of other structural parameters. Recovering the group membership of every agent facilitates, and is often needed for, identifying the remaining structural elements in models with strategic interactions. A classical example is English auctions among bidders with unobserved types (in the sense that bidders' private values are drawn independently from the distributions ``labeled'' by bidder types) and where the data only report the transaction price and the identity of the winner.	\citet{AtheyHaile2007} show that the distributions of private values cannot be identified in this model if the type of the auction winner is unknown. We provide details and additional examples in the supplemental note to this paper.

Finally, our approach offers a way to estimate games with UIH with low computational cost, compared with the alternative approach of estimating the fixed effects jointly with other structural parameters. For example, consider an environment with a large number of players where many independent games (each containing only a small subset of players) are observed in the data. The numerical optimization (such as simulated GMM or MLE) requires evaluating the objective function which involves computing equilibrium for each value of the fixed effects and other structural parameters. This can be computationally prohibitive in practice. In contrast, our method requires the game to be solved only for the estimated configuration of the agents’ group memberships rather than for every possible configuration as is required under the joint estimation.\footnote{\citet{KrasnokutskayaSongTang2020JPE} used this classification method as a first step in the structural analysis of an online service market for computer coding.}

Our method is advantageous especially in settings where the number of agents is moderately large but each market (observation) in the data involves only a small subset of participants. For example, the total number of participants may be several hundreds but each market may contain only several participants. In this case, despite the large number of markets observed, the researcher may have only a small number of markets which contain the same set of participants. We call this issue \textit{the problem of the sparsely common set of agents}. In such settings, the researcher cannot build inference on the conditional moments \textit{given the full set of participating agents} in a market, as typically done in the structural empirical literature, because we do not have many such observations. Hence the researcher needs to ``aggregate'' the markets or the agents in order to conduct reliable inference with sufficiently many observations. Pairwise restrictions are often testable with accuracy since the number of markets where a given pair of agents is present tends to be large even if the number of markets with the same set of participants is small. Thus, pairwise restrictions and the classification procedure offer a natural way to aggregate agents into groups which permits estimation of other primitives. 

We investigate the finite-sample performance of our classification method in Monte Carlo simulation. The data-generating process (DGP) is a lowest-price procurement auction among asymmetric bidders whose independent private values are drawn from distributions with different means. 
We report the outcome of classification for DGPs with various numbers of bidders and group structures. Our classification method works well. Its performance is better when the number of bidders and groups are smaller relative to the number of observed markets/games, and when the differences between groups are larger. We also find that the impact of classification errors on subsequent estimation of other structural parameters in the game is non-substantial.

We analyze the California highway procurement market by applying our classification method to a model of asymmetric lowest-price procurement auction.
Existing empirical studies of auction markets typically emphasized asymmetry in bidders' private values associated with their \textit{observable} characteristics.\footnote{For example, \citet{AtheyLevinSeira2011}, \citet{RobertsSweeting2013} and \citet{AradillasGandhiQuindt2013} accounted for the bidder heterogeneity associated with size in the timber market (`mills' vs `loggers'); \citet{KrasnokutskayaSeim2011}, \citet{Jofre-BonetPesendorfer2003} and \citet{GentryKomarovaShiraldi2016} incorporated bidder participation differences in highway procurement market (`regular' vs `fringe' bidders); \citet{ConleyDecarolis2016}, \citet{Asker2010} and \citet{Pesendorfer2000} allowed for bidder heterogeneity in collusive behaviors.}
In comparison, we allow the bidders' private values to be drawn from heterogeneous distributions with different means. To account for other sources of cost heterogeneity, we control for bidders' distance to the project site. We also accommodate possible endogeneity in the competitive structure. We use the classification method to recover the unknown group structure (i.e., partition bidders into groups with different mean costs). Then, using this estimated group structure, we estimate group-specific cost distributions using GMM. 

Our estimates indicate that the bidders in the data come from several unobserved groups with substantial differences in mean costs. We also find that ignoring such unobserved bidder heterogeneity would lead to biased estimates of how bidders' costs depend on various factors. \smallskip

\noindent \textbf{Related Literature.} One of the popular methods of accounting for UIH in structural modeling is to adopt finite mixtures. (See \citet{Hu2008}, \citet{HuSchennach2008}, \citet{KasaharaShimotsu2009}, \citet{HuShum2012}, \citet{HuShiu2013},  \citet{HuMcAdamsShum2013}, and \citet{HenryKitamuraSalanie2014}. See also See \cite{KasaharaShimotsu2014} and \cite{KasaharaShimotsu2015} for estimating and testing for the number of mixture components in finite mixture models.) The finite mixture modeling assumes that the UIH is a random variable drawn from some unknown distribution. The goal is to identify this distribution and estimate it from data. It does not require each individual to appear in many independent games. In contrast, our approach aims to \textit{classify individual agents in the sample into disjoint groups defined by their realized unobserved types}, using their participation outcomes in many games. Thus the two approaches are fundamentally different both in their aims and their data requirements. While general identification results have been developed in this literature of finite mixture models (see, e.g., \citet{HuSchennach2008}, \cite{BonhommeJochmansRobin2016}), implementation of the finite-mixture method is impractical in our set-up due to the issue of sparse commonality, and technical issues associated with high dimensionality.\footnote{When each market is drawn from a finite mixture distribution, and there are $n$ agents with each having a type from $S$ values, the number of the mixture components becomes $S^n$ which can be very large in pratice, even when $n$ is a moderate number such as five or ten. In addition, to construct a likelihood or moment condition, one would need to solve for a different market equilibrium for each component.}

The classification algorithm we propose is related to the clustering method in statistics. (See, e.g., Chapter 14 of \citet{HastieTibshiraniFriedman2009}.) The main difference is that the clustering methods aim to group individuals based on the similarity of their observed attributes, whereas in our setting, a researcher's objective is to group individuals according to testable implications of their \textit{unobserved attributes}. To accomplish this, we exploit the relationship between endogenous outcomes and the unobserved types of individuals implied by an economic model. Our method also requires a data structure different from clustering methods. The literature of clustering methods mostly considers a set-up in which each cross-sectional unit is observed once, whereas our method uses many observations per individual in the sample.

Also related to our approach is the literature of panel models with group-level heterogeneity. For example, \cite{Sun2005} introduced a linear panel model where parameters take values in a finite set according to a logistic probability, and offered methods of estimating the group structure. \citet{Song2005} considered a panel model with finite-valued nonstochastic parameters and produced an algorithm to recover the unobserved group structure in large panel models. \cite{LinNg2012} provided a method of estimating a panel model using threshold variables when the group membership is unknown. \citet{SuShiPhillips2016} developed a new Lasso method to recover the unknown group-specific parameters.  \citet{BonhommeManresa2014} proposed a k-means clustering algorithm to recover the group structure in a linear panel model. These papers often focus on models which admit a reduced form for the dependent variable in which its functional relation to UIH is made explicit. In contrast, our method targets a set-up where the dependence of the outcome variables on the UIH arises only implicitly through equilibrium contraints in games, and the group structure of UIH is identified only through pairwise inequality restrictions. Thus, the approaches developed in the panel literature are not applicable in settings our proposal focuses on.\smallskip

\noindent \textbf{Roadmap.} This paper is organized as follows. Section 2 introduces the basic environment and defines pairwise inequality restrictions. This section also provides several examples from various contexts to motivate our classification method. 

Section 3 establishes identification of the unobserved group structure using pairwise inequality restrictions.  Section 4 proposes a consistent estimator of the group structure. Section 5 provides results from Monte Carlo simulation. Section 6 presents the empirical application. Section 7 concludes. Section 8 contains the proof of some results in the paper. Further examples and mathematical proofs are provided in the supplemental note of this paper. The note also contains further simulation results and details in our empirical application.

\section{The Model and Examples}

\subsection{Pairwise Inequalities in Game Models}

We consider a setting where the econometrician observes $L$ games, and in each game, a set of agents interact with each other. Each agent $i$ is associated with a non-stochastic type, $q_i$, which is not observed by the researcher. We assume that the type is finite-valued so that $q_i \in Q_0 = \{\bar{q}_{1},...,\bar{q}_{K_{0}}\}$, with $\bar{q}_{1}<\cdots <\bar{q}_{K_{0}}$. This induces an \textit{(ordered) partition} $(N_{1},N_{2},...,N_{K_{0}})$ of the set $N$ of agents such that for each $k=1,2,...,K_0$, $N_k$ consists of agents with higher type than those in $N_{k-1}$. The group structure is characterized by a function $\tau: N \rightarrow \{1,...,K_{0}\}$ that links the identity of a player to his unobserved type so that $q_{i}=\bar{q}_{\tau(i)}$ and for each $ k=1,...,K_{0}$, \[N_{k}=\left\{ i\in N :\tau(i)=k\right\}.\] The group structure defined by $\tau$ is represented as an ordered collection of sets $N_k$:
\begin{eqnarray}
\label{def T}
	T = (N_1,N_2,...,N_{K_0}).
\end{eqnarray}

The data available to the researcher contain for every observation $\ell = 1,...,L$: a vector of observable characteristics for all the agents involved, $\{X_{j,\ell}\}_{j\in S_\ell}$, as well as at least one but possibly multiple vectors of outcome variables, $Y_\ell=\{Y_{j,\ell}\}_{j \in S_\ell}$, where $S_\ell$ is the set of players involved in the $\ell$-th observation (e.g., a game or a market). Our main focus is on recovering the ordered partition $(N_{1},N_{2},...,N_{K_{0}})$ of the agents from data.

The main insight of our paper begins with the observation that in many structural models, the ordering among $q_i$'s (or equivalently $\tau(i)$'s) coincides with the ordering between indexes that can be estimated consistently even in the presence of sparse commonality of agents. Specifically, we use pairwise indexes $\delta_{ij}$ and $\delta_{ij}^0$ that satisfy the following relations:\footnote{For the sake of concreteness, our exposition in the paper focuses on this form of indexes. Our procedures rely on the indexes only through the availability of consistent tests of pairwise inequality restrictions: $\delta_{ij} >0$ and $\delta_{ij}^0 = 0$. As long as such consistent tests are available, one can use our method for other forms of pairwise indexes.}
\begin{eqnarray}
\label{pairwise inequalities}
\delta_{ij} &>& 0 \textnormal{ if and only if } \tau (i)>\tau (j)
\label{general_setup}; \\ 
\delta_{ij}^{0} &=& 0 \textnormal{ if and only if } \tau (i)=\tau (j),
\nonumber
\end{eqnarray} 
where the indexes $\delta_{ij}$ and $\delta_{ij}^0$ can be consistently estimated using the sample. In many applications, we can take the indexes as (a variant of) the following form:
\begin{eqnarray}
\label{index}
   \delta_{ij} &=& \int \max\{\mathbf{E}[Y_{i,\ell}|X_{i,\ell} = x] -  \mathbf{E}[Y_{j,\ell}|X_{j,\ell} = x],0\}dF(x), \text{ and }\\ \notag
   \delta_{ij}^0 &=& \int \left|\mathbf{E}[Y_{i,\ell}|X_{i,\ell} = x] -  \mathbf{E}[Y_{j,\ell}|X_{j,\ell} = x]\right|dF(x),
\end{eqnarray}
where $F$ is a known distribution or the distribution of an observable random vector.

For example, suppose that the outcome $Y_{i,\ell}$ admits the following reduced form:
\begin{eqnarray*}
	Y_{i,\ell} = g(\tau(i),X_{i,\ell},\eta_{i,\ell}),
\end{eqnarray*}
where $g$ is a function that is strictly increasing in $\tau(i)$ and $\eta_{i,\ell}$ is an unobserved component that is independent of $X_{i,\ell}$ and does not depend on $\tau(i)$. Then under regularity conditions that ensure that $\delta_{ij}$ and $\delta_{ij}^0$ are well defined, we obtain the pairwise relations (\ref{pairwise inequalities}) with (\ref{index}). The main advantage of our approach is that we do not require an explicit characterization of the reduced form $g$. Due to this flexibility, our approach is most useful for analyzing UIH in structural models where the reduced form for outcomes arises only implicitly through equilibrium constraints. In such a setting, the sign of the indexes $\delta_{ij}$ represents the pairwise relation which says that between any two agents, one agent's type is higher than the other if and only if his outcome tends to be higher than that of the other. As we demonstrate through examples below, many structural models imply such pairwise relations through indexes $\delta_{ij}$ and $\delta_{ij}^0$.

The main goal of this paper is to develop a statistical procedure to recover the group structure $\tau$ from data. Our method relies only on the pairwise inequality restrictions in (\ref{pairwise inequalities}). Thus so far as the group structure is concerned, the pairwise comparison indexes $\delta_{ij}$ and $\delta_{ij}^0$ play the role of a sufficient statistic; the recovery of the group structure does not rely on other details of the structural model.

\subsection{Examples \label{sec:otherexamples}}

We now provide examples of pairwise inequality restrictions which arise as equilibrium implications in a variety of commonly studied empirical contexts. 
 
\subsubsection{Unobserved Quality in Multi-attribute Auctions}

Consider a simplified version of multi-attribute auctions in \citet{KrasnokutskayaSongTang2020JPE} that abstracts away from observed auction and seller heterogeneity. Let $N$ denote the total set of sellers and $S_\ell$ the set of sellers who submitted bids for a project $\ell$. Each seller has a discrete unobservable quality: $q_i \in \{\bar{q}_{1},..., \bar{q}_{K_{0}}\}$, with $\bar{q}_k < \bar{q}_{k'}$ whenever $ k < k'$. Such a quality is known to buyers but not reported in data.
The buyer for project $\ell$ selects a seller among those who submitted bids or chooses an outside option to maximize his payoff. The payoff to the buyer from engaging services of seller $i\in S_\ell$ is given by
$ U_{i,\ell}=\alpha _{\ell}q_{i}+\epsilon _{i,\ell}-B_{i,\ell}$ whereas the payoff from an outside option is $U_{0,\ell}$. Here $ \alpha_\ell $ is a non-negative weight the buyer gives to the seller's quality relative to the seller's bid, whereas $ \epsilon_{i,\ell} $ reflects a buyer-seller match-specific stochastic component. 

Let us suppress the auction subscript $\ell$ and define for any two sellers $i,j$, 
\[
\rho_{ij}(b)=P\left\{\,i\,\mbox{wins}\,|\,B_{i,\ell}=b,\,i\in S_\ell,\,j\not\in S_\ell\right\},
\]
for all $b$ on the intersection of the supports of $B_{i}$ and $B_{j}$. 
Suppose that $\alpha, S_\ell,\{B_{i,\ell}, \epsilon_{i,\ell}\}_{i\in S_{\ell}}$ are mutually independent.\footnote{This is plausible, for example, if sellers are not informed of the weights or outside option of the buyer, or the identities of other sellers in $S_\ell$.} Proposition 1 of \citet{KrasnokutskayaSongTang2020JPE} showed that
\begin{equation}
  \text{sign}(\rho_{ij}(b)-\rho_{ji}(b))= \text{sign}(q_{i}-q_{j}),  \label{res}
\end{equation}%
for any $b$ in the intersection of bid supports. 

On the basis of this property the comparison indexes can be constructed as follows:
$\delta_{ij} \equiv \int \max \{\rho_{ij}(b)-\rho_{ji}(b),0\}db$ and $\delta _{ij}^{0} \equiv \int \left\vert \rho_{ij}(b)-\rho_{ji}(b)\right\vert db$. Note that the comparison indexes do not depend on other details of the structural model such as specific parametric assumptions for the distribution of buyers' tastes.
\medskip

\subsubsection{Firms' Cost Efficiency and Pricing Decisions}

Consider a population of $n$ firms or brands, each of which produces a single brand of product. The data consists of independent markets indexed by $\ell = 1,...,L$. The marginal cost for firm $ i $ on market $\ell$ is $c_{i,\ell}=\varphi(w_{i,\ell},q_{i},\eta_{i,\ell})$, where $w_{i,\ell}$ are observable cost shifters, $q_i$ a brand-specific unobserved heterogeneity that is fixed across markets, and $\eta_{i,\ell}$'s are i.i.d. idiosyncratic noises independent of $ w_{i,\ell}$ and $q_i $.
We may interpret $q_i$ as a measure of firm $ i $'s cost efficiency. Firms have complete information about each others' cost efficiencies.\footnote{This assumption is plausible in certain industries where production efficiency is mostly determined by firms' technology or equipment that is publicly observable.} Firms in the population are partitioned into groups with different levels of $q_i$: $N=\cup_k N_k$ where $i \in N_k$ if $\tau(i)=k$.

Let $\sigma_{i,\ell}(\mathbf{x}_{\ell},\,\mathbf{p}_{\ell},\Omega_\ell)$ denote firm $i$'s market shares, which is a function of product attributes ($\mathbf{x}_{\ell}=\{x_{i,\ell}\}_{i\in S_\ell}$, where $S_\ell$ denotes the set of brands in market $\ell$) and prices ($\mathbf{p}_{\ell}=\{p_{i,\ell}\}_{i\in S_\ell})$  conditional on the set of products available in market $\ell$ and other market factors denoted by $\Omega_\ell$.
The profit for firm $ i $ in market $\ell$ is: $\pi_{i,\ell}=(p_{i,\ell}-c_{i,\ell})\sigma_{i,\ell}(\mathbf{x}_{\ell},\mathbf{p}_{\ell},\Omega_\ell)M_\ell$, where $M_\ell$ is a measure of potential consumers in market $\ell$.
In any pricing equilibrium with an interior solution, the first-order condition implies
\begin{equation}\label{foc} 
c_{i,\ell}=p_{i,\ell}+\frac{\sigma_{i,\ell}}{\partial \sigma_{i,\ell}/\partial p_{i,\ell}}. 
\end{equation} 
Notice that if $ \eta_{i,\ell} $ is independent of $q_i$ and $w_{i,\ell}$, and $\varphi(w_{i,\ell},q_{i},\eta_{i,\ell})$ is strictly monotone in $q_i$ then so is the right-hand side of (\ref{foc}), which can be constructed from estimates of the demand system. 
Hence, for any pair $i,j\in N$, $q_{i}\geq q_{j}$ if and only if $\mathbf{E}[z _{i,\ell}|w_{i,\ell}=w_0] \geq \mathbf{E}[z _{j,\ell}|w_{j,\ell}=w_0]$, for all $w_0$, where $z_{i,\ell}$ is defined as the quantity on the right-hand side of (\ref{foc}). The statement is also true when both inequalities are strict. 
Thus we can define a pairwise comparison index
\begin{equation}\label{index3} 
\delta_{ij}\equiv \int \max\{\mathbf{E}[z_{i,\ell}|w_{i,\ell}=w_0] - \mathbf{E}[z _{j,\ell}|w_{j,\ell}=w_0],0\}dF(w_0), 
\end{equation} 
where $F$ is the distribution of $w_{i,\ell}$. In equilibrium $ \delta_{ij} > 0 $ if and only if $q _i > q_j $. 
Likewise, define $\delta_{ij}^0 $ by replacing $\max\{\cdot,0\}$ in the integral in $ \delta_{ij}$ with the absolute value. These pairwise comparison indexes do not condition on specific identities of firms in a market. 

\subsubsection{Assortative Matching in Labor Market} 

Sorting of heterogeneous employees across heterogeneous firms has been studied in \citet{LentzMortensen2010}, \citet{AbowdKramarzMargolis1999}, and \citet{LiseMeghirRobin2011}. 
In a typical setting, firms are heterogeneous in the productivity from a given worker \textit{ceteris paribus}. 
Workers differ in their unobservable ability $q_i$. 
Under further restrictions (see \citet{EeckhoutKircher2011} and \citet{HagedornLawManovski2016}), workers with higher ability would in equilibrium earn higher wages than co-workers at the same firm, holding other things equal.

This forms a basis for pairwise comparisons. Specifically, let $w_{i,f,t}=W(q_i,X_{i,t},\Omega_{f,t})$ denote the wage worker $i$ earns at time $t $ while employed by firm $f$, where $W$ is a non-stochastic function. Here $\Omega_{f,t}$ captures all the relevant firm-specific unobservable factors while $X_{i,t}$ reflects worker $i$'s observable characteristics other than $q_i$. Here we assume that $(X_{i,t},\Omega_{f,t})$ is identically distributed across $f$'s and $t$'s. Using $N_{f,t}$ to denote the set of workers employed by firm $f$ at time $t$, we define the comparison index as
\begin{eqnarray*}
	\delta_{ij}=\int \max\{\mathbf{E}[w_{i,f,t}|X_{i,t} = x] - \mathbf{E}[w_{j,f,t}|X_{j,t} = x] ,0\}dF(x), \text{  } i,j\in N_{f,t},
\end{eqnarray*}
where $F$ is the distribution of $X_{i,t}$. Then $ \delta_{ij} > 0 $ if and only if $ q_i > q_j $, under regularity conditions such as strict mononicity of $W$ in $q_i$. Similar to before, define $\delta^0_{ij}$ by replacing the max operator in $ \delta_{ij} $ with its absolute value. In this setting comparison of workers is complicated by the (unobserved) firm heterogeneity and sorting of workers across firms. Pairwise comparisons allow researchers to circumvent these issues by focusing on workers' wages earned while they are employed by the same firm. 

\subsection{Sparsely Common Set of Agents and Pairwise Inequalities}

Our pairwise comparison method is most useful in settings where players appear in markets only sparsely. When most distinct sets of players appear only once or twice in the data, it is challenging to study the competition pattern in the market, because we cannot recover from data the conditional probability of their choices conditional on the set of competitors.  

\begin{figure}[t]
	\begin{center}
		\includegraphics[scale=0.55]{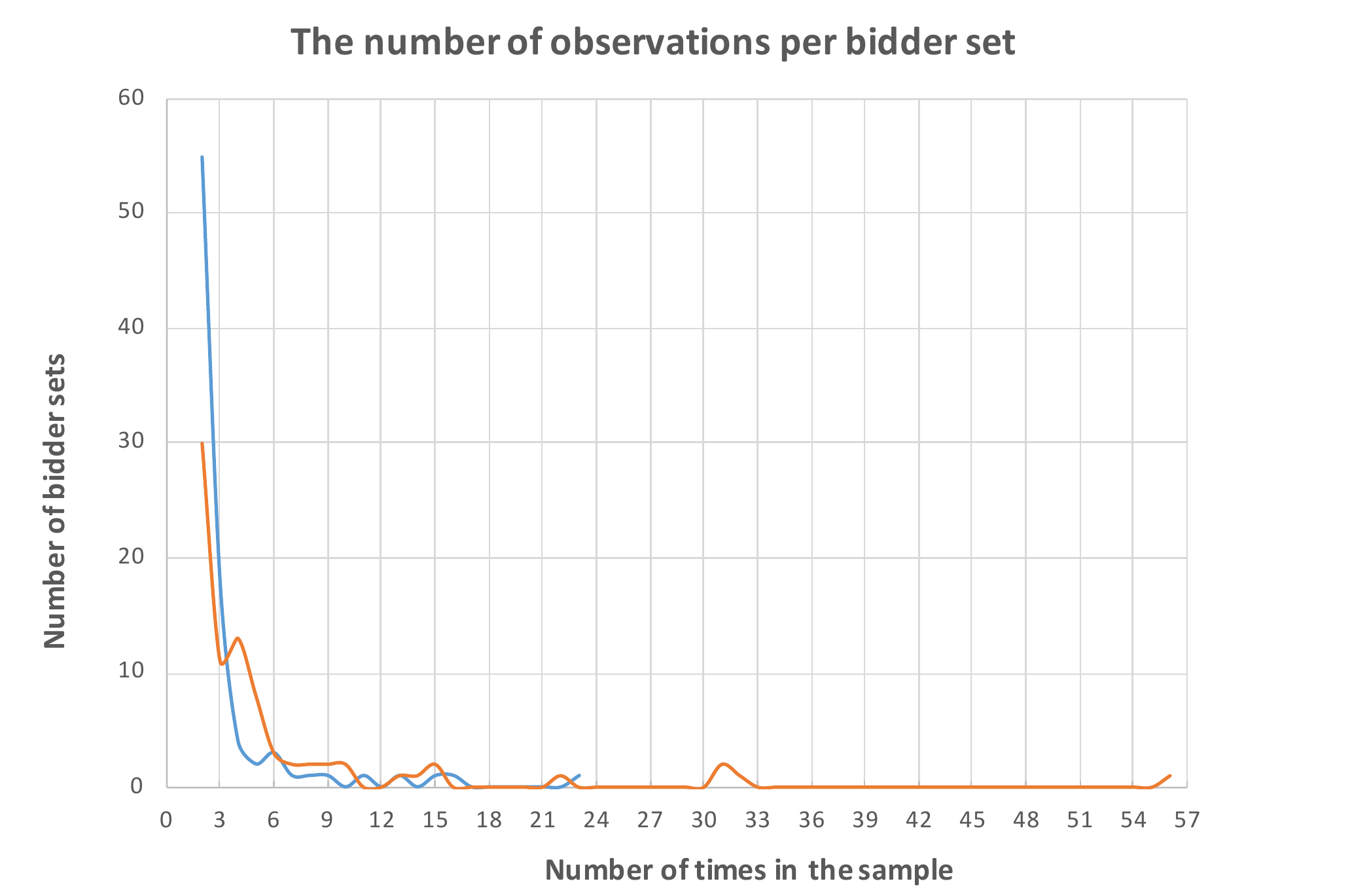} \includegraphics[scale=0.55]{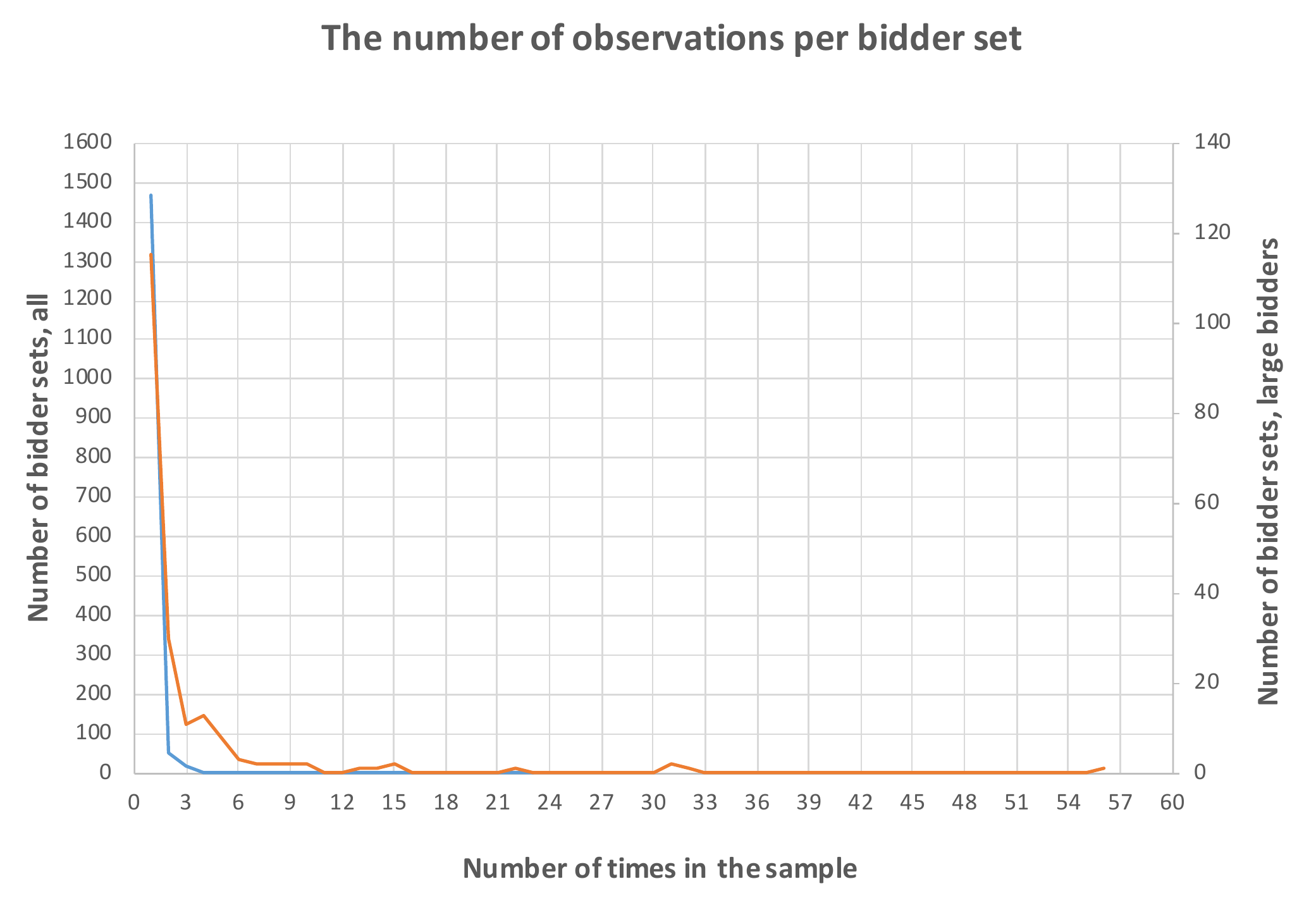}
		
		\caption{ \footnotesize The panels show sparse commonality of bidder sets in the auction data we use for the empirical application later. In both panels, the $x$-axis denotes the number of the times that a particular bidder set appears in the data and the $y$-axis represents the number of distinct bidder sets in the data. For example, $(x,y) = (2,250)$ represents that there are 250 distinct bidder sets that appear twice in the data. The upper panel shows the number of distinct bidder subsets consisting of regular bidders submitting bids in an auction. (We define a regular bidder to be one who submits at least 5 bids per year in this market.) The lower panel shows all data. As the figure shows, most bidder sets appear only once or twice. Notice that in the upper panel, there is a single bidder set that appears in the data 56 times and this is the largest occurences that we observe in the data.}
		\label{Figure1}
	\end{center}
\end{figure}

Such a data feature is not uncommon in practice. In Figure \ref{Figure1}, we show the structure of the auction data used for our empirical application. The $x$-axis in each panel denotes the number of the times that a generic bidder set appears in the data, and the $y$-axis represents the number of distinct bidder sets in the data. For example, $(x,y) = (2,250)$ represents that there are 250 distinct bidder sets that appear twice in the data. The upper panel shows the number of distinct bidder subsets consisting of regular bidders submitting bids in an auction.\footnote{We define a regular bidder to be one who submits at least 5 bids per year in this market.} The lower panel shows all data. As the figure shows, most bidder sets appear only once or twice. Notice that in the upper panel, there is a single bidder set that appears in the data 56 times and this is the largest occurences that we observe in the data. Furthermore, this set consists only of two bidders, making it hard to take this set as representing the whole auction market.

To express this data feature, define for any $S \subset N$,
\begin{eqnarray*}
	\mathcal{L}(S) =\{1 \le \ell \le L: S_{\ell} = S \}.
\end{eqnarray*}
Thus $\mathcal{L}(S)$ represents the set of markets where the set of participants in a market $S_\ell$ is precisely $S$. 
In this paper, we refer to the setting as that of a 
\textbfit{sparsely common set of agents}, if the proportion $\max_{S \subset N} |\mathcal{L}(S)| / L$ is negligible in finite samples. In other words, only a small fraction of the markets in the sample share exactly the same set of participants.
\begin{figure}[t]
	\begin{center}
		\includegraphics[scale=0.5]{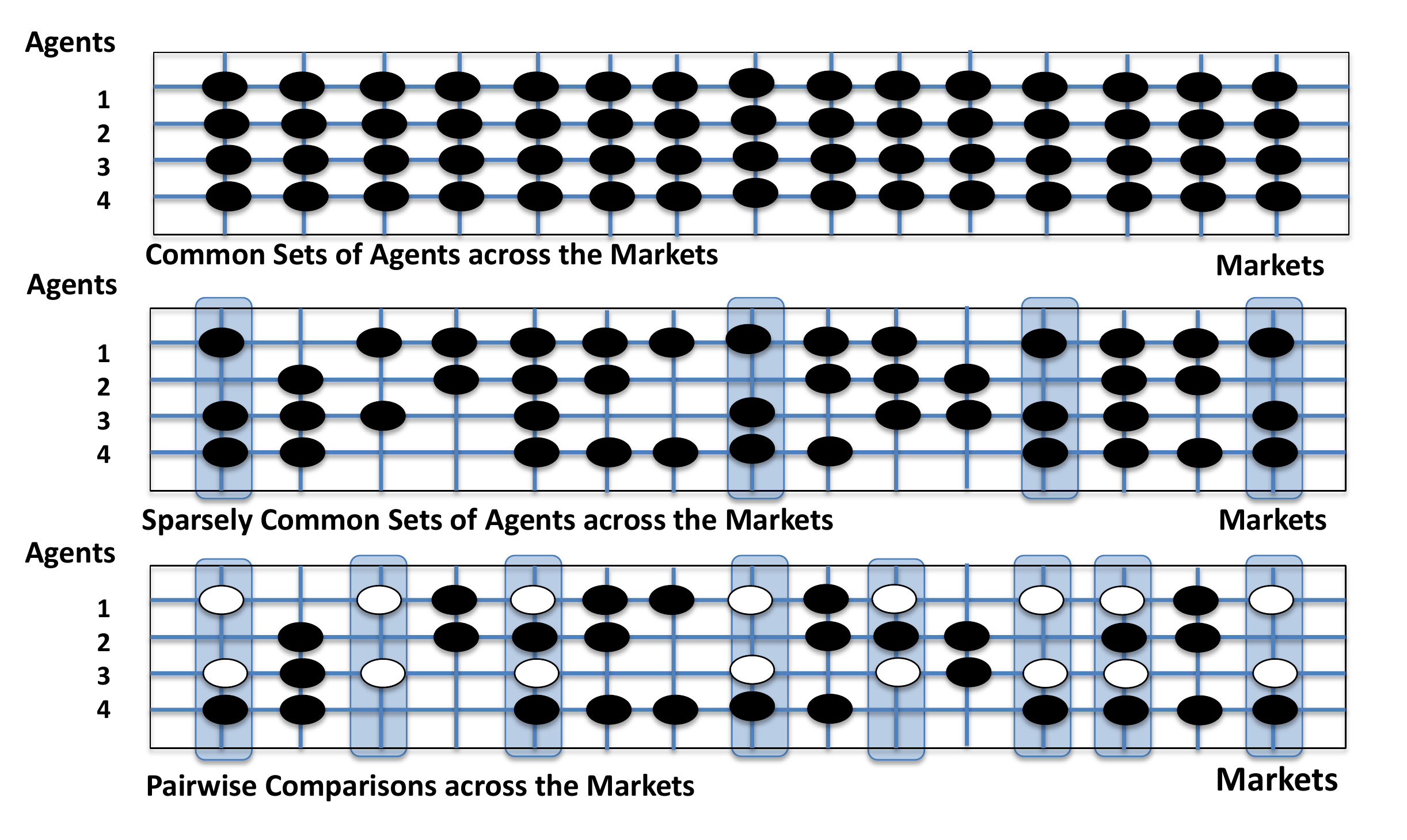}
		\caption{ \footnotesize The first panel shows an example of a data set where the set of participants is the same across all markets. The second and third panels show another example of data where only a small fraction of markets share the same set of participants. As illustrated here, there are only three markets where the set of participants is precisely $\{1,3,4\}$. The third panel shows that in the same data represented in the second panel, there are many more markets where both agents $1$ and $3$ (represented by white ellipses) participate. Thus a researcher can estimate population quantities that condition on the joint participation of the two agents, $1,3$, with better accuracy than quantities that condition on the whole set of participants or the triple $\{1,3,4\}$.
		}
	\label{Figure2}
	\end{center}
\end{figure}

We illustrate the advantage of using pairwise comparisons in Figure \ref{Figure2}. Each column in the figure symbolizes a ``market" and each row an individual agent. The ellipses in each column represent agents participating in a market. The first panel shows a standard set-up where all the agents appear in all the markets. The second panel shows an example of a data set where only very few markets share exactly the same set of participants $\{1,3,4\}$. As illustrated previously using real data, this feature of data is more realistic than the first panel. In this case, the conditional choice probability given the same set of agents simultaneously participating in the market cannot be accurately estimated. However, if we focus on only subsets with two agents $\{1,3\}$, there are many more markets in which the two agents participate. If each pair of agents appear in many markets simultaneously, we may aggregate over these markets, and infer accurately the ordering between the two agents using an inequality test. Given the p-values from inequality tests across pairs of agents, it remains to recover the whole group structure of the agents from these pairwise $p$-values. We develop an algorithm that recovers the group structure from the pairwise $p$-values consistently.

\section{Identification of the Ordered Group Structure}

Let us discuss conditions for the identification of the group structure. First, let $P$ be the distribution of observed random variables that belong to a market. (We assume that $P$ is the same across the markets.) We say that agents $(i,j)$ are \textbfit{comparable} if there exist pairwise indexes $\delta_{ij}$ and $\delta_{ij}^0$ such that the indexes are identified by $P$ and (\ref{general_setup}) holds. In this identification analysis, we assume that a researcher knows whether each pair of agents is comparable through some pairwise comparison index or not. The determination of such comparability can be done in practice by checking whether the data contains sufficiently many markets in which both $i$ and $j$ participate so that the pairwise indexes may be accurately estimated.

Let $\mathcal{E}$ be the collection of pairs $(i,j)$ that are comparable. We refer to comparable agents as \textbfit{adjacent}, so that the set $\mathcal{E}$ forms the set of edges in a graph on the set of agents $ N $. We call this graph (denoted by $G = (N,\mathcal{E})$) the \textbfit{comparability graph}.\footnote{In a graph (or network) $G=(N,\mathcal{E})$ the set $N$ represents the set of vertices (or nodes) and $\mathcal{E}$ consists of some pairs $ij$, with $i,j \in N$, where each pair $ij$ is called an edge (or link). 
Thus, if $(i,j) \in \mathcal{E}$, $i$ and $j$ are adjacent. A \textbfit{path} is a set of vertices $\{i_1,i_2,...,i_M\}$ such that $i_1i_2,i_2i_3,...i_{M-1}i_{M} \in \mathcal{E}$. 
Two vertices are called \textbfit{connected} if there is a path having $i$ and $j$ as end vertices. A graph is called \textbfit{connected} if all pairs of vertices are connected in the graph.} 
We say a group structure $\tau$ is \textbfit{identified} if it is uniquely determined once the comparability graph $G$ and the vectors of pairwise indexes $ (\delta_{ij},\delta_{ij}^0)_{ij \in \mathcal{E}}$ are known. Thus when $\tau$ is identified, it is only through the identification of the comparability graph $G$ and the pairwise indexes $ (\delta_{ij},\delta_{ij}^0)_{ij \in \mathcal{E}}$, not through other specification details of the structural model. 
 
\begin{figure}[t]
	\begin{center}
		\includegraphics[scale=0.4]{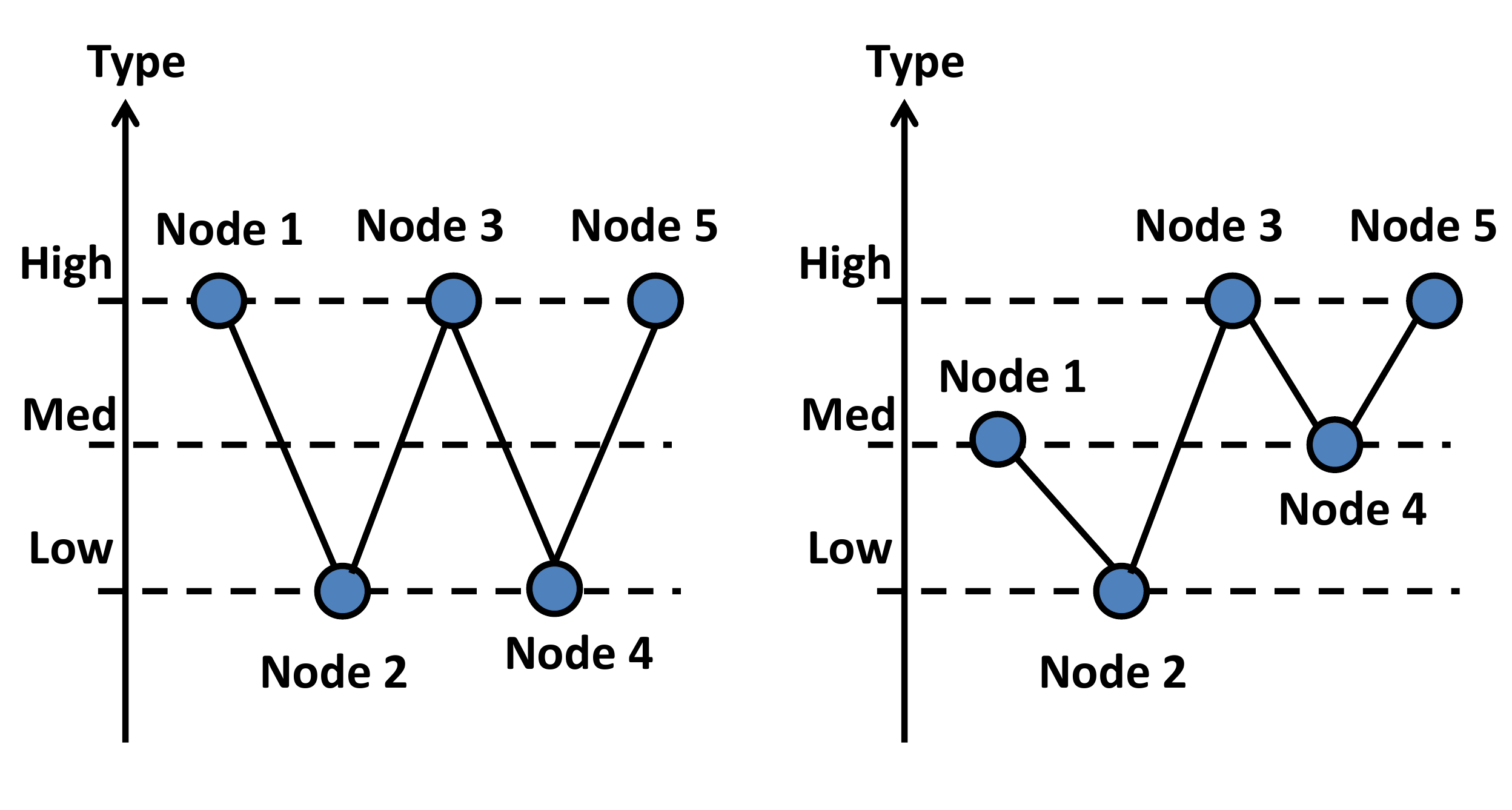}
		\caption{ \footnotesize This figure illustrates an example where the group structure is not identified even when all nodes are connected. The comparability graph is $G=(N,\mathcal{E})$, where $N=\{1,2,...,5\}$ and $\mathcal{E}
			=\{12,23,34,45\}$. 
		Pairwise comparison is feasible only between nodes linked by solid black lines (a.k.a. links).			
		The two different group structures in this figure are compatible with the same pairwise ordering.
		Therefore we cannot identify the group structure from pairwise orderings in this case.		
		}
		\label{fig:figure_illustration}
	\end{center}
\end{figure}
Let us explore the identification of $\tau$ given the comparability graph $G$ and the vector of pairwise indexes. 
It is easy to see that if $\mathcal{E}$ contains only a small subset of possible pairs, we may not be able to identify the group structure. The identification of the ordered group structure $\tau$ is not guaranteed even when many pairs of agents are comparable. For example, even if $G$ is a connected graph (where any two agents are connected at least indirectly), the ordered group structure $\tau$ may not be identified. This is illustrated in a counter-example in Figure \ref{fig:figure_illustration}. Certainly, when $G$ is a complete graph, i.e., every pair of agents are adjacent in the graph $G$,  the ordered group structure $\tau$ is identified.\footnote{
	If all pairs of agents are comparable, we can split the set of agents into one group with the lowest type and the other group with the remaining agents. Then we split these remaining agents into one group with the lowest type within these agents and the remaining agents. By continuing this process, we can identify the whole group structure.} 

Below we establish a necessary and sufficient condition for the group structure to be identified from a potentially incomplete graph $G$ and the pairwise comparison indexes. Let us introduce some definitions.

\begin{figure}[t]
	\begin{center}
		\includegraphics[scale=0.45]{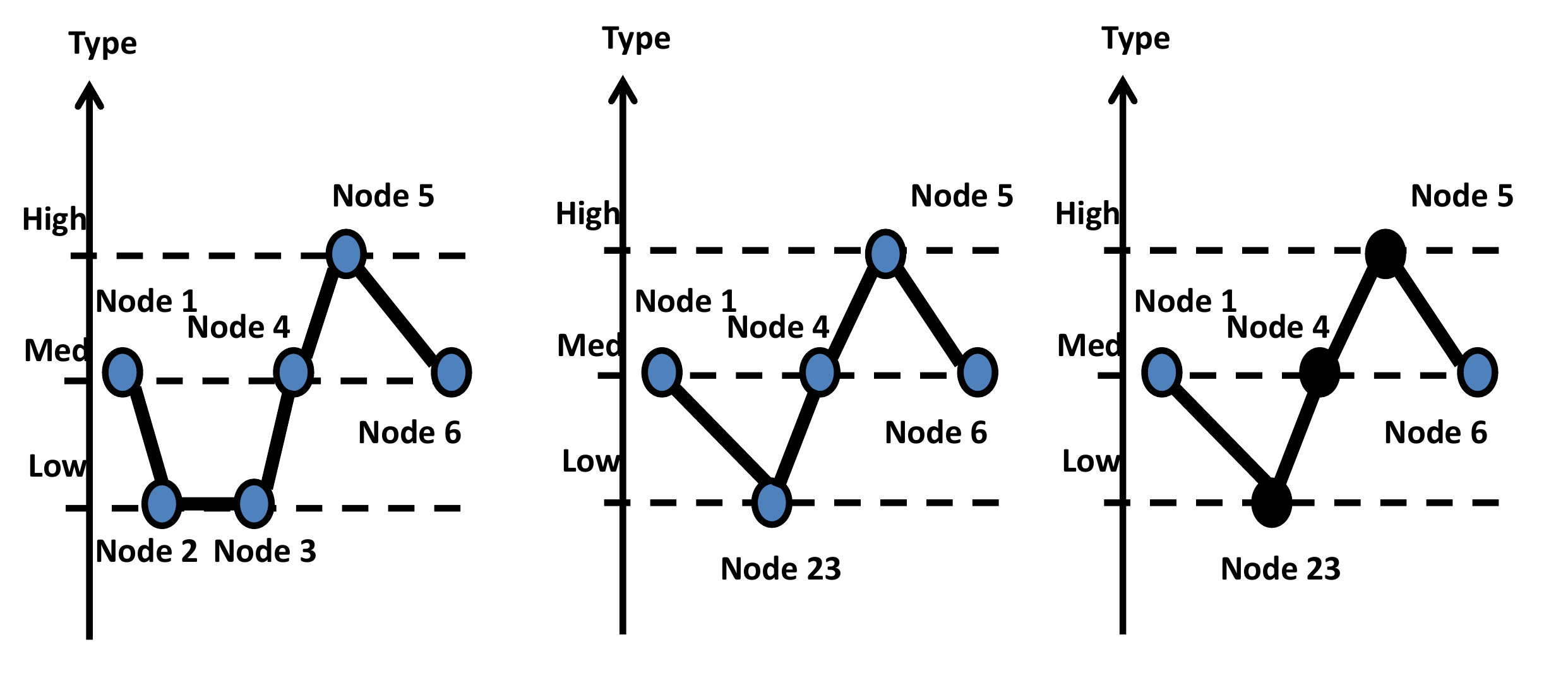}
		\caption{ \footnotesize This figure shows an example where the condition $N = N^*$ in Theorem 3.1 is violated. The first panel depicts the comparability graph as one connecting 6 vertices (or nodes). The second panel shows the $\tau$-collapsed graph where the two comparable nodes 2 and 3 that have the same type are collapsed into one node named 23. The last panel shows that Nodes 23, 4, and 5 (expressed as solid black nodes) are identified, because they are on a monotone path of length $K_0 - 1 = 2$. In this example, the group membership of Nodes 1 and 6 are not identified and thus the comparable graph does not lead to the identification of the group structure.}
		\label{Figure4}
	\end{center}
\end{figure}

\begin{definition}
	\noindent (i) A graph $G_{\tau}$ is the \textbfit{$\tau$-collapsed graph} of $G$ if (a) any two adjacent vertices $i$ and $j$ in $G$ with $\tau(i)=\tau(j)$ collapse to a single vertex (denoted by $(ij)$) in $G_{\tau}$, (b) any edge in $G$ joining a vertex $k$ to either $i$ or $j$ joins vertex $k$ to $(ij)$ in $G_{\tau}$ when $\tau(i) = \tau(j)$ and (c) all the remaining vertices and edges in $G_{\tau}$ consist of the remaining vertices and edges in $G$.\\	
	\noindent (ii) A path in $G_{\tau}$ is \textbfit{monotone} if $\tau(i)$ is monotone as $i$ runs along the path.\\
	\noindent (iii) A vertex $i$ is said to be \textbfit{identified} if its type $\tau(i)$ is identified.
\end{definition}

The $\tau$-collapsed graph of $ G $ is constructed by reducing any comparable pair of agents in $ G $ who have the same type to a single ``agent", and retaining edges as in the original graph of $ G $. 
Certainly, a $\tau$-collapsed graph $G_{\tau}$ is uniquely determined by $\delta_{ij}^0$'s and $G$. Any pair of adjacent agents in the $\tau$-collapsed graph must have different types, and hence the types of agents on a monotone path are strictly monotone. This means that every vertex on a monotone path in $G_{\tau}$ of length $K_0-1$ is identified.\footnote{The length of a path is defined as the number of the edges in the path.} Also by similar logic, every vertex on a monotone path with end vertices $i_H$ and $i_L$ is identified if the path has length $\tau(i_H) - \tau(i_L)$ and the end vertices $i_H$ and $i_L$ are identified. Using these two facts, we can recover the set of vertices that are identified as follows. 

First, let $N_{[1]} \subset N$ denote the set of vertices such that each vertex in the $\tau$-collapsed graph $G_\tau$ is on a monotone path in $G_{\tau}$ of length $K_0-1$. For $j \ge 1$ generally, let $N_{[j+1]}$ be the set of vertices each of which belongs to a  monotone path, say, $P$, such that its end vertices $i_H$ and $i_L$ are from $N_{[j]}$ and $\tau(i_H) - \tau(i_L)$ is equal to the length of the monotone path $P$. Then define
\begin{eqnarray*}
	N^* \equiv \bigcup_{j \ge 1} N_{[j]}.
\end{eqnarray*}
Given $G_\tau$, $N^*$ is uniquely determined as a subset of $N$. It is not hard to see that if $N = N^*$ and $K_0$ is identified, the type structure $\tau$ is identified. The following theorem shows that this condition is in fact necessary for the identification of $\tau$ as well. The proof of the theorem is given in the appendix.

\begin{theorem}
	\label{identification}
	Let $G$ be a given comparability graph and $G_{\tau}$ be its $\tau$-collapsed graph. The group structure $\tau$ is identified if and only if there exists a monotone path in $G_{\tau}$ whose length is equal to $K_0 - 1$ and $N = N^*$.
\end{theorem}

No monotone path in $G_{\tau}$ can have length greater than $K_0-1$. Note that there exists a monotone path in $G_{\tau}$ whose length is equal to $K_0 - 1$ if and only if $K_0$ is identified.\footnote{If there exists a monotone path in $G_\tau$ whose length is equal to $K_0 - 1$, then $K_0$ is identified, because through the comparability indexes, $\delta_{ij}$ and $\delta_{ij}^0$, we can identfy a longest monotone path and the length of this path should be $K_0 - 1$.} The conditions in the theorem are obviously satisfied if $G$ contains a monotone path that is monotone and covers all the vertices. The latter condition is trivially satisfied when $G$ is a complete graph. Figure \ref{Figure4} gives a counterexample where the condition that there exists a monotone path in $G_{\tau}$ whose length is equal to $K_0 - 1$ is satisfied, but $N \ne N^*$ so that the comparability graph does not lead to the identification of the group structure. 

\section{Consistent Estimation of the Ordered Group Structure}
\subsection{Pairwise Hypothesis Testing Problems}

In this section, we develop a method to estimate the group structure consistently for the case where the comparability graph is complete, so that we take $\mathcal{E}$ to be all $ij$ with $i,j \in N, i \ne j$. We first formulate three pairwise hypothesis testing problems for each comparable pair $ij \in \mathcal{E}$:
\begin{align}
H_{0,ij}^{+} &:\delta_{ij}\leq 0\text{ against }H_{1,ij}^{+}:\delta_{ij}>0\text{,}  \label{d} \\
H_{0,ij}^{0} &:\delta _{ij}^{0}=0\text{ against }H_{1,ij}^{0}:\delta_{ij}^{0}\neq 0\text{ and}  \notag \\
H_{0,ij}^{-} &:\delta_{ji}\leq 0\text{ against }H_{1,ij}^{-}:\delta_{ji}>0\text{.}  \notag
\end{align}
In most examples, we have various tests available. Instead of committing
ourselves to a particular method of hypothesis testing, let us assume
generally that we are given $p$-values $\hat{p}_{ij}^{+}$, $\hat{p}_{ij}^{0}$
and $\hat{p}_{ij}^{-}$ from the testing of $H_{0,ij}^{+}$, $H_{0,ij}^{0}$
and $H_{0,ij}^{-}$, against $H_{1,ij}^{+}$, $H_{1,ij}^{0}$ and $H_{1,ij}^{-}$
respectively. Let $L$ be the size of the sample (i.e., the number of the markets or games) that is used to construct these $p$-values. We will present conditions for the p-values later and explain how we construct $p$-values using bootstrap in Section 4.3.

\subsection{The Classification Method}

\subsubsection{The Selection-Split Algorithm}

Let us introduce a method of obtaining an ordered partition $(\hat N_1',\hat N_2')$ of a given subset $N' \subset N$ using $p$-values $\hat p_{ij}^s$, $s \in \{+,0,-\}$.

\begin{definition} For a subset $N' \subset N$, we say that the ordered partition of $N'$ into $(\hat N_1',\hat N_2')$ is obtained by \textbfit{the Split Algorithm} if it is obtained as follows. For each $i\in N'$, we let 
\begin{align*}
\hat N_{1}'(i) &= \{j\in N'\backslash \{i\}: \log \hat{p}%
_{ij}^{+}\le \log \hat{p}_{ij}^{-} - r_L \}\text{ and } \\
\hat N_{2}'(i) &= \{j\in N'\backslash \{i\}: \log \hat{p}%
_{ij}^{-}\le \log \hat{p}_{ij}^{+} - r_L \},
\end{align*}
where $r_L \rightarrow \infty$ satisfies Assumption \ref{assump: consistency2} below.\footnote{In many cases, it suffices to consider a sequence such that $r_L/\log L \rightarrow 0$. In practice, we propose $r_L = (\log L)^{1/3}$ which satisfies Assumption \ref{assump: consistency2} below under lower level regularity conditions. See Section C.3 in the supplemental note for details.} Set 
$i^*= \text{argmin}_{i\in N'}\min\{s_1(i),s_2(i)\}$, where
\begin{eqnarray*}
	s_1(i) =\frac{1}{|\hat N_1'(i)|}\sum_{j\in \hat N_1'(i)}\log \hat{p}_{ij}^{+}, \text{ and } s_2(i) =\frac{1}{|\hat N_2'(i)|}\sum_{j\in \hat N_2'(i)}\log \hat{p}_{ij}^{-}.
\end{eqnarray*} 
(We set $s_1(i) = 0$ if $\hat N_1'(i)$ is empty, and similarly with $s_2(i)$.) Then we take
\begin{eqnarray*}
   (\hat N_1',\hat N_2') &=& (\hat N_1'(i^*), N' \setminus \hat N_1'(i^*)), \text{ if } s_1(i^*) \le s_2(i^*);\\
   (\hat N_1',\hat N_2') &=& (N' \setminus \hat N_2'(i^*),\hat N_2'(i^*)), \text{ if } s_1(i^*) > s_2(i^*).
\end{eqnarray*}
\end{definition}

The set $\hat N_{1}'(i)$ estimates the set of agents of lower type than $i$, and the set $\hat N_{2}'(i)$ estimates the set of agents of higher type than $i$. Let
\begin{eqnarray*}
	N_1'(i) = \{j \in N'\setminus\{i\}: \tau(i) > \tau(j) \}, \text{ and } N_2'(i) = \{j \in N'\setminus\{i\}: \tau(i) < \tau(j) \}.
\end{eqnarray*}
A necessary condition for $\hat N_1'(i)$ to coincide with $N_1'(i)$ is that $i$ has higher type than those in $\hat N_1'(i)$. The more negative the quantity $s_1(i)$ is, the more likely that this necessary condition is met. A similar observation applies to $s_2(i)$ as well. Thus we choose a partition based on $i^*$ that minimizes $\min\{s_1(i),s_2(i)\}$ over $i$.

Suppose that we are given an ordered partition $(\hat N_1',...,\hat N_s')$ of $N$. The Selection-Split Algorithm that we propose produces an ordered partition $(\hat N_1'',...,\hat N_{s+1}'')$ of $N$ from $(\hat N_1',...,\hat N_s')$ using two steps, the Selection Step and the Split Step, as follows.\medskip

\textbfit{1. The Selection Step}: Let $\hat p_k = \min_{i,j \in \hat N_k': i \ne j}\hat p_{ij}^0$, $k=1,...,s$, and select $\hat N_{k^*}'$ with $k^*$ such that
\begin{align*}
	\hat p_{k^*} = \min_{1 \le k \le s} \hat p_k.
\end{align*}

\textbfit{2. The Split Step}: We split $\hat N_{k^*}'$ into $(\hat N_{k^*,1}',\hat N_{k^*,2}')$ using the Split Algorithm, and relabel the partition: $(\hat N_1',...,\hat N_{k^*-1}',\hat N_{k^*,1}',\hat N_{k^*,2}',\hat N_{k^*+1}',...,\hat N_s') = (\hat N_1'',...,\hat N_{s+1}'')$.
\medskip

The Selection Step chooses a group $\hat N_{k^*}'$ that is most likely to contain agents with heterogeneous types and the Split Step splits this group into two sets using the Split Algorithm. The Selection-Split algorithm depends on the data only through the p-values $\hat p_{ij}^s$, $s \in \{+,0,-\}$.

\subsubsection{The Classification Method}
For a given positive integer $ K $, partition $ N $ into $ K $ groups as follows.
First, split $N$ into $(\hat N_1^{[2]},\hat N_2^{[2]})$ using the Split Algorithm to $N$, and apply the Selection-Split Algorithm sequentially to obtain $(\hat N_1^{[3]},\hat N_2^{[3]},\hat N_3^{[3]})$, $(\hat N_1^{[4]},...,\hat N_4^{[4]})$, and so on, until we have $(\hat N_1^{[K]},...,\hat N_K^{[K]})$ for a given number $K$. For each $ K $, we define
\begin{equation*}
\hat V(K)=\frac{1}{K}\sum_{k=1}^{K}\left\vert \min_{i,j\in \hat N_{k}^{[K]}} \log \hat{p}_{ij}^{0}\right\vert,
\end{equation*}
and then select 
\begin{equation*}
\hat{K}=\text{argmin}_{1 \le K \le n} \hat V(K)+Kg(L),
\end{equation*}
where $g(L)$ is positive and slowly increasing in $L$.\footnote{The choice of $g(L)=\log \log L$ appears to work very well from our numerous Monte Carlo simulation experiments.} We take
\begin{align}
	\label{hat T hat K}
	\hat T_{\hat K} = (\hat N_1^{[\hat K]},...,\hat N_{\hat K}^{[\hat K]})
\end{align}
to be our estimated group structure. The component $\hat V(K)$ measures the goodness-of-fit of the classification, and the second component $Kg(L)$ represents a penalty term that prevents overfitting. 
We show that $\hat T_{\hat K}$ is consistent for the underlying group structure $T$ defined in (\ref{def T}) under regularity conditions.

\subsection{Constructing $p$-Values Using Bootstrap}
\label{bootstr}

In most applications, we can use bootstrap to construct $p$-values for
testing the inequality restrictions of (\ref{d}).\footnote{%
	See \citet{Bugni2010}, %
	\citet{AndrewsShi2013}, \citet{ChernozhukovLeeRosen2013}, %
	\citet{LeeSongWhang2013}, and \citet{LeeSongWhang2018}, among many others,
	and references therein.} For the sake of concreteness, we explain the bootstrap procedure along the proposal made by \citet{LeeSongWhang2018}. Suppose that we are given observations $\{Z_{\ell}\}_{\ell = 1}^{L}$,
where $Z_{\ell}=(Z_{i,\ell})_{i=1}^{n}$ denotes the observations pertaining to
market $\ell$ and $Z_{i,\ell}$ denotes the vector of observations specific to agent $i$.
Suppose that for each pair of agents $i$ and $j$, there exists a nonparametric
function, say, $r_{ij}(x)$ such that $\tau(i) \ge \tau(j)$ if and only if $r_{ij}(x) \ge 0$ for all $x \in \mathcal{X}$, where $\mathcal{X}$ is the common domain of the function $r_{ij}(\cdot)$, $i,j \in N$.

To construct a test statistic, we first estimate $r_{ij}(x)$ using the
sample $\{Z_{\ell}\}_{\ell = 1}^{L}$ to obtain $\hat{r}_{ij}(x)$ (e.g., using a kernel regression estimator). Then we construct
the following indexes:%
\begin{eqnarray}
\label{delta hats}
\hat \delta_{ij} =\int \max \left\{\hat{r}_{ij}(x),0\right\} dx \text{ and } \hat \delta_{ij}^{0} = \int \left\vert \hat{r}_{ij}(x)\right\vert dx.
\end{eqnarray}
For $p$-values, we re-sample $\{Z_{\ell}^*\}_{\ell = 1}^{L}$ (with replacement)
from the empirical distribution of $\{Z_{\ell}\}_{\ell = 1}^{L}$ and construct a
nonparametric estimator $\hat{r}_{ij}^*(x)$ for each pair $(i,j)$ in
the same way as we did using the original sample. Using these bootstrap
estimators, we construct the following bootstrap test statistics:%
\begin{eqnarray}
\label{delta stars}
\hat \delta_{ij}^{\ast } =\int \max \left\{ \hat{r}_{ij}^*(x)-\hat{r}%
_{ij}(x),0\right\} dx \text{ and  } \hat \delta_{ij}^{0\ast } &=&\int \left\vert \hat{r}_{ij}^*(x)-\hat{r}%
_{ij}(x)\right\vert dx.  \notag
\end{eqnarray}%
Note that the bootstrap test statistic involves recentering to impose the
null hypothesis. Now, the $p$-values, $\hat{p}_{ij}^{+}$, $\hat{p}_{ij}^{-},$
and $\hat{p}_{ij}^{0}$ can be constructed from the bootstrap distributions
of $\hat \delta_{ij}^{\ast }$, $\hat \delta_{ji}^{\ast }$, and $\hat \delta_{ij}^{0\ast }$ respectively, using $\hat \delta_{ij}, \hat \delta_{ji}$ and $\hat \delta_{ij}^{0}$ as test statistics.

\subsection{Consistency of Classification}
We prove consistency of the estimated classification $\hat T_{\hat K}$ as $L\rightarrow \infty$ while $n$ fixed. (Consistency results and the proof for the case of both $n$ and $L$ increasing to infinity are found in the supplemental note.) Let $\mathcal{P}$ be the collection of the distributions $P$ of the whole vector of the observations in each market $\ell$. For each $\varepsilon>0$, $ij \in \mathcal{E}$ and $s \in \{+,0,-\}$, we define
\begin{eqnarray*}
	\mathcal{P}_{0,ij}^s = \{P \in \mathcal{P}: \delta_{ij}^s(P) \le 0 \}, \text{ and } 
	\mathcal{P}_{\varepsilon,ij}^s = \{P \in \mathcal{P}: \delta_{ij}^s(P) \ge \varepsilon \},
\end{eqnarray*}
where we write the pairwise indexes $\delta_{ij}^s$ as $\delta_{ij}^s(P)$ to reflect that the pairwise indexes depend on $P$. Thus $\mathcal{P}_{0,ij}^s$ is the collection of probabilities under the pairwise null hypothesis $H_{0,ij}^s$, and $\mathcal{P}_{\varepsilon,ij}^s$ is the collection of probabilities under the pairwise alternative hypotheses $H_{1,ij}^s$ such that $\delta_{ij}^s(P)$ is away from zero at least by $\varepsilon$. Then we define
\begin{eqnarray*}
	\mathcal{P}_{0,\varepsilon} = \bigcup_{s \in \{+,0,-\}} \bigcup_{ij \in \mathcal{E}} (\mathcal{P}_{0,ij}^s \cup \mathcal{P}_{\varepsilon,ij}^s).
\end{eqnarray*}

We assume that the $p$-value takes the following form:
\begin{eqnarray*}
	\hat p_{ij}^s = 1 - \tilde F_{ij}^s(\tilde T_{ij}^s), s \in \{+,0,-\},
\end{eqnarray*}
where $\tilde F_{ij}^s$ is a CDF and $\tilde T_{ij}^s$ is a random variable both of which depend on the data. Typically, $\tilde T_{ij}^s$ represents an appropriately normalized test statistic and $\tilde F_{ij}^s$ represents the CDF of the bootstrap distribution of the test statistic after recentering. We make the following assumption.
\begin{assumption}
	\label{assump: consistency2}
	There exist sequences $\lambda_L \rightarrow \infty$ and $\rho_L \rightarrow 0$ and constants $c_{ij}^s$, $s \in \{+,0,-\}$  such that along each sequence of probabilities $P_L \in \mathcal{P}_{0,\varepsilon}$ and for each pair $i,j \in N$, the following holds for all $s \in \{+,0,-\}$, as $L \rightarrow \infty$.
	
	(i) If $\tau(i) = \tau(j)$, $\tilde T_{ij}^s \rightarrow_d W_{ij}^s,$ for some random variable $W_{ij}^s$.
	
	(ii) If $\tau(i) > \tau(j)$, $\tilde T_{ij}^+/\lambda_L \rightarrow_P c_{ij}^+$ and $\tilde T_{ij}^- = O_P(1)$. 
	
	(iii) If $\tau(i) < \tau(j)$, $\tilde T_{ij}^-/\lambda_L \rightarrow_P c_{ij}^-$ and $\tilde T_{ij}^+ = O_P(1)$. 
	
	(iv) If $\tau(i) \ne \tau(j)$, $\tilde T_{ij}^0/\lambda_L \rightarrow_P c_{ij}^0$. 
	
	(v) $\sup_{t \in \mathbf{R}} |\tilde F_{ij}^s(t) - F_{ij,\infty}^s(t) | = O_P(\rho_L)$, where $F_{ij,\infty}^s$ is the CDF of $W_{ij}^s$.
	
	(vi) $r_L^{-1} \log(1 - F_{ij,\infty}^s(c_1 \lambda_L) + c_2 \rho_L) \rightarrow - \infty$, for all constants $c_1,c_2>0$.
	
\end{assumption}
Here we are assuming that for each pair of agents we have enough observations on markets in which the pair of agents participate so that we can perform consistent tests based on pairwise comparison. Assumption \ref{assump: consistency2} is a high level assumption and is typically satisfied for various choices of test statistics that arise in the literature of moment inequality testing. We provide lower level conditions in the case of nonparametric tests based on \citet{LeeSongWhang2018} in the next subsection.

\begin{theorem}
	\label{thm: consistency 2}
	Suppose that Assumption \ref{assump: consistency2} holds, and that $g(L) \rightarrow \infty$ and $g(L)/r_L \rightarrow 0$ as $L\rightarrow \infty$. Then, for any $\varepsilon>0$, along a sequence of probabilities $P_L$ from $\mathcal{P}_{0,\varepsilon}$, 
	\begin{equation*}
	P_L\{\hat{K}=K_{0}\}\rightarrow 1, \text{ as } L \rightarrow \infty,
	\end{equation*}
	and the estimated group structure $\hat{T}_{\hat{K}}$ in (\ref{hat T hat K}) satisfies that as $L\rightarrow \infty,$
	\begin{equation*}
	P_L \{ \hat{T}_{\hat{K}} = T \} \rightarrow 1.
	\end{equation*}
\end{theorem}

The proof of Theorem \ref{thm: consistency 2} is in the supplemental note. It proceeds in two steps. First, we show that $\hat T_{K_0}$ is consistent for $T$. Second, we show that $\hat{K}$ is consistent for $K_{0} $. To see the intuition for this second step, note that when $K\ge K_{0}$, the component $\hat V(K)$ is $O_P(1)$, and when $K < K_{0}$, the component $\hat V(K)$ diverges at a rate faster than $g(L)$. From this, we obtain that $\hat{K}$ is consistent for $K_{0}$.

\subsection{Lower Level Conditions for Assumption \ref{assump: consistency2}}

Lower level conditions for Assumption \ref{assump: consistency2} can be found from the literature of testing for moment inequality restrictions. For the sake of concreteness, we focus on the situation where nonparametric function $r_{ij}(x)$ introduced in Section \ref{bootstr} arises from difference between two nonparametric regression functions and the testing procedure is done by the method proposed in \cite{LeeSongWhang2018}. Suppose that we have
\begin{align}
\label{general_setup2}
m_i(x) > m_j(x), \forall x \in \mathcal{X} & \textnormal{ if and only if } \tau (i)>\tau (j); \\ \notag
m_i(x) = m_j(x), \forall x \in \mathcal{X} & \textnormal{ if and only if } \tau (i)=\tau (j); \\\notag
m_i(x) < m_j(x), \forall x \in \mathcal{X} & \textnormal{ if and only if }\tau (i)<\tau (j),
\end{align}
where $m_i(x) = \mathbf{E}[Y_{i,\ell}|X_{i,\ell} = x], i \in N$, and $\ell = 1,...,L$, is the sample unit index. We take $r_{ij}(x) = m_i(x) - m_j(x)$. Let us define a kernel estimator of $m_i(x)$ as follows:
\begin{eqnarray*}
	\hat m_i(x) = \frac{\displaystyle \sum_{\ell = 1}^L Y_{i,\ell} K_h(X_{i,\ell} - x)}{\displaystyle \sum_{\ell = 1}^L K_h(X_{i,\ell} - x)},
\end{eqnarray*}
where $K_h(x) = K(x/h)/h$, and $K(\cdot)$ is a multivariate kernel and $h$ is a bandwidth. We let
\begin{eqnarray*}
	\hat r_{ij}(x) = \hat m_i(x) - \hat m_j(x).
\end{eqnarray*}
Then the test statistics we use are defined as
\begin{align}
\hat \delta_{ij}^+ &= \int_\mathcal{X} \max\{\hat r_{ij}(x),0\} dx, \quad \hat \delta_{ij}^- = \int_\mathcal{X} \max\{\hat r_{ji}(x),0\} dx, \text{ and }\\ \notag
\hat \delta_{ij}^0 &= \int_\mathcal{X} |\hat r_{ij}(x)| dx.
\end{align}
As for the bootstrap test statistics, we first obtain bootstrap samples $\{(Y_{i,\ell}^*,X_{i,\ell}^*)_{i \in N}\}_{\ell = 1}^L$ by resampling the vector $(Y_{i,\ell}^*,X_{i,\ell}^*)_{i \in N}$ from the empirical distribution of $\{(Y_{i,\ell},X_{i,\ell})_{i \in N}\}_{\ell = 1}^L$ with replacement. Using the bootstrap sample, we construct
\begin{eqnarray*}
	\hat r_{ij}^*(x) = \hat m_i^*(x) - \hat m_j^*(x),
\end{eqnarray*}
where
\begin{eqnarray*}
	\hat m_i^*(x) = \frac{\displaystyle \sum_{\ell = 1}^L Y_{i,\ell}^* K_h(X_{i,\ell}^* - x)}{\displaystyle \sum_{\ell = 1}^L K_h(X_{i,\ell}^* - x)}.
\end{eqnarray*}
Then the bootstrap test statistics we use are defined as\footnote{One could also use alternative bootstrap statistics using estimated contact sets as in \cite{LeeSongWhang2018} to enhance the power. For simplicity of exposition, here we present the case where we use the least favorable configurations.}
\begin{align}
\hat \delta_{ij}^{+*} &= \int_\mathcal{X} \max\{\hat r_{ij}^*(x) - \hat r_{ij}(x),0\} dx, \quad \hat \delta_{ij}^{-*} = \int_\mathcal{X} \max\{\hat r_{ji}^*(x) - \hat r_{ji}(x),0\} dx, \text{ and }\\ \notag
\hat \delta_{ij}^{0*} &= \int_\mathcal{X} |\hat r_{ij}^*(x) - \hat r_{ij}(x)| dx.
\end{align}
Let the CDF of the bootstrap distribution of $\hat \delta_{ij}^{s*}$ be denoted by $F_{ij}^s$. Then we set the pairwise $p$-value to be $\hat p_{ij}^s = 1 - F_{ij}^s(\hat \delta_{ij}^s)$. In this situation, we provide a low level condition for Assumption \ref{assump: consistency2}.

Let $a_{ij,L}^s$ and $\sigma_{ij,L}^s$ be sequences of constants such that
\begin{eqnarray}
\label{seqs}
a_{ij,L}^s = O(1), \text{ and } \sigma_{ij,L}^s \rightarrow \sigma_{ij}^s >0,
\end{eqnarray}
as $n,L\rightarrow \infty$. We take 
\begin{align}
\tilde T_{ij}^s &= (\sqrt{L}\hat \delta_{ij}^s - h^{-d/2} a_{ij,L}^s)/\sigma_{ij,L}^s, \text{ and }\\ \notag
\tilde T_{ij}^{s*} &= (\sqrt{L}\hat \delta_{ij}^{s*} - h^{-d/2} a_{ij,L}^s)/\sigma_{ij,L}^s.
\end{align}
(The researcher does not need to know, estimate, or use the constants $a_{ij,L}^s$ and $\sigma_{ij,L}^s$ for the  construction of the pairwise $p$-values and for the implementation of the classification algorithm of this paper.) Then we can rewrite
\begin{eqnarray*}
	\hat p_{ij}^s = 1 - \tilde F_{ij}^s(\tilde T_{ij}^s),
\end{eqnarray*}
where $\tilde F_{ij}^s$ is the CDF of the bootstrap distribution of $\tilde T_{ij}^{s*}$. Let $\|\cdot\|_\infty$ denote the sup norm, i.e., $\|f\|_\infty = \sup_x |f(x)|$ for any real function $f$. Let us consider the following set of assumptions. 
\begin{assumption}
	\label{assump: lower level}
	(i) For each $s \in \{+,0,-\}$ and each $i,j 
	\in N$, there exist sequences of constants $a_{ij,L}^s$ and $\sigma_{ij,L}^s$ such that the conditions in (\ref{seqs}) hold, and under $H_{0,ij}^0$ in (\ref{d}),
	\begin{eqnarray*}
		\tilde T_{ij}^s \rightarrow_d N(0,1),
	\end{eqnarray*}
	and $\|\tilde F_{ij}^s - \Phi\|_\infty = O_P(L^{-\alpha})$ for some $\alpha>0$, and $\Phi$ is the CDF of $N(0,1)$.
	
	(ii) $\sup_{x \in \mathcal{X}}|\hat r_{ij}(x) - r_{ij}(x)| = o_P(1)$, as $L\rightarrow \infty$.
	
	(iii) Suppose that $L h^d \rightarrow \infty$ while $h \rightarrow 0$ as $L\rightarrow \infty$.
\end{assumption}

The lower level conditions for Condition (i) can be found in \cite{LeeSongWhang2018}. Condition (ii) follows if the kernel regression estimators $\hat g_i(x)$ are uniformly consistent. (See, e.g., \cite{Hansen2008}.) Condition (iii) is a standard bandwidth condition in the literature of kernel estimators. Then, we obtain the following lemma.

\begin{lemma}
	\label{lemm: suff cond}
	Suppose that Assumption \ref{assump: lower level} holds, and that the sequence $r_L \rightarrow \infty$ is such that $r_L/\log L \rightarrow 0$ as $L\rightarrow \infty$. Then Assumption \ref{assump: consistency2} holds.
\end{lemma}

The proof of this lemma is given in the appendix.

\subsection{Two-Step Estimation Using the Estimated Group Structure}

The estimated group structure can be used as a first-step estimator in a two-step procedure for estimating a structural parameter. Recall that $\tau:N \rightarrow \{1,...,K_0\}$ defines the group structure. Let us denote $\tau(\cdot;P)$ to indicate that the group structure is identified. Suppose that $\theta_0$ is a structural parameter that is identified as follows: $Q(\overline \theta, \tau(\cdot;P);P)$ as a function of $\overline \theta$ has a unique minimizer in a parameter space $\Theta$ and
\begin{eqnarray*}
	\theta_0 = \arg \min_{\overline \theta \in \Theta} Q(\overline \theta, \tau(\cdot;P);P).
\end{eqnarray*}
In many applications $Q(\overline \theta, \tau(\cdot;P);P)$ is a population objective function that arises from Generalized Method of Moment (GMM) estimation or Maximum Likelihood Estimation (MLE). The two step estimator $\hat \theta$ of $\theta_0$ is defined as
\begin{eqnarray*}
	\hat \theta = \arg \min_{\overline \theta \in \Theta} \hat Q(\overline \theta, \hat \tau(\cdot)),
\end{eqnarray*}
where $\hat \tau(\cdot)$ is the first-step estimator of the group structure that is consistent, i.e., for all $\varepsilon>0$,
\begin{eqnarray}
\label{cons}
	P\left\{\max_{1 \le i \le n}|\hat \tau(i) - \tau(i;P)|> \varepsilon\right\} \rightarrow 0,
\end{eqnarray}
as $L \rightarrow \infty$. As we saw before, one can obtain such estimator $\hat \tau$ using our Select-Split algorithm. Let $\tilde \theta$ be such that
\begin{eqnarray*}
	\tilde \theta = \arg \min_{\overline \theta \in \Theta} \hat Q(\overline \theta, \tau(\cdot;P)),
\end{eqnarray*}
so that $\tilde \theta$ is an infeasible estimator when one uses the true group structure $\tau(\cdot;P)$ rather than the estimated version $\hat \tau(\cdot)$. The asymptotic normality of $\sqrt{L}(\tilde \theta - \theta_0)$ can be derived using the standard arguments, for example, using the general results in \cite{NeweyMcFadden1994}. Then it is not hard to see that $\sqrt{L}(\tilde \theta - \theta_0)$ has the same limit distribution. To see this, by taking $\varepsilon<1$, we find from (\ref{cons}) that
\begin{eqnarray*}
	P\left\{\max_{1 \le i \le n}|\hat \tau(i) - \tau(i;P)| > 0 \right\} \rightarrow 0.
\end{eqnarray*}
Therefore,
\begin{eqnarray*}
	P\{\hat \theta \ne \tilde \theta \} &=& P\left\{\arg\min_{\overline \theta \in \Theta}\hat Q(\overline \theta, \hat \tau(\cdot)) \ne \arg\min_{\overline \theta \in \Theta} \hat Q(\overline \theta, \tau(\cdot;P))\right\}\\
	&\le& P\left\{\max_{1 \le i \le n}|\hat \tau(i) - \tau(i;P)| > 0 \right\} \rightarrow 0,
\end{eqnarray*}
as $L \rightarrow \infty$. Hence $\hat \theta = \tilde \theta$ with probability approaching one, as $L \rightarrow \infty$. This means that if the inference based on $\tilde \theta$ is asymptotically valid, so is that based on $\hat \theta$.

Note that this asymptotic validity holds pointwise in $P$. Given the known failure of uniform validity (uniform in $P$) for post-model selection inference (\citet{LeebPotcher2005}), it is likely that the inference based on this two-step estimator $\hat \theta$ fails to satisfy asymptotic validity uniformly in $P$, unless one modifies the procedure appropriately. In this general set-up, it is far from trivial to find such a modification. We leave it to future research.

\section{Monte Carlo Simulations}

\subsection{Finite Sample Performance of the Classification}
We use a model of a first-price procurement auction with asymmetric independent private costs to study performance of our classification procedure. (See Appendix B.1 of the supplemental note.) 
Bidders are classified into $K_{0}\ $groups. We abstract away from the formation of equilibrium strategies, and draw bids from a normal distribution $N(\mu _{k},\sigma ^{2})$ for each group $k=1,...,K_0$.
Let $L$ denote the number of auctions in which any given pair of bidders participate. 
We consider two specifications of $\mu_k$'s. In one specification, $\mu_1=2.0,\mu_2=2.6,\mu_3=3.2,$ and $\mu_4=3.8$ with increment $D_\mu=0.6$, and in the other specification,  $\mu_1=2.0,\mu_2=2.2,\mu_3=2.4,$ and $\mu_4=2.6$ with increment $D_\mu=0.2$. The variance $\sigma^2$ is taken to be $0.25$.

Table 1 summarizes the designs of group structures in our simulation. 
The first two structures involve a total of 12 bidders and the last two 40 bidders. 
The first and third are designed to be coarser group structures than the second and fourth respectively. 
We construct $p$-values using the procedure in Section \ref{bootstr} and obtain group classification from 500 simulated samples. 
For each estimate, we used 200 bootstrap iterations to calculate $p$-values.
\begin{table}[t]
	\small 
	\begin{center}
		Table 1: Group Structure in Experiments\medskip
		\par
		\begin{tabular}{cccc}
			\hline\hline
			{\small Structure} & ${\small \ \ \ \ }n{\small \ \ \ \ }$ & ${\small \ \ \
				\ K}_{0}{\small \ \ \ \ }$ & ${\small \ \ \ \ n}_{k}{\small \ \ \ \ }$ \\ 
			\hline
			{\small S1} & {\small 12} & {\small 2} & {\small 6} \\ 
			{\small S2} & {\small 12} & {\small 4} & {\small 3} \\ 
			{\small S3} & {\small 40} & {\small 2} & {\small 20} \\ 
			{\small S4} & {\small 40} & {\small 4} & {\small 10} \\ \hline\hline
		\end{tabular}%
	\end{center}
	\par
	\medskip 
	\begin{flushleft}
		\noindent {\footnotesize Note:} $n$ {\small \ denotes the total number of
			the bidders; }$K_{0}${\small \ denotes the number of the groups; }$n_{k}$%
		{\small \ denotes the number of actual bidders from group }${\small k}$%
		{\small . For each structure in the simulation design, groups all have the
			same number of bidders.}
	\end{flushleft}
\end{table}

To evaluate the performance of our classification method, we define a measure of discrepancy between two ordered partitions $T_1$ and $T_2$:
\begin{eqnarray}
\label{delta}
\delta \left( T_{1},T_{2}\right) &=& \frac{1}{K_1}\sum_{k=1}^{K_1} \min_{1 \le j \le K_2 } |N_k^1 \triangle N_j^2|,
\end{eqnarray}
where $T_{1}=(N_{1}^{1},...,N_{K_1}^{1})$ and $T_{2}=(N_{1}^{2},...,N_{K_2}^{2})$ are ordered partitions of $N$ and $\triangle$ denotes set-difference: $A \triangle B = (A\setminus B) \cup (B \setminus A)$.
We evaluate our classification method using two criterion: (1) Expected Average Discrepancy (EAD) defined as $ \mathbf{E} (\delta (T,\hat{T}_{\hat{K}}) ) $ and (2) HAD$(\lambda) \equiv P\{\delta (T,\hat{T}_{\hat{K}})>\lambda n\} $ for $0 < \lambda < 1$.

Table 2 reports estimates when there is no unobserved heterogeneity among bidders ($ K_0 = 1 $). In this case, our procedure detects the absence of unobserved heterogeneity effectively. 
For a given $n$, there is a moderate increase in the accuracy of classification as $L$ increases, both in terms of EAD and HAD($\lambda$).
\fontsize{11}{13}\selectfont 
\begin{table}[t]
	\small
	\begin{center}
		Table 2: Performance of the Classification with One Group ($K_{0} = 1$ and unknown)
		\par
		\medskip
		\par
		\begin{tabular}{ccc|cccccc}
			\hline\hline
			\small
			& $n$	&	$L$	&	$\hat K_0$	&	EAD	&	HAD(.10)	&	HAD(.25)	&	HAD(.50)	&	\\ 
			\cline{2-8} 
			& 12	&	400	&	1.002	&	0.012	&	0.001	&	0.000	&0.000	&\\
			&	12	&	200	&	1.003	&	0.014	&	0.002	&	0.000	&0.000	&\\
			&	12	&	100	&	1.003	&	0.018	&	0.002	&	0.001	&0.000 &\\
			\cline{2-8} 
			&	40	&	400	&	1.003	&	0.082	&	0.005	&	0.003	&	0	&\\
			&	40	&	200	&	1.006	&	0.084	&	0.008	&	0.002	&	0	&\\
			&	40	&	100	&	1.008	&	0.096	&	0.010	&	0.004	&	0	&\\
			\hline\hline
		\end{tabular}
	\end{center}
	\par
	\medskip 
	\begin{flushleft}
		\noindent {\footnotesize Note: $n$ is the number of bidders in data; and $L$ the number of markets. $\hat K_0$ is the average number of estimated groups in 500 simulation samples. EAD is the average number of mismatched bidders across true groups and simulated samples. HAD($\lambda$) is the hazard rate of average discrepancy. For example, HAD(.10) = 0.002 means that in 499 simulated samples (out of a total of 500) the average number of mismatched bidders is less than 10 percent of the total number of bidders.}
	\end{flushleft}
\end{table}

\fontsize{11}{13}\selectfont 
\begin{table}[t]
	\small 
	\begin{center}
		Table 3: Performance of the Classification with Multiple Groups ($K_{0} \ge 2$ and unknown)
		\par
		\medskip
		\par
		\begin{tabular}{ccc|cccc|ccccc}
			\hline\hline
			&  &  & \multicolumn{4}{|c|}{$K_{0}=2$} & \multicolumn{4}{|c}{$K_{0}=4$} \\ 
			$n$ & $L$ & $D_{\mu }$ & $\hat{K}_{0}$ & EAD & HAD(.25) & HAD(.75) & $\hat{%
				K}_{0}$ & EAD & HAD(.25) & HAD(.75) \\ \hline
			12	&	400	&	0.6	&	2.00	&	0.00	&	0.00	&	0.000	&	3.96	&	0.03	&	0.01	&	 0.00\\ 	
			12	&	400	&	0.2	&	2.00	&	0.01	&	0.00	&	0.000	&	3.94	&	0.04	&	0.03	&	 0.00 \\ 	
			12	&	100	&	0.6	&	2.00	&	0.00	&	0.00	&	0.000	&	3.98	&	0.01	&	0.01	&	 0.00 \\  	
			12	&	100	&	0.2	&	2.03	&	0.52	&	0.07	&	0.004	&	3.24	&	1.53	&	0.24	&	 0.00 \\ \hline	
			40	&	400	&	0.6	&	2.00	&	0.01	&	0.00	&	0.000	&	3.97	&	0.08	&	0.02	&	 0.00 \\	
			40	&	400	&	0.2	&	2.01	&	0.01	&	0.00	&	0.000	&	3.83	&	0.43	&	0.09	&	 0.00 \\ 	
			40	&	100	&	0.6	&	2.01	&	0.01	&	0.00	&	0.000	&	3.95	&	0.13	&	0.03	&	 0.00 \\  	
			40	&	100	&	0.2	&	2.18	&	1.91	&	0.02	&	0.000	&	3.06	&	1.93	&	0.49	&	 0.11 \\ \hline\hline	
		\end{tabular}
	\end{center}
	\par
	\medskip 
	
	\begin{flushleft}
		\noindent {\footnotesize Note: $\hat K_0$, EAD and HAD($\lambda$) are defined as in Table 2. $D_\mu$ is the difference between group means $\mu_1$ and $\mu_2$. Conditional on the number of markets ($L$) and the number of bidders in population ($n$), the classification task is harder when the difference between group means $D_\mu$ is smaller.} 
	\end{flushleft}
\end{table}	

\normalsize

Table 3 reports results for $ K_0 = 2 $ and $ K_0 = 4 $. In both cases, the estimates for $K_0$ are mostly correct. Estimation accuracy increases with the difference between group means. For a given number of groups, the performance in terms of EAD and HAD are both better with greater group differences and larger sample sizes. 

Misclassification errors tend to arise less often when the number of true groups is smaller, with the exception of $D_{\mu} = 0.2 $ and $ L = 100$. Intuitively, this is because when the same number of bidders is partitioned into fewer groups, we can use more pairwise inequalities for classification.

\normalsize

\subsection{Two-Step Estimation in a Structural Model}
In this section we use a simple structural model of procurement auctions to investigate the impact of classification errors on subsequent estimation of structural parameters. 
A set of $N$ providers (bidders) is partitioned into $K_0$ groups, each with a distinct distribution of private costs. Let $N_{k} $ denote the set of providers in $N$ with type $k \in \{1,2,...,K_0\}$, and let $|N_{k}|$ denote its cardinality. The cost for a provider $i$ with type $\tau(i)\in \{1,2,...,K_0\}$ is given by $ c_{i,\ell}=\mu_{\tau(i)}+\epsilon_{i,\ell}$, 
where $\epsilon_{i,\ell}$ follows $N(0,\sigma)$ with the support $[\underline{c},\,\bar{c}]$.\footnote{We set the upper and lower bounds of costs to $ \underline{c}=\frac{1}{K_0}\sum_k (\mu_{k}-1.96\times \sigma) $ and $\bar{c}=\frac{1}{K_0}\sum_k (\mu_{k}+1.96\times \sigma) $.
	True parameters are chosen so that $\underline{c}$ is strictly positive.}

Auction participants are determined in two steps. First, two out of $K_0$ groups, $\tau_{l,1}$ and $\tau_{l,2}$, are chosen at random. 
Next, $n_{1}$ and $n_{2}$ providers are randomly drawn from the corresponding groups $N_{\tau_{l,1}}$ and $N_{\tau_{l,2}}$ and their costs were constructed as above. Here $n_{1}$ and $n_{2}$ denote the numbers of actual participants (those who submitted bids in the auction). Then participants bid based on their realized costs. 
The participant with the lowest bid wins. The identity of each participant and its bid are both reported in data.
We consider two specifications: ($S_1$) with $K_0=4$, $|N_k|=4$ for all $k$, and ($S_2$) with $K_0=4$, $|N_k|=10$ for all $k$. 
For both specifications, we set $\mu=(2,\,2.4,\,2.8,\,3.2)$ and $\sigma=0.5$. 
We run the following four experiments with different specification and sample sizes:  (A) $S_1$, $L=200$; (B) $S_2$, $L=200$; (C) $S_1$, $L=400$; and (D) $S_2$, $L=400$. We set $n_{1} = 1$ and $n_{2}=1$ so that each auction has $2$ actual participants chosen from $K_0$ different types. 

Structural parameters $K_0,\,\tau(\cdot),\, \theta=\{(\mu_{k})_{k=1}^{K_0}, \sigma\}$ are estimated via two steps. 
First, the group structure ($\hat{\tau}(\cdot)$ and $\hat{K}$) are estimated using our classification algorithm. 
Next, we apply GMM to estimate the remaining structural parameters using the following moments for $k = 1,...,K_0$: (1) within-group means of bids: 
$\sum_{i=1}^n \mathbf{E}[B_{i,\ell}-\mu_{B,k}(\theta;\,I)]1\{\hat \tau(i) = k\} = 0$; and (2) within-group second moment of bids:  $\sum_{i=1}^n \mathbf{E}[B_{i,\ell}^2-(\mu_{B,k}(\theta;\,I)^2+\sigma_{B,k}(\theta;\,I)^2)]1\{\hat \tau(i) = k\} = 0$, where $\mu_{B,k}(\theta;\,I)$ and $\sigma_{B,k}(\theta;\,I)$ denote the mean and standard deviation of equilibrium bid distribution for bidders from group $k$, $\theta$ the vector of parameters and $I$ the profile of participant types. 
Standard errors are computed from the analytic expression for the covariance matrix in asymptotic distribution.

To compute $\mu_{B,k}(\theta_0;\,I)$ and $\sigma_{B,k}(\theta_0;\,I)$ for a given $\theta_0$ and profile of participant types $I=(\tau_{l,1}$, $\tau_{l,2}$, $n_1$, $n_2$) we simulate the equilibrium bidding functions.\footnote{
	Specifically, we start from the analytical bidding function when all participants belong to the same group, and use a modified version of the numerical method in \citet{MarshallMeurerRichardStromquist1994} to solve for 	bidding strategies in the presence of multiple groups.
	We impose a sample version of $\underline{c}$ and $\bar{c}$ by replacing $\sigma$ with its sample analog.}
The bidding functions are then combined with the cost distributions implied by a vector of trial parameters, $\theta_0$, to obtain the distribution of bids: $F_{B,k}(b|\,\theta_0,\,I)=F_{C,k}(\beta_k^{-1}(b)|\,\theta_0)$. Here $F_{C,k}(.|\,\theta_0)$ denotes the distribution of project's cost for a bidder belonging to group $k$ which is corresponds to a parameter vector $\theta_0$ and $\beta_k(c)$, $\beta_k^{-1}(b)$ are the bid and the inverse bid functions used by such bidder. We then compute the mean and the standard deviation of this bid distribution.

\fontsize{11}{13}\selectfont 
\begin{table}[t]
	\begin{center}
		\small
		Table 4: Simulation Results from Specifications A and B
		\par
		\begin{tabular}{rrrr|ccccc}
			\hline\hline
			\small
			&  &	& &	$\mu_1$	& $\mu_2$ & $\mu_3$ & $\mu_4$ & $\sigma$ \\ 
			\hline\hline
			Spec A & \multicolumn{2}{r}{\small Using True Groups}      &                       	&	&	&	&	&	\\
			& & & Rej. Prob.                	&  0.0150  &  0.0515 &	 0.0523 &  0.0546 &	0.0149 \\
			& & & Bias                        & -0.0189  & -0.0252 &  -0.0610 & -0.0511 &	0.0242 \\	
			& & & MSE                         &  0.0005  &  0.0008 &   0.0035 &  0.0039 & 0.0039 \\
			\cline{2-9} 
			& \multicolumn{2}{r}{Using Est'd Groups}                          	&	&	&	&	&	\\
			& & & Rej. Prob.                	&  0.0148  & 0.0542 &  0.0510 &  0.0485	 &  0.0151 \\
			& & & Bias                      	&  0.0059  & 0.0329	& -0.0241 & -0.0225  & -0.0549	\\
			& & & MSE                      	  &  0.0041  & 0.0083	&  0.0035 &	 0.0027  &  0.0383 	\\
			\hline
			\hline 
			Spec B & \multicolumn{2}{r}{Using True Groups}                        	&	&	&	&	&	\\			
			& & & Rej. Prob.                	&  0.0120  &  0.0515 &   0.0512 &  0.0514 &	0.0111 \\
			& & & Bias                        & -0.0211  & -0.0233 &  -0.0621 & -0.0622 &	0.0236 \\	
			& & & MSE                         &  0.0005  &  0.0007 &   0.0039 &  0.0039 & 0.0034	\\
			\cline{2-9} 
			& \multicolumn{2}{r}{Using Est'd Groups}                 	&	&	&	&	&	\\
			& & & Rej. Prob.                	&  0.0131  &  0.0550  &  0.0540 &  0.0530	&  0.0160 \\
			& & & Bias                      	& -0.0213  & -0.0218  & -0.0763 & -0.0765   &  0.0211 \\
			& & & MSE                      	  &  0.0004  &  0.0015  &  0.0411 &  0.0441   &  0.0023 \\
			\hline\hline
		\end{tabular}%
		\small
	\end{center}
	\par
	\medskip 
	\begin{flushleft}
		\noindent {\footnotesize Note: Specification A uses $K_{0} = 4$, $n_k=4$, and $L=200$ and Specification B uses $K_{0} = 4$, $n_k=10$, and $L=200$. Here $n_k$ is the number of bidders in group $k$, and $L$ the number of markets. The rejection probabilities are from $t$-tests for the individual parameters. The nominal rejection probability is set to 0.05. }
	\end{flushleft}
\end{table}

\fontsize{11}{13}\selectfont 
\begin{table}[t]
	\begin{center}
		\small
		Table 5: Simulation Results from Specifications C and D
		\begin{tabular}{rrrr|ccccc}
			\hline\hline
			\small
			& 	&	& & $\mu_1$	& $\mu_2$ & $\mu_3$ & $\mu_4$ & $\sigma$ \\ 
			\hline\hline		
			Spec C & \multicolumn{2}{r}{Using True Groups}                        	&	&	&	&	&	\\			
			& & & Rej. Prob.                	&  0.0149  &  0.0500 &	 0.0513 &  0.0520 &	0.0149 \\
			& & & Bias                        & -0.0214  & -0.0222 &  -0.0615 & -0.0713 &	0.0237 \\	
			& & & MSE                         &  0.0005  &  0.0007 &   0.0039 &  0.0039 & 0.0006 \\
			\cline{2-9} 
			& \multicolumn{2}{r}{Using Est'd Groups}                 	&	&	&	&	&	\\
			& & & Rej. Prob.                	&  0.0151  & 0.0485 &  0.0526 &  0.0545	 &  0.0151 \\
			& & & Bias                      	&  0.0131  & 0.0412	& -0.0625 & -0.0656  & -0.0241	\\
			& & & MSE                      	  &  0.0060  & 0.0008	&  0.0053 &	 0.0031  &  0.0054 	\\
			\hline
			\hline
			Spec D & \multicolumn{2}{r}{Using True Groups}                        	&	&	&	&	&	\\			
			& & & Rej. Prob.                	&  0.0133  &  0.0480 &   0.0510 &  0.0520 &	0.0108 \\
			& & & Bias                        & -0.0227  & -0.0219 &  -0.0611 & -0.0231 &	0.0236 \\	
			& & & MSE                         &  0.0005  &  0.0007 &   0.0039 &  0.0039 & 0.0006	\\
			\cline{2-9}
			& \multicolumn{2}{r}{Using Est'd Groups}                 	&	&	&	&	&	\\
			& & & Rej. Prob.                	&  0.0126  &  0.0498  &  0.0520 &  0.0520	&  0.0128 \\
			& & & Bias                      	& -0.0229  & -0.0123  & -0.0761 & -0.0361   &  0.0098	\\
			& & & MSE                      	  &  0.0093  &  0.0056  &  0.0068 &  0.0061   &  0.0007 	\\
			\hline\hline
		\end{tabular}%
	\end{center}
	\par
	\medskip 
	\begin{flushleft}
		\noindent {\footnotesize Note: Specification C uses $K_{0} = 4$, $n_k=4$, and $L=400$, and Specification D, $K_{0} = 4$, $n_k=10$, and $L=400$. Here $n_k$ is the number of bidders in group $k$ and $L$ the number of markets. The nominal rejection probability is set to 0.05.}
	\end{flushleft}
\end{table}
\normalsize

Tables 4 and 5 report the bias and mean squared errors (MSEs) of two estimators for $(\mu_{k})_{k=1}^{K_0}$ and $\sigma$.  The first is an ``infeasible" estimator that uses the knowledge of the true group structure. The second is the two-step estimator we propose, which requires bidder classification in the first step. 
These two tables also report the rejection probabilities from t-tests of individual parameters. 

Table 4 reports results for a smaller sample with $L=200$. 
The rejection probabilities are close to the nominal rejection rate 0.05, except for parameters $\mu_1$ and $\sigma$.  The MSE and bias for all parameters are reasonably small. Table 4 also shows that the rejection probabilities for the infeasible estimator using the true groups and those for the actual estimator using the estimated groups are very similar. There is some minor difference between these two estimators in the MSE of some group means. The discrepancy seems more pronounced when the size of each group is increased from $|N_k|=4$ to $|N_k|=10$. 

Table 5 reports the same results for larger samples with $ L=400 $. The performance of the estimators improves slightly relative to Table 4. Again, the rejection probabilities for the two estimators are similar. 
There is also evidence that with a larger number of the markets, our classification method performs better given the same number of within-group bidders.
Overall, Table 4 and 5 provide simulation evidence that the classification errors in the first-step do not have any major impact on the finite sample performance of the two-step estimators for structural parameters. 

\section{Empirical Application: California Market for Highway Procurement}

We apply our methodology to analyze procurement auctions conducted by the California Department of Transportation (CalTrans) to allocate projects for highway repair work. Our goal is to demonstrate the performance of our method in the empirical setting, and to highlight the consequences of ignoring agent unobserved heterogeneity in estimation.\footnote{Please note that under traditional approach the most straightforward way to account for possible cost asymmetries would be to
estimate firm-specific cost distributions. This approach, however, is not
feasible in most auction studies. This is because the primitives of an
auction game (cost distributions) are linked to the observed auction
outcomes (bid distributions) through a set of non-linear bidding strategies
which have to be obtained by solving a system of differential equations that
has a degeneracy on the boundary. Such system needs to be solved for every possible configuration of the set of participating bidders.
If we define such sets taking into account bidders' identities, the number of such sets will be very large, i.e. $2^N$ (N is the number of bidders).\medskip 

Such concerns do not arise in non-parametric studies since the bidding
strategy and the underlying cost distribution could be recovered from the
first-order conditions by applying them to appropriate bid distributions
(see \citet{GuerrePerrigneVuong2000}). However, as before, the estimation
has to be implemented conditional on the composition of the set of
participants which summarizes the competitive structure of an auction known
to all market participants and is reflected in bidding strategies. Thus, the
afore-mentioned procedure is likely to be infeasible due to data limitations
which are demonstrated in Figure 1.}\medskip

\paragraph{\textbf{Model.}}

We follow the literature in modeling the auction market so that each project attracts a set of potential bidders who decide whether to participate in the auction and, if deciding to participate, choose a bid to submit. Our main innovation is to allow for contractors participating in this market to differ in a way that is not observed by the researcher. Specifically, each contractor is characterized by a
contractor-specific cost factor (invariant across projects) $q_{i}$ which takes discrete values in $\{\bar{q}_{0},\,\bar{q}_{1},\,...,\,\bar{q}_{K_0}\}$. 
This unobserved cost factor captures the difference in cost efficiencies
across firms generated perhaps by the differences in managerial ability or
other factors associated with the firm organization. 
As in our basic set up this cost factor induces partitioning of the population of firms participating in this market into the groups: 
$N=\cup_{k=1}^{K_0} N_{k}$ with $N_{k}=\{i:\,q_i=\bar{q}_k\}$ and so that $\tau(i)=k$ if and only if $i\in N_{k}$.

Following the convention in the empirical auction literature, we assume that each project $\ell$ auctioned in this market is summarized by a set of observable characteristics $X_{\ell}$ and an unobservable factor $U_{\ell}$. The latter is distributed according to normal distribution with mean zero and standard deviation $\sigma _{U}$. The set of firms which are potentially interested in project $\ell$ (potential bidders), denoted here by $S_\ell$, is exogenously drawn from $N$. A contractor $i$ that is a potential bidder for project $\ell$ is characterized by private entry costs, $E_{i,\ell}$, and the private cost of completing the
project, $C_{i,\ell}$. We assume that private costs vary independently across
bidders and auctions. The entry costs additionally are independent of $U_\ell$, and are distributed according to the exponential
distribution with a rate parameter $\lambda_{i,\ell}$. The costs of
completing the work are drawn from a log normal distribution with mean $\mu_{i,\ell}$ and standard deviation $\sigma _{C}$. The mean of the cost distribution depends on project characteristics including the distance between the project and the bidder's locations, $D_{i,\ell}$, as well as an unobserved cost factor $q_{i}$.\footnote{The groups reflect differences in the contractors' cost efficiencies related to the project work. While entry costs may also vary across groups, there is no reason for the group differences in project costs to coincide with the group differences in entry costs. For this reason, we explicitly distinguish between the parameters capturing the former ($\bar{q}_k$) and the latter ($\tilde{q}_k$) effects.}$^{,}$\footnote{
	Following the literature, we distinguish between the bidders who regularly participate in the procurement market (regular bidders) and those who only appear in a very
	small number of auctions (fringe bidders). We assume that all fringe bidders
	are associated with the same fixed level of the unobserved cost factor $\bar{q}_{0}$.} Reflecting these features,  we set
\begin{eqnarray*}
	\mu_{i,\ell}=X_{\ell}\alpha_1 +D_{i,\ell}\alpha_2 +\sum_{k=1}^{K_0}\bar{q}_{k}1\{\tau(i)=k\}+U_{\ell}, \text{ and }
	\lambda_{i,\ell} = X_{\ell}\gamma_1+\sum_{k=1}^{K}\tilde{q}_{k}1\{\tau(i)=k\}.
\end{eqnarray*}
A potential bidder decides to participate in the auction for project $\ell$ if his ex-ante expected profit conditional on participation exceeds entry costs.\footnote{The expected profit reflects his expectation over the participation decisions of other potential bidders, the expectation over his costs of completing the project, and reflects expected probability of winning the project which depends on the costs draws of his competitors.} The set of such bidders is denoted by $A_\ell$. A bidder who decides to participate observes realization of his costs and the identities of other contractors who decided to participate. He chooses a bid to maximize his interium profit which reflects the probability of winning the project conditional on his costs and the set of competitors.  
We assume that the observed outcomes reflect a type-symmetric pure-strategy Bayesian Nash
equilibrium (psBNE).\footnote{In such an equilibrium, participants who are \textit{ex ante} identical in an auction $\ell$ (i.e. $i,j\in S_{\ell}$ such that $q_{i}=q_{j},$ and $D_{i,\ell}=D_{j,\ell}$) adopt the same strategies.}
\smallskip

\paragraph{\textbf{Estimation Details.}}
The estimation methodology consists of two steps. In the first step we use the pairwise comparison indexes to recover the unobserved group structure. In the second step the parameters of the model are estimated through a GMM procedure while imposing the group structure recovered in the first step. We assume bids are rationalized by a single equilibrium.

In the first step we use the pairwise comparison indexes derived in Appendix B.1 in the supplemental note to recover the unobserved group structure.\footnote{The pairwise comparison indexes are derived using Corollary 3 of \citet{Lebrun1999} which for $G_{ij}(b)$ defined as $P\{B_{i,\ell}\geq b|\,i,j\in S_\ell\}$ establishes that $G_{ij}(b)\leq G_{ji}(b)$ for all $b$ in the common support of $B_{i,\ell}$ and $B_{j,\ell}$ whenever $ \tau(i)\leq \tau(j)$. The inequality holds strictly at least over some interval with positive Lebesgue measure and holds unconditionally when aggregated over bidder identities and auction characteristics.} Specifically, in accordance with the notation used in the paper, we define $\delta_{ij}\equiv \int \max \left\{r_{ij}(b),0\right\}db$ and $\delta _{ij}^{0}\equiv \int |r_{ij}(b)|db$ with
$r_{ij}(b)=G_{ji}(b|\,d)-G_{ij}(b|d)$ and $G_{ij}(b|d)=P\{B_{i,\ell}\geq b|\,D_{i,\ell}=d,\,D_{j,\ell}=d,\,i\in A_\ell,\,j\in A_\ell\}$.\footnote{We recover group structure on the basis of the indexes which aggregate over the values of the distance $d$. As a robustness check we also compute groupings on the basis of subsets of distances. We find that the results of classification are very similar across these approaches.}
We obtain empirical counterparts of these indexes by replacing $G_{ij}(b|d)$ with its sample analog $\hat{G}_{ij}(b|d)$. 
We implement classification using the bootstrap testing procedure described previously.

In the second stage, we consider the following moments: (a) the first and the second moment of bid distribution for a given level of $d$ and for a given group of bidders; (b) the
covariance between bids and the observable project characteristics; (c) the
covariance between any two bids submitted in the same auction; (d) the
expected number of participants in any given auction for every $(d,\,q)$-group; (e) the covariance between the number of participants and the observable project characteristics. We search for the set of parameters which minimizes the distance between the empirical and theoretical counterparts of these moments subject to participation constraints.\footnote{We do not explicitly solve for participation strategies. Instead, we discretize the support of auction characteristics $(X_\ell,U_\ell)$ and treat the probabilities of participation for bidders of various $(q,d)-$types corresponding to these grid values as auxiliary parameters. We follow the spirit of \citet{DubeFoxSu2012} by maximizing a moment-based objective function subject to the constraints that the optimality of the participation strategies is satisfied on the grid of the project characteristics' values.}
\smallskip

\paragraph{\textbf{Estimation Results.}}
We implement the analysis using the data for California Highway Procurement
projects auctioned between 2002 and 2012. The projects in our sample are worth
\$523,000 and last for around three months on average; 38\% of these projects are partially supported through federal funds. There are 25 firms that participate regularly in this market. The other firms are referred to as ``fringes". An average auction attracts six regular potential
bidders and eight fringe bidders. Since only a fraction of potential bidders
submits bids, an entry decision plays an important role in this market.
Finally, the distance to the company location varies quite a bit and is
around 28 miles on average for regular potential bidders.

In the first step, we obtain through our classification method the grouping of the bidders into eight groups that consist of 2, 3, 8, 3, 2, 3, 2 and 2 bidders respectively.
The parameter estimates obtained in the second stage of our estimation procedure and their standard errors are summarized in Table 6. We normalize bids by the engineer's estimate in the estimation. Therefore all the parameters measure
the effects relative to the project size.
\begin{table}[!t]
	\begin{center}
		\small
		Table 6. Parameter Estimates\medskip
		\par
		\begin{tabular}{cc|cc|cc|cc}
			\hline\hline
&                 & Estimate                   & Std. Error & Estimate                   & Std. Error & P-value&\\ \hline
\multicolumn{7}{l}{The Distribution of Project Costs}  \\ 
&	 Constant ($\bar{q}_0$) 	 &	 0.127$^{\ast \ast \ast }$ 	 &	(0.0129)	  &	 0.113$^{\ast \ast \ast }$    &	(0.0119)	  & 0.216 &	\\ 
&	 Eng. Estimate 	           &	-0.0004$^{\ast \ast \ast }$  &	(0.0002)	  &	-0.0005$^{\ast \ast \ast }$  &	(0.0002)	  &	0.392&\\ 
&	 Duration 	               &	 0.00026$^*$ 	               &	(0.00036)	  &	0.00022$^{\ast }$            &	(0.00027)  &	0.212&\\ 
&	 Distance 	               &	 0.0012$^{\ast \ast \ast }$  &	(0.00022)	  &	0.00086$^{\ast \ast \ast }$   &	(0.00019)	  &	0.041&\\ 
&  Bridge                    & -0.0092$^{\ast \ast \ast }$  &  (0.0018)    & -0.012$^{\ast \ast \ast }$   &  (0.0011)    & 0.074&\\
&	 Federal Aid	             &	-0.043$^{\ast \ast \ast }$   &	(0.0103)	  &	-0.078$^{\ast \ast \ast }$   &	(0.009)	    & 0.012 &\\
&  Regular Bidder&                               &              & -0.035$^{\ast \ast \ast }$   &  (0.003)     &  &\\
&	 $\sigma _{C}$ 	&	 0.087$^{\ast \ast \ast }$ 	 &	(0.032)	    &	0.112$^{\ast \ast \ast }$    &	(0.022)	    &	0.087&\\ 
&	 $\sigma _{U}$ 	&	 0.021$^{\ast \ast \ast }$ 	 &	(0.009)	    &	0.0207$^{\ast \ast \ast }$   &	(0.008)	    &	0.452&\\\hline 
\multicolumn{7}{l}{The Distribution of Entry Costs}  \\ 
&	 Constant ($\tilde{q}_0$)  &	 -0.0114$^{\ast }$ 	       &	(0.0078)	&	-0.0161$^{\ast }$	&	(0.0091)	   &	0.212 &  \\ 
&	 Eng. Estimate             &	 0.0055$^{\ast \ast \ast }$&	(0.0016)	&	 0.0051$^{\ast \ast \ast }$	   &	(0.0012)	&	0.333 &  \\ 
&	 Number of Items           &	 0.0018$^{\ast }$	         &	(0.0011)	&	 0.0011$^{\ast \ast \ast }$	   &	(0.0005)	&	0.082 &  \\ 
&  Regular Bidder            &                             &            & -0.022 $^{\ast \ast \ast }$    &   (0.004)   &   \\
\hline\hline
\end{tabular}%
\end{center}
\par
\medskip 
\begin{flushleft}
		\noindent {\footnotesize Note: In the results above the distance is measured in miles. The fringe bidders are the reference group. The results are based on the data for 1,054 medium-sized projects that involve paving and bridge work. Standard errors are computed using bootstrap. The first two columns correspond to the specification which allows for the unobserved bidder heterogeneity; the next two columns correspond to the specification without unobserved bidder heterogeneity. The last column reports the p-value of the bootstrap-based test of the equality of coefficients estimated under the specifications with and without unobserved bidder heterogeneity. Details of the test can be found in the Supplemental Note to the paper. }
\end{flushleft}
\end{table}

The first two columns present the estimates which are obtained when the unobserved group structure is taken into account in the estimation. The results indicate significant differences in bidders' costs across the
groups. Specifically, fringe bidders (the reference group) tend to have the highest
costs whereas the difference in costs between the group of fringe bidders and the groups of regular bidders is comparable in impact to the shortening of the distance to the project site by 42.5 (i.e., by 0.051/0.0012), 10.1, 26.67, 48.33, 11.67, 6.67, 7.5, and 41.67 miles respectively.\footnote{The estimates of the group-specific fixed effects are omitted for brevity. The full table that contains these estimates is found in the supplemental note to the paper.} 
The distance increases project costs (additional 8.33 miles result in costs which are 1\% higher on average).\footnote{Recall that the coefficients reflect the impact on costs in terms of the fraction of the engineer's estimate. The distance resulting in 0.01 increase of average costs can thus be computed as 0.01/0.012. } The entry costs of regular bidders are significantly lower than entry costs of fringe bidders. However, they appear to be quite similar across the groups of regular bidders.

The next two columns of Table 6 show the parameter estimates under the specification when the unobserved group structure of the regular bidders is ignored in the estimation. The parameter estimates are obtained by the GMM estimation procedure using the same set of moments by imposing that only two groups of sellers are present in the data: fringe and regular bidders. Under this specification, the cost reduction due to the federal aid is estimated to be much higher (7.8\% rather than 4.3\%), the impact of the distance is estimated to be lower (the distance to the project has to be 11.67 miles higher in order to increase the average cost by 1\%). Additionally, the entry costs are estimated to be lower relative to the baseline specification. 

The last column reports the results of the bootstrap-based test of the equality of coefficients estimated under the specifications with and without unobserved bidder heterogeneity. The results indicate that the difference is significant for the mean parameters in front of the distance, the indicator for the federal aid, the indicator that a project entails bridge-related work, and the standard deviation for the distribution of project costs. The effect of the number of items on the distribution of entry costs is also significant. Our results thus confirm that regular participants in the highway procurement market are characterized by important unobserved cost differences that persist in the data.

\section{Conclusion}

This paper makes a number of contributions to the literature. First, for models with strategic interdependence between multiple agents, we develop a method to classify these agents based on their discrete unobserved individual heterogeneity, using pairwise inequalities implied by an economic model. Second, we show such pairwise inequalities arise in a number of game-theoretical settings where identification of model primitives is challenging. Third, we propose a computationally feasible method which consistently estimates the group structure defined by unobserved heterogeneity. 
We apply this method to California highway procurement data to show that unobserved bidder heterogeneity plays an important role in this procurement market.

The classification method proposed in this paper is especially useful in settings where the analysis of unobserved individual heterogeneity is complicated by the presence of strategic interdependence in the model. We offer new insights into the identification and estimation of such models. 
Specifically, classification could be used as a first step in the structural studies of many environments where analyses would otherwise be infeasible due to the identification or computational challenges.

\section{Appendix: Mathematical Proofs}

\noindent \textbf{Proof of Theorem \ref{identification}}: Sufficiency is obvious. We focus on necessity. Let us assume that $\tau$ is identified. First consider the two facts:\medskip

\noindent Fact 1: If $G_{\tau}$ does not contain a monotone path of length $K_0-1$, $\tau$ is not identified.

\noindent Fact 2: A vertex $i$ is identified if and only if there is a monotone path $P$ containing $i$ such that its end vertices $i_H$ and $i_L$ are identified and
\begin{eqnarray*}
	\tau(i_H) - \tau(i_L) = \ell(P),
\end{eqnarray*}
where $\ell(P)$ denotes the length of $P$.
\medskip

By Fact 1, the necessity of $G_{\tau}$ containing a monotone path of length $K_0-1$ follows, and Fact 2 completes the proof of the necessity part of the theorem.

Now let us prove Fact 1. Suppose that $G_{\tau}$ does not contain a monotone path of length $K_0-1$. Let $N_{max}$ be the set of vertices such that for each vertex $i$ in $N_{max}$, all his $G_{\tau}$-neighbors have lower type than the vertex $i$. Then there is no edge in $G_{\tau}$ which joins any two vertices from the set $N_{max}$. Choose a vertex $i^*$ from $N_{max}$ which is an end vertex of a longest monotone path, say, with length $K-1 < K_0-1$. This identifies a lower bound for $K_0$ but there is no upper bound for $K_0$ that we can obtain from $G_{\tau}$. Take any $\tau'$ such that $\tau'(i^*) > \tau(i^*)$ and $\tau'(i) = \tau(i)$ for all $i \in N\setminus\{i^*\}$. Then $\tau'$ is compatible with $G_{\tau}$ and the given comparison indexes, proving that $\tau$ is not identified from $G$ and the comparison indexes.

Let us prove Fact 2. Sufficiency is trivial. Let us focus on necessity. 
First, suppose to the contrary that there is no monotone path with identified end vertices which contains $ i $. Then obviously $ i $ is not identified. Therefore, if $i$ is identified, there exists a monotone path with identified end vertices which contains $i$. So it suffices to show that if $i$ is identified, it is necessary that such a monotone path has to have length equal to $ \tau(i_H) - \tau(i_L)$. 

Suppose to the contrary that the following condition holds.\medskip

Condition A: Every monotone path $P$ that contains $i$ and has identified end vertices $i_H$ and $i_L$ also satisfies $\tau(i_H) - \tau(i_L) > \ell(P)$. Then we will show that $i$ is not identified.
\medskip

First, assume that there exists a monotone path which contains $i$ but not as one of its end vertices. Let $i_H^*$ be a lowest type vertex among all the identified vertices each of which is on a monotone path that contains $i$ and is of higher type than $i$. Also, let $i_L^*$ be a highest type vertex among all the identified vertices each of which is on a monotone path that contains $i$ and is of lower type than $i$. Let $P$ be a monotone path between $i_H^*$ and $i_L^*$ that passes through $i$. Then by construction, the type difference $\tau(i_H^*) - \tau(i_L^*)$ between the two end vertices is smallest among all the monotone paths that go through $i$. Furthermore, $i_H^*$ and $i_L^*$ are adjacent to $i$ in $G_\tau$. By Condition A, we have $\tau(i_H^*) - \tau(i_L^*) > 2$. 
Therefore, we have multiple different ways to assign $\tau(i_H^*)-1,\tau(i_H^*)-2,...,\tau(i_L^*)+1$ to the vertex $ i $ on the path $P$. 
Hence $i$ is not identified.

Second, assume that all the monotone paths that contain $i$ have $i$ as one of their end vertices. Then either all neighbors of $i$ are of higher type than $i$ or all neighbors of $i$ are of lower type than $i$. Suppose that we are in the former case. (The latter case can be dealt with similarly.) Let $i_H^*$ be a lowest type vertex among all the vertices each of which is on a monotone path that contains $i$ and is of higher type than $i$. Then $i_H^*$ is adjacent to $i$ in $G_\tau$, and by Condition A, $\tau(i_H^*) - \tau(i) > 1$. 
Thus we have multiple different ways to assign $\tau(i_H^*)-1,\tau(i_H^*)-2,...,\tau(i)+1,\tau(i)$ to the vertex $i$. Hence $ i $ is not identified.   $\blacksquare$\medskip

\noindent \textbf{Proof of Lemma \ref{lemm: suff cond}}  Conditions (i) and (v) of Assumption \ref{assump: consistency2} follow from Condition (i) of Assumption \ref{assump: lower level} with $W_{ij}^s$ being a standard normal random variable, and $\rho_L = L^{-\alpha}$. As for Conditions (ii)-(iv) of Assumption \ref{assump: consistency2}, we focus on only (ii), because the proof for the other two statements is similar. Observe that when $\tau(i) > \tau(j)$, so that
\begin{align}
\tilde T_{ij}^+/\sqrt{L} &= (\sigma_{ij,L}^+)^{-1}\left( \int \max\{\hat r_{ij}(x) - r_{ij}(x) + r_{ij}(x),0\}dx - L^{-1/2} h^{-d/2} a_{ij,L}^+\right)\\ \notag
&= (\sigma_{ij}^+)^{-1} \int \max\{r_{ij}(x),0\}dx + o_P(1),
\end{align}
by Assumption \ref{assump: lower level} (ii)(iii) and Condition (\ref{seqs}). Hence Assumption \ref{assump: consistency2}(ii) follows with
\begin{eqnarray*}
	c_{ij}^+ = (\sigma_{ij}^+)^{-1} \int \max\{r_{ij}(x),0\}dx,
\end{eqnarray*}
and $\lambda_L = \sqrt{L}$. 

As for Condition (vi) of Assumption \ref{assump: consistency2}, note that $F_{ij,\infty}^s = \Phi$, the standard normal CDF. Hence  there exists $C>0$ such that for all $t > C$,
\begin{eqnarray*}
	1 - \Phi(t) \le C \exp\left( -\frac{C t^2}{2}\right).
\end{eqnarray*}
Therefore, for any constants $c_1,c_2>0$, taking $\lambda_L = \sqrt{L}$ and $\rho_L = L^{-\alpha}$, (from some large $L$ on)
\begin{eqnarray*}
	r_L^{-1}\log \left(1 - \Phi(c_1\sqrt{L}) + c_2 L^{-\alpha} \right) \le r_L^{-1} \log\left(C \exp(-C c_1^2L/2) + c_2 L^{-\alpha}\right) \rightarrow - \infty,
\end{eqnarray*}
as $L\rightarrow \infty$, by the condition that $r_L /\log L \rightarrow 0$. $\blacksquare$
\medskip

\putbib[refs_comp]
\end{bibunit}

\pagebreak
\appendix
\makeatletter
\def\section{\@startsection{section}{1}
	\z@{0.7\linespacing}{.7\linespacing}{\large}}

\def\subsection{\@startsection{subsection}{2}
	\z@{.4\linespacing}{.6\linespacing}{\normalfont\bfseries}}

\def\subsubsection{\@startsection{subsubsection}{3}
	\z@{.4\linespacing}{.6\linespacing}{\normalfont}}
\makeatother

\begin{bibunit}[econometrica] 
\fontsize{12}{12} \selectfont
\vspace*{3ex minus 1ex}
\begin{center}
	\Large \textsc{Supplemental Note for ``Estimating Unobserved
		Agent Heterogeneity Using Pairwise Comparisons"}
	\bigskip
\end{center}

\date{%
	\today%
}

\vspace*{3ex minus 1ex}
\begin{center}
	Elena Krasnokutskaya, Kyungchul Song, and Xun Tang\\
	\textit{Johns Hopkins University, University of British Columbia, Rice University}
	\bigskip
\end{center}
\maketitle

\section{Introduction}
This note is a supplemental note to \citet{KrasnokutskayaSongTang2020}. It consists of five parts. Appendix B gives details about how to derive pairwise comparison index in examples of first-price and English auctions where bidders have asymmetric private values, or collusive behavior. Appendix C gives the proof of the consistency result of the group structure. Appendix D discusses a bootstrap method to construct a confidence set for the group structure. Appendix E presents further simulation results regarding the performance of our classification algorithm proposed in \citet{KrasnokutskayaSongTang2020}, and Appendix F reports summary statistics for the data used in the empirical application in the paper and some further results.

\section{Bidders with Asymmetric Values or Collusive Behavior}

\subsection{First-Price Auctions with Asymmetric Bidders}
\label{sec: First Price Auctions with Asymmetric Bidders}

Let the population of bidders, $N$, be partitioned into $K_0$ groups, each of which is characterized by a distinct distribution of private values $F_{k}(\cdot)$. 
To fix ideas, assume that $F_{k}(\cdot)$ has the same shape of distribution, but differs only in their location (means) $\bar{q}_{1}<\bar{q}_{2}<...<\bar{q}_{K_0}$, with $\bar{q}_{k}$ being the mean of $F_{k}$. 
(Our method applies in a more general setting when $F_{k}$'s are stochastically ordered.)
In this case, for a bidder $i$ with $\tau(i)=k$ the mean of the private value distribution is given by $q_i=\bar{q}_{k}$.
Let $S_\ell$ denote the set of participants in auction $\ell$ and $B_\ell=\{B_{i,\ell}\}_{i \in S_{\ell}}$ a vector of bids. In a type-symmetric equilibrium $B_{i,\ell}=\beta_{k}(V_{i,\ell})$ if $\tau(i)=k$, with $\beta_{k}(\cdot)$ being a bidding strategy and $V_{i,\ell}$ the private value for $i$ in auction $\ell$.
Define $G_{ij}(b)=P\{B_{i,\ell}\leq b|\,i,j\in S_\ell\}$, where $S_\ell$ denotes the set of bidders in auction $\ell$. We assume bids are rationalized by a single equilibrium.

Corollary 3 of \citet{Lebrun1999} showed that $G_{ij}(b)\ge G_{ji}(b)$ for all $b$ in the common support of $B_{i,\ell}$ and $B_{j,\ell}$ whenever $ \tau(i)\leq \tau(j)$.
The inequality holds strictly at least over some interval with positive Lebesgue measure. 
This inequality holds unconditionally when aggregated over bidder identities and auction characteristics.\footnote{A similar property holds in the settings where allocations are implemented through first-price procurement auctions.
	The only difference is that in these settings $G_{ij}(b)$ should be defined as $G_{ij}(b)=P\{B_{i,\ell}\geq b|\,i,j\in S_\ell\}$.}
Thus pairwise comparison indexes can be constructed as follows:%
\begin{eqnarray}
\label{delta ij private value auc}
\delta_{ij} &\equiv &\int \max \left\{ G_{ji}(b)-G_{ij}(b),0\right\}
db.  \label{indexes} 
\end{eqnarray}%
Likewise, define $ \delta_{ij}^0$ by replacing the integrand in $ \delta_{ij} $ by the absolute value of $G_{ij}(b) - G_{ji}(b)$. These indexes do not condition on the specific identities of bidders participating in each auction $\ell$. Thus it allows us to utilize observations from a large number of auctions when constructing a comparison index for any generic pair $i$ and $j$. 

For the rest of this subsection, we derive the pairwise comparison inequalities in a general model of asymmetric first-price auctions where independent private values are drawn from distributions that are stochastically ordered. 
That is, $F_{k^{\prime }}$ first-order stochastically dominates $F_{k}$ whenever $k^{\prime }>k$. 
Also assume that the ordering of the distributions is strict ($F_{1}(v)>F_{2}(v)>$ $...>F_{K_0}(v)$) at least for $v$ within some non-degenerate interval on the support.
Let $N_{(k)}$ denote the set of all agents in group $k$.

For simplicity, suppose that a bidder from group $k$ becomes active with a fixed probability that is exogenously given.
Let $A$ denote the set of entrants in a given auction and $\lambda(A)$ denote the structure, or the profile, of entrants. 
That is, $\lambda(A)$ is a $K_0$-vector of integers $(|A_{(1)}|,...,|A_{(K_0)}|)$, with $A_{(k)}$ being the set of entrants from group $k$.
An entrant $i$ submits bid $B_{i}$ according to his private value $v_{i}$, taking into account the competitive structure of an auction $\lambda(A)$ which he observes at the time of bidding. 
Across auctions in the data, $A$ and $\{v_{i}\}_{i\in A}$ are independent draws from the same population distribution. 

Let $G_{k}(\cdot ;\lambda )$ be the distribution of $B_{i}$ when $i\in N_{(k)}$. 
The private values are independent of $\lambda_{A}$ under exogenous entry. 
Part (i) of Corollary 3 in \citet{Lebrun1999} 
showed that, given any realization of $\lambda(A)$, the supremum of the support of bids is the same for all bidder types. 
That is, for any $\lambda$, $\beta_{1}(\overline{v}|\lambda)=\beta _{2}(\overline{v}|\lambda )=...=\beta _{K}(\overline{v}|\lambda )\equiv \eta (\lambda )<\infty $ for some $\eta(\lambda )\in (\underline{v},\overline{v})$, where $ \beta_k $ denotes the equilibrium bidding strategy for a bidder from group $ k $. 
Furthermore, the corollary also showed that for any $\lambda(A)$,
\[F_{k^{\prime }}(\beta _{k^{\prime }}^{-1}(b|\lambda(A) ))\leq F_{k}(\beta_{k}^{-1}(b|\lambda(A) )),\] for all $b\in \lbrack \underline{v},\eta (\lambda(A) )]$ and $k<k^{\prime }$, and the inequality holds strictly at least over some interval on $[\underline{v},\eta (\lambda(A) )]$.
Consider $i\in N_{(k^{\prime })}$ and $j\in N_{(k)}$ with $k^{\prime }>k$. 
It then follows that 
\begin{eqnarray*}
	P\left\{ B_{i}\leq b|i,j\in A\right\} &=&\sum\nolimits_{\lambda(A)}F_{k^{\prime }}(\beta _{k^{\prime }}^{-1}(b|\lambda(A) ))P\{\lambda(A) |i,j\in A\}
	\\
	&\leq &\sum\nolimits_{\lambda(A) }F_{k}(\beta _{k}^{-1}(b|\lambda(A) ))P\{\lambda(A) |i,j\in A\}
	=P\{B_{j}\leq b|i,j\in A\},
\end{eqnarray*}%
with the inequality holding strictly over some non-degenerate interval in the shared bid support. 
The inequality does not condition on the identities of the entrants other than $i$ and $j$.

Finally, note that by a symmetric argument, a similar inequality holds in first-price procurement auctions with $ P\left\{ B_{i}\geq b|i,j\in A\right\} \leq P\{B_{j}\geq b|i,j\in A\} $ (with inequality being strict over some non-degenerate interval in the shared bid support), whenever the private cost distribution for $ i $ is stochastically lower than that of $ j $.

\subsection{English Auctions with Asymmetric Bidders} 

Consider the setting in Section \ref{sec: First Price Auctions with Asymmetric Bidders}, except that the auction format is English (ascending). The data report the identities of entrants in $ A $ and the transaction price $W$ in each auction. In a dominant strategy equilibrium, the price in an auction equals the second highest private value among all entrants.

With independent private values, we show below that %
\begin{equation}
P\{W\leq w|i\in A,j\not\in A\}\leq P\{W\leq w|j\in A,i\not\in A\},  \label{english_auction_1}
\end{equation}%
for all $ w $ over the intersection of support, whenever $\tau(i) > \tau(j)$. 
Furthermore, the inequality holds strictly for some $w$ over a set of positive measure in common support. 
This implies%
\begin{equation}
\mathbf{E}[W|i\in A,j\not\in A]>\mathbf{E}[W|j\in A,i\not\in A]\text{.}  \label{english_auction}
\end{equation}

The intuition behind (\ref{english_auction_1}) is as follows. 
Given any structure of entrants who compete with $i$ or $j$ (but not both), the distribution of the transaction price is stochastically higher when $i$ is present but $j$ is not than when $j$ is present but $i$ is not. Loosely speaking, when $j$ is replaced by the stronger type $i$ in the set of entrants, the overall profile of value distributions becomes \textquotedblleft stochastically higher\textquotedblright. 
Then the law of iterated expectations implies (\ref{english_auction_1}) and (\ref{english_auction}). 

To infer the group structure, define the following indexes:%
\begin{eqnarray*}
	\delta_{ij} &=& \max \{\mathbf{E}[W|i\in A,j\not\in A]-%
	\mathbf{E}[W|j\in A,i\not\in A],0\},\text{ and} \\
	\delta_{ij}^{0} &=&\left\vert \mathbf{E}[W|i\in A,j\not\in 
	A]-\mathbf{E}[W|j\in A,i\not\in A]\right\vert.
\end{eqnarray*}%
One can then use our procedure proposed in the main text to classify the bidders based on pairwise comparison.

We now derive (\ref{english_auction_1}) formally. Let $V_{i}$ denote the private value for bidder $i$. 
Consider the case where $i\in N_{(k^{\prime })}$ and $j\in N_{(k)}$ with $k^{\prime }>k$. 
Let $ \lambda(A) $ denote the $K_0$-vector of integers that summarizes the group structure of the set of entrants $ A $. 
Let $1_{k}$ denote the unit vector with the $k$-th component being $1$. 
Then define
\begin{eqnarray*}
	H_{j,i}(w;\lambda ^*) &\equiv &P\{W\leq w|j\in A,i\not\in 
	A,\lambda (A\backslash \{j\})=\lambda ^*\} \\
	&=&P\left\{ \left. \max_{s\in A}V_{s}\leq w\right\vert \lambda (%
	A)=\lambda ^*+1_{k}\right\} \\
	&& + P\left\{ \left. \max_{s\in A}V_{s}>w,W\leq w\right\vert \lambda (%
	A)=\lambda ^*+1_{k}\right\} ,
\end{eqnarray*}%
where the first term on the right-hand side equals $F_{k}(w)\left(
\prod\nolimits_{\ell = 1}^{K_0}F_{\ell}(w)^{\lambda _{\ell}^*}\right) $, and the
second on the right-hand side is
\begin{eqnarray*}
	&&P\left\{ \left. \max_{s\in A\backslash \{j\}}V_{s}\leq
	w,V_{j}>w\right\vert \lambda (A\backslash \{j\})=\lambda ^{\ast
	}\right\} \\
 && +P\left\{ \left. V_{j}\leq w, W \le w, \max_{s\in A\backslash
		\{j\}}V_{s}>w \right\vert \lambda (A\backslash
	\{j\})=\lambda ^*\right\} \\
	&=&[1-F_{k}(w)]\left( \prod\nolimits_{\ell = 1}^{K_0}F_{\ell}(w)^{\lambda _{\ell}^{\ast
	}}\right) +F_{k}(w)\varphi (w;\lambda ^*),
\end{eqnarray*}%
where $\varphi (w;\lambda ^*)$ denotes the probability that the
maximum value in $A\backslash \{j\}$ is strictly greater than $w$
while the second highest value in $A\backslash \{j\}$ is less than
or equal to $w$ conditional on the classification $\lambda (A%
\backslash \{j\})=\lambda ^*$. Therefore%
\[
H_{j,i}(w;\lambda ^*)=\left( \prod\nolimits_{\ell = 1}^{K_0}F_{\ell}(w)^{\lambda
	_{\ell}^*}\right) +F_{k}(w)\varphi (w;\lambda ^*)\text{.}
\]%
By the same argument, 
\begin{eqnarray*}
	H_{i,j}(w;\lambda ^*) &\equiv &P\{W\leq w|i\in A,j\not\in 
	A,\lambda (A\backslash \{i\})=\lambda ^*\} \\
	&=&\left( \prod\nolimits_{\ell = 1}^{K_0}F_{\ell}(w)^{\lambda _{\ell}^*}\right)
	+F_{k^{\prime }}(w)\varphi (w;\lambda ^*)\text{.}
\end{eqnarray*}%
It is then straightforward to show that for any $\lambda ^*$, that $%
F_{k^{\prime }}\succeq _{F.S.D.}F_{k}$ implies $H_{i,j}(w;\lambda ^{\ast
})\leq H_{j,i}(w;\lambda ^*)$ over the union of the $K_0$ supports of $%
\{F_{\ell}:1\leq l\leq K_0\}$, and the inequality holds strictly at least for
some $w$ in an interval on the intersection of the $K_0$ supports of $%
\{F_{\ell}:1\leq l\leq K_0\}$. Under exogenous entry, we get 
\[
P\{W\leq w|i\in A,j\not\in A\}\leq P\{W\leq w|j\in A,i\not\in A\}
,\]%
after integrating out $\lambda ^*$. The inequality holds strictly for some $w$ over common support. 

One may wonder whether we can recover the classification of bidders in the
English auction example through a \textquotedblleft
global\textquotedblright\ approach when the identity of the winner is reported in the data. 
That is, by comparing the distribution
of transaction prices when $i$ is the winner versus that when $j$ is the
winner, as opposed to the pairwise comparison approach proposed above.
Let us explain why this is not feasible.

For any $i\in N_{(k^{\prime })}$ and $j\in N_{(k)}$ and $%
F_{k^{\prime }}\succeq _{F.S.D.}F_{k}$, let $A\backslash \{i,j\}$
denote the set of entrants out of $N\backslash \{i,j\}$ and let $M(%
A\backslash \{i,j\})\equiv \max \{V_{s}:s\in A\backslash
\{i,j\}\}$. Let $\phi (w;\lambda ^*)$ denote the distribution of $M(A
\backslash \{i,j\})$ conditional on $\lambda (A\backslash
\{i,j\})=\lambda ^*$. Let $D$ denote the identity of the winner in the
auction; and $S_{k}$ denote the survival function for the private value of a
type-$k$ bidder. Then,
\begin{eqnarray*}
	&&P\left\{ W\leq w,D=i|i\in A\right\} \\
	&=&p_{j}P\{W\leq w,D=i|i,j\in A\}+(1-p_{j})P\left\{ W\leq
	w,D=i|i\in A,j\not\in A\right\} ,
\end{eqnarray*}%
where $p_{j}$ is shorthand for $j$'s entry probability. Also note that, by
construction, once conditioned on the realized set of entrants from $A\backslash \{i,j\}$, we have
\begin{eqnarray*}
	&&P\left\{ W\leq w,D=i|i,j\in A,\lambda (A\backslash
	\{i,j\})=\lambda ^*\right\} \\
	&=&\int_{-\infty }^{w}F_{k}(t)\phi (t;\lambda ^*)dF_{k^{\prime
	}}(t)+S_{k^{\prime }}(w)F_{k}(w)\phi (w;\lambda ^*),
\end{eqnarray*}%
and%
\begin{eqnarray*}
	&&P\left\{ W\leq w,D=i|i\in A,j\not\in A,\lambda (A\backslash \{i,j\})=\lambda ^*\right\} \\
	&=&\int_{-\infty }^{w}\phi (t;\lambda ^*)dF_{k^{\prime
	}}(t)+S_{k^{\prime }}(w)\phi (w;\lambda ^*)\text{.}
\end{eqnarray*}%
Likewise $P\{W\leq w,D=j|j\in A\}$ can be written by swapping the roles of $i$ and $j$ and swapping the roles of $k$ and $k^{\prime }$ respectively. Then it can be shown that
\begin{equation}
P\{W\leq w,D=i|i\in A,j\in A\}>P\{W\leq w,D=j|i\in A,j\in A\}\text{.\footnote{%
		This inequality (\ref{ineq1}) also allows us to make pairwise comparison
		between $i$ and $j$. However, when the population of potential bidders is
		large and the entry probability of each individual bidder is small, there
		may be few observations with $i,j\in A$ in a finite sample. Thus it
		is not feasible to implement the inference using the inequality in (\ref%
		{ineq1}).}}  \label{ineq1}
\end{equation}%
To see why the inequality in (\ref{ineq1}) holds, note for any $\lambda
^*$,%
\begin{eqnarray*}
	P\{W &\leq &w,D=i|i,j\in A,\lambda (A\backslash
	\{i,j\})=\lambda ^*\} \\
	-P\{W &\leq &w,D=j|i,j\in A,\lambda (A\backslash
	\{i,j\})=\lambda ^*\},
\end{eqnarray*}%
where the difference is written as%
\begin{eqnarray*}
	&&\left[ \int_{-\infty }^{w}F_{k}(t)\phi (t;\lambda ^*)dF_{k^{\prime
	}}(t)-\int_{-\infty }^{w}F_{k^{\prime }}(t)\phi (t;\lambda ^*)dF_{k}(t)%
	\right] \\
	&&+\phi (w;\lambda ^*)\left[ S_{k^{\prime
	}}(w)F_{k}(w)-S_{k}(w)F_{k^{\prime }}(w)\right] .
\end{eqnarray*}%
The first square bracket in the display above is positive because
\[
\int_{-\infty }^{w}F_{k}(t)\phi (t;\lambda ^*)dF_{k^{\prime
}}(t)>\int_{-\infty }^{w}F_{k^{\prime }}(t)\phi (t;\lambda ^{\ast
})dF_{k^{\prime }}(t)>\int_{-\infty }^{w}F_{k^{\prime }}(t)\phi (t;\lambda
^*)dF_{k}(t)\text{.}
\]%
Furthermore, the second square bracket in the display is also positive
because \textquotedblleft $F_{k^{\prime }}\succeq _{F.S.D.}F_{k}$%
\textquotedblright\ implies 
\[
S_{k^{\prime }}(w)\geq S_{k}(w)\text{ and }F_{k}(w)\geq F_{k^{\prime }}(w)%
\text{ for all }w
\]%
and these inequalities hold strictly for some set of $w$ with positive
measure. Integrating out $\lambda ^*$ on both sides of the inequality%
\begin{eqnarray*}
	&&P\{W\leq w,D=i|i,j\in A,\lambda (A\backslash
	\{i,j\})=\lambda ^*\} \\
	&>&P\{W\leq w,D=j|i,j\in A,\lambda (A\backslash
	\{i,j\})=\lambda ^*\}\text{,}
\end{eqnarray*}%
yields the first inequality in (\ref{ineq1}).

Similarly, the difference between $P\{W\leq w,D=i|i\in A,j\not\in 
A,\lambda (A\backslash \{i,j\})=\lambda ^*\}$ and $%
P\{W\leq w,D=j|j\in A,i\not\in A,\lambda (A%
\backslash \{i,j\})=\lambda ^*\}$ equals 
\[
\left[ \int_{-\infty }^{w}\phi (t;\lambda ^*)dF_{k^{\prime
}}(t)-\int_{-\infty }^{w}\phi (t;\lambda ^*)dF_{k}(t)\right] +\phi
(w;\lambda ^*)[S_{k^{\prime }}(w)-S_{k}(w)]
\]%
which must be positive because the two terms in the square brackets are
positive.

Now we write 
\begin{eqnarray}
&&P\{W\leq w,D=i|i\in A\}  \label{eq1} \\
&=&p_{j}P\{W\leq w,D=i|i,j\in A\}+(1-p_{j})P\{W\leq w,D=i|i\in 
A,j\not\in A\}  \nonumber
\end{eqnarray}%
where $p_{j}\equiv P(j\in A)$. A similar expression exists for $%
P\{W\leq w,D=j|j\in A\}$ by swapping the roles of $i$ and $j$ in (%
\ref{eq1}). Therefore
the difference between $P\{W\leq w,D=i|i\in A\}$ and $P\{W\leq
w,D=j|j\in A\}$ is also indeterminate in the absence of knowledge about $p_{i},p_{j}$.

\subsection{Bidding Cartel in First-Price Procurement Auctions}

Our method can be used to detect the identities of cartel members in a model
of first-price procurement auctions in which a bidding cartel competes with
competitive non-colluding bidders (\citet{Pesendorfer2000}). Let the
population of bidders/firms $N$ be partitioned into a set of colluding firms $%
N_{(c)}$ and non-colluding firms $N_{(nc)}$. In each auction, the
set of potential bidders (who are interested in bidding for the contract) $%
A$ is partitioned into $A_{(c)}$ and $A_{(nc)}$. The
cardinality of $A_{(c)}$ is common knowledge among the bidders. The
potential bidders in $A_{(c)}$ collude by refraining from
participation except for one bidder $i^*$ who is chosen among them to
submit a bid.\footnote{%
	The cartel is sustained through side payments among its members.} 

In an efficient truth-revealing mechanism considered in \citet{Pesendorfer2000}, the cartel member that has the lowest cost is selected to be the sole bidder from the cartel. 
That is, $i^*(A_{(c)})=\arg \min_{j\in A_{(c)}}C_{j}$ where $C_{j}$ is the private cost of bidder $j$. Thus, the set of final entrants who are observed to submit bids in the data is $A^*\equiv \{i^*(A_{(c)})\}\cup A_{(nc)}$. (The set of colluding potential bidders is not reported in the data available to the researcher.)

We maintain that across the auctions in the data bidders' private costs are
independent draws from the same distribution. Bidders are ex ante symmetric
in that each bidder's private cost is drawn independently from the same
distribution. Entrants know that a representative of the cartel is
participating in bidding, and all follow Bayesian Nash equilibrium bidding
strategies. 

We are interested in detecting the identities of the set of colluding firms in $N_{(c)}$ from the reported bidding and participation decisions. Let $N_{(c)}^{\prime }\subset N$ denote the set of bidders such that no two bidders in $N_{(c)}^{\prime }$ are ever observed to compete with each other in the bidding stage. By construction, $N_{(c)}\subseteq N_{(c)}^{\prime }$
so the latter should be interpreted as a set of suspects for collusion.
However, the set $N_{(c)}^{\prime }$ could also contain innocent
non-colluding bidders who are never observed to compete with each other in
the data because of random entry in finite sample. Our goal is to use bidding
data to separate $N_{(c)}$ from $N_{(c)}^{\prime }\backslash 
N_{(c)}\equiv N_{(nc)}\cap N_{(c)}^{\prime }$. 


\citet{Pesendorfer2000} (Remark 3) shows that in any given auction with
participants $A_{(c)}\cup A_{(nc)}$, the distribution of bids
from a non-colluding bidder $j$ first-order stochastically dominates the
distribution of the bids from the sole bidder representing the cartel $%
i^*$.\footnote{\citet{Pesendorfer2000} proved this result using the
	implicit assumption that the distribution of costs for non-colluding bidders
	and that for the sole cartel is common knowledge among all participants in
	an auction. (See proof of Remark 3 in \citet{Pesendorfer2000}.) This
	assumption is consistent with the informational environment that the
	partition of $N$ into $N_{(c)}$ and $N_{(nc)}$ is
	common knowledge among all bidders.} Specifically, for any such $i^*$ and $j$,
\begin{equation}
P\{B_{i^*}\leq b|i^*\in A^*,|A^{\ast
}|\}>P\{B_{j}\leq b|j\in A^*,|A^*|\}
\label{collude_ineq}
\end{equation}%
for all $b$ on the common support of the two distributions.\footnote{%
	Note that the statement is conditional since the bidding strategies depend
	on the cardinality of the final set of bidders $|A^*|$.} 

The intuition, as is noted in \citet{Pesendorfer2000}, is that the sole bidder representing a cartel has a higher hazard rate than a non-colluding bidder. 
That is, relative to a competitive bidder, the cartel representative has a higher probability of having a low cost conditional on the costs being above any fixed threshold. 
Besides, ex ante symmetry among bidders implies that
\[
P\{B_{i}\leq b|i\in A^*,|A^*|\}=P\{B_{j}\leq
b|j\in A^*,|A^*|\}
\]%
whenever $i,j\in N_{(c)}$ or $i,j\in N_{(nc)}\cap N%
_{(c)}^{\prime }$.

We can then construct pairwise comparison indexes $\delta_{ij}, \delta_{ij}^{0}$ by replacing $G_{ij}$ and $G_{ji}$ in (\ref{delta ij private value auc}) with the left- and right-hand side of (\ref{collude_ineq}).

\section{Mathematical Proofs}

\subsection{Proof of Consistency of Classification}

We provide a proof of consistency of the classification allowing both $n$ and $L$ to diverge to infinity jointly. This requires a more elaborated set of assumptions than Assumption \ref{assump: consistency2}. It is not hard to use the same proof to prove Theorem \ref{thm: consistency 2}.

\begin{assumption}
	\label{assump: consistency2b}
	There exist sequences $\rho_L,\kappa_L \rightarrow 0$ and $\lambda_L \rightarrow \infty$ and constants $c_{ij}^s>0$, $s \in \{+,0,-\}$ such that along each sequence of probabilities $P_L \in \mathcal{P}_{0,\varepsilon}$, the following conditions hold for all $s \in \{+,0,-\}$, $s' \in \{+,-\}$, as $n,L\rightarrow \infty$.\medskip
	
	(i) 
	\begin{eqnarray*}
		\max_{i,j \in N: i \ne j, \tau(i) = \tau(j)}\left|P\left\{\tilde T_{ij}^s \le t \right\} - F_{ij,\infty}^s(t)\right| \rightarrow 0,
	\end{eqnarray*}
    where $F_{ij,\infty}^s$ is a continuous CDF.\medskip
	
	(ii)(a)
	\begin{eqnarray*}
		\max_{i,j \in N: i \ne j, \tau(i) > \tau(j)} P\left\{|\tilde T_{ij}^+/\lambda_L - c_{ij}^+| + |\tilde T_{ij}^0/\lambda_L - c_{ij}^0| \ge \kappa_L\right\} = O(\rho_L).
	\end{eqnarray*}
	\medskip
	
	(ii)(b) For each pair $i,j \in N$ such that $i \ne j$, if $\tau(i) \ge \tau(j)$, there exists a CDF $F_{ij}^-$ such that for all $t \in \mathbf{R}$,
	\begin{eqnarray*}
		P\{\tilde T_{ij}^- \le t  \} \ge F_{ij}^-(t).
	\end{eqnarray*}
	\medskip
	
	(iii)(a)
	\begin{eqnarray*}
		\max_{i,j \in N: i \ne j, \tau(i) < \tau(j)} P \left\{|\tilde T_{ij}^-/\lambda_L - c_{ij}^-| + |\tilde T_{ij}^0/\lambda_L - c_{ij}^0| \ge \kappa_L \right\} = O(\rho_L).
	\end{eqnarray*}
	\medskip
	
	(iii)(b) For each pair $i,j \in N$ such that $i \ne j$, if $\tau(i) \le \tau(j)$, there exists a CDF $F_{ij}^+$ such that for all $t \in \mathbf{R}$,
	\begin{eqnarray*}
		P\{\tilde T_{ij}^+ \le t  \} \ge F_{ij}^+(t).
	\end{eqnarray*}
\end{assumption}
\medskip

\begin{assumption}
	\label{assump: consistency3b}
	Suppose that for each pair $i,j \in N$ such that $i \ne j$, $F_{ij}^{s'}$ and $F_{ij,\infty}^s$, $s,s' \in \{+,0,-\}$, are CDFs and $\rho_L$, $\kappa_L$, and $\lambda_L$ are sequences in Assumption \ref{assump: consistency2b}. Then the following holds as $n, L \rightarrow \infty$.
	\medskip
	
	(i)
	\begin{eqnarray*}
		\max_{i,j \in N: i \ne j} P\left\{\|\tilde F_{ij}^s - F_{ij,\infty}^s \|_\infty > \varepsilon_L \right\} = O(\rho_L)
	\end{eqnarray*}
     and
     \begin{eqnarray*}
     	\max_{i,j \in N: i \ne j} \left\|F_{ij}^{s'} - F_{ij,\infty}^{s'} \right\|_\infty = O(\varepsilon_L),
     \end{eqnarray*}
     where $\varepsilon_L \rightarrow 0$.
	\medskip
	
	(ii) $r_L^{-1} \log(1 - F_{ij,\infty}^s((c - \kappa_L) \lambda_L) + \varepsilon_L) \rightarrow - \infty$, for any constant $c>0$.
	\medskip
	
	(iii) $n^2 (\exp(-r_L/2) + \varepsilon_L + \rho_L) \rightarrow 0$.
\end{assumption}

Assumptions \ref{assump: consistency2b} and \ref{assump: consistency3b} are variants of Assumption \ref{assump: consistency2} which require the rate of convergences explicitly. Assumption \ref{assump: consistency3b}(iii) is new here because we now allow both $n$ and $L$ to increase to infinity. This assumption says that for the consistency of the estimated group structure, we need to have $n$ increase sufficiently faster than $L$. This condition is trivially satisfied when $n$ is fixed and $L$ increases to infinity. Note that the conditions are high level conditions suited to the generic set-up here. With a more detailed specification of the pairwise comparison indexes $\delta_{ij}$ and $\delta_{ij}^0$, test statistics and $p$-values, one may improve the conditions. By using high level conditions, we can focus only on more novel aspects of the mathematical proofs.

While existence of such sequences as $\kappa_L$, $\varepsilon_L$, and $\rho_L$ is fairly expected, obtaining their precise forms using lower level conditions require substantial yet tedious arguments depending on the way the test statistics and the $p$-values are constructed. Essentially what one needs to obtain for lower level conditions is the rate in the convergence of the test statistics to the limiting distribution both under the null hypothesis and under the local alternatives. For example, if one follows the approach of \cite{LeeSongWhang2018}, the rate of convergence is ultimately delivered by a Berry-Esseen bound (for a sum of independent random variables) used to obtain the asymptotic normality of test statistics (under appropriate mean shifts). However, the final result comes, in combination with this, only with carefully derived rate results on asymptotically negligible terms that arise, among other things, due to the Poissonization technique used in the paper. While these additional developments are feasible, they do not add insights to the main idea of the paper. Hence we do not pursue details in this direction here.

Let us first state the theorem.
\begin{theorem}
	\label{thm: consistency 2 app}
	Suppose that Assumptions \ref{assump: consistency2b} - \ref{assump: consistency3b} hold, and that $g(L) \rightarrow \infty$ and $g(L)/r_L \rightarrow 0$ as $L\rightarrow \infty$. Then, for any $\varepsilon>0$, along a sequence of probabilities $P_{n,L} \in \mathcal{P}_{0,\varepsilon}$, 
	\begin{equation*}
	P_{n,L}\{\hat{K}=K_{0}\}\rightarrow 1, \text{ as } n,L \rightarrow \infty,
	\end{equation*}
	and the estimated group structure $\hat{T}_{\hat{K}}$ satisfies that as $n,L\rightarrow \infty,$
	\begin{equation*}
	P_{n,L}\{ \hat{T}_{\hat{K}} = T \} \rightarrow 1.
	\end{equation*}
\end{theorem}

The proof of this theorem is long. We first prepare some auxiliary lemmas. Throughout this section, we assume that Assumptions \ref{assump: consistency2b} and \ref{assump: consistency3b} hold. First, for any subset $N' \subset N$, we define $N'(i) = N' \setminus \{i\}$, and let
\begin{align}
\label{def N(i)}
	N_1'(i) &= \{j \in N'(i): \tau(i) > \tau(j) \},\\ \notag
	N_2'(i) &= \{j \in N'(i): \tau(i) < \tau(j) \},\\ \notag
	\overline N_1'(i) &= \{j \in N'(i): \tau(i) \ge \tau(j) \}, \text{ and }\\ \notag
	\overline N_2'(i) &= \{j \in N'(i): \tau(i) \le \tau(j) \}.
\end{align}

We also define
\begin{align*}
\hat N_{1}'(i) &= \{j\in N'(i): \log \hat{p}%
_{ij}^{+} \le \log \hat{p}_{ij}^{-} - r_L \}\text{ and } \\
\hat N_{2}'(i) &= \{j\in N'(i):\log \hat{p}%
_{ij}^{-}\le \log \hat{p}_{ij}^{+} - r_L \}.
\end{align*}
Following the convention, given a CDF $G$, we define its generalized inverse  $G^{-1}$ as $G^{-1}(t) = \inf\{s \in \mathbf{R}: G(s) \ge t \}$, $ t \in (0,1)$.

\begin{lemma}
	\label{lemm: tail bounds}
	(i)
	\begin{eqnarray*}
		\max_{i \in N} P\left\{ \min_{j \in \overline N_1'(i)} \log \hat p_{ij}^- < - r_L/2 \right\} &=& O\left(n\omega_{n,L}  \right)\\
		\max_{i \in N} P\left\{ \max_{j \in N_1'(i)} \log \hat p_{ij}^+ \ge  - 3r_L/2 \right\} &=& O\left(n  \rho_L  \right)\\
		\max_{i \in N} P\left\{ \min_{j \in \overline N_2'(i)} \log \hat p_{ij}^+ < - r_L/2 \right\} &=& O\left(n\omega_{n,L} \right), \text{ and }\\
		\max_{i \in N} P\left\{ \max_{j \in N_2'(i)} \log \hat p_{ij}^- \ge - 3r_L/2 \right\} &=& O\left(n  \rho_L  \right),
	\end{eqnarray*}
where
\begin{eqnarray*}
	\omega_{n,L} = \exp\left(-\frac{r_L}{2} \right) + \varepsilon_L + \rho_L.
\end{eqnarray*}

(ii)
\begin{eqnarray*}
	\max_{i,j \in N: \tau(i) \ne \tau(j)} P\left\{\log \hat p_{ij}^0 \ge  - r_L \right\} = O\left(\rho_L  \right),
\end{eqnarray*}
and
\begin{eqnarray*}
	\max_{i,j \in N: i \ne j, \tau(i) = \tau(j)} P\left\{\log \hat p_{ij}^0 \le  - r_L \right\} = O\left( \omega_{n,L} \right).
\end{eqnarray*}
\end{lemma}

\noindent \textbf{Proof: } (i) We will show the first and the second statements only. The third and the fourth statements can be proved similarly. Let us prove the first statement first. For all $i \in N$,
\begin{eqnarray*}
	P\left\{ \min_{j \in \overline N_1'(i)} \log \hat p_{ij}^- < - \frac{r_L}{2} \right\} \le \sum_{j \in \overline N_1'(i)}  P\left\{\log \hat p_{ij}^- \le - \frac{r_L}{2} \right\}.
\end{eqnarray*}
Note that for each $j \in \overline N_1'(i)$,
\begin{eqnarray*}
	P\left\{\log \hat p_{ij}^- < - \frac{r_L}{2} \right\} &=& P\left\{ 1 -  \exp \left(-\frac{r_L}{2}\right) < \tilde F_{ij}^-(\tilde T_{ij}^-) \right\}\\
	&\le& P\left\{ 1 -  \exp\left(-\frac{r_L}{2}\right) < F_{ij,\infty}^-(\tilde T_{ij}^-) + \varepsilon_L \right\}
	  + P\{\| \tilde F_{ij}^- - F_{ij,\infty}^-\|_\infty \ge \varepsilon_L \}\\
	&\le& P\left\{ 1 -  \exp\left(-\frac{r_L}{2}\right) < F_{ij,\infty}^-(\tilde T_{ij}^-) + \varepsilon_L  \right\} + O(\rho_L ),
\end{eqnarray*}
where the last $O(\rho_L)$ term is uniform over $i,j \in N$ such that $\tau(i) \ne \tau(j)$ and is due to Assumption \ref{assump: consistency3b}(i). As for the leading probability on the right end side, we use the fact that for any CDF $G$ and any $t \in (0,1)$ and $x \in \mathbf{R}$, $G^{-1}(t) \ge x$ if and only if $t \ge G(x)$, (e.g. Lemma A.1.1. of \cite{Reiss89}), p.318, and bound
\begin{eqnarray*}
	&& P\left\{ 1 -  \exp\left(-\frac{r_L}{2}\right) < F_{ij,\infty}^-(\tilde T_{ij}^-) + \varepsilon_L  \right\}\\
	&=& 1 - P\left\{\tilde T_{ij}^- \le (F_{ij,\infty}^{-})^{-1}\left(1 -  \exp\left(-\frac{r_L}{2}\right) - \varepsilon_L \right) \right\}\\
	&\le& 1 - F_{ij}^-\left((F_{ij,\infty}^{-})^{-1}\left(1 -  \exp\left(-\frac{r_L}{2}\right) - \varepsilon_L \right) \right),
\end{eqnarray*}
by Assumption \ref{assump: consistency2b}(ii)(b) and because $j \in \overline N_1'(i)$. By Assumption \ref{assump: consistency3b}(i), the last term is equal to
\begin{eqnarray*}
	&& 1 - F_{ij,\infty}^-\left((F_{ij,\infty}^{-})^{-1}\left(1 -  \exp\left(-\frac{r_L}{2}\right) - \varepsilon_L \right) \right) +  O(\varepsilon_{L})\\
	&\le& \exp\left(-\frac{r_L}{2}\right) + \varepsilon_L + O(\varepsilon_L)	 = O(\omega_{n,L}),
\end{eqnarray*}
uniformly over $i,j \in N$ such that $\tau(i) \ne \tau(j)$. The inequality above is due to the definition of $(F_{ij,\infty}^-)^{-1}$. Thus we obtain the first statement.

Let us consider the second statement. Suppose that $\tau(i) > \tau(j)$. We let
\begin{eqnarray*}
A_{1,L} &=& \left\{\|\tilde F_{ij}^+ - F_{ij,\infty}^+\|_\infty \le \varepsilon_L \right\} \text{ and }\\
A_{2,L} &=& \left\{ |\tilde T_{ij}^+/\lambda_L - c_{ij}^+| \le \kappa_L  \right\},
\end{eqnarray*}
and let $A_L = A_{1,L} \cap A_{2,L}$. Note that
\begin{eqnarray*}
	P \{\log \hat p_{ij}^+ \ge - 3r_L/2 \} &\le& P \{\log \hat p_{ij}^+ \ge - 3 r_L/2 \} \cap A_L + P A_L^c\\
	&\le& P \{\log \hat p_{ij}^+ \ge - 3 r_L/2 \} \cap A_L + O(\rho_L),
\end{eqnarray*}
where the last $O(\rho_L)$ term is due to Assumptions \ref{assump: consistency2b}(ii)(a) and \ref{assump: consistency3b}(i), and is uniform over $i,j \in N$ such that $i \ne j$. For the leading probability on the right hand side, note that
\begin{eqnarray*}
P \{\log \hat p_{ij}^+ \ge - 3 r_L/2 \} \cap A_L
&=& P \{\log (1 - \tilde F_{ij}^+(\tilde T_{ij}^+)) \ge - 3 r_L/2 \} \cap A_L\\
&\le& P \{\log(1 - F_{ij,\infty}^+(\tilde T_{ij}^+) + \varepsilon_L)  \ge - 3 r_L/2 \} \cap A_L\\
&\le& P A_L \cdot 1\{ 1 - F_{ij,\infty}^+((c_{ij}^+ - \kappa_L) \lambda_L) + \varepsilon_L \ge \exp(- 3 r_L/2 ) \}.
\end{eqnarray*}
The last indicator becomes zero from some large $L$ on, with this large $L$ chosen to be independent of $i,j$, by Assumption \ref{assump: consistency3b}(ii). Thus, we obtain the second statement.\medskip

(ii) The proof of the first statement is the same as that of the second statement of (i), and the proof of the second statement is the same as that of the first statement of (i). (As for the proof of the second statement, we have the weak inequality instead of the strict inequality, but this does not make difference to the arguments, because $F_{ij,\infty}^0$ is assumed to be continuous.) Details are omitted. $\blacksquare$

\begin{lemma}
	\label{lemm: consistency}
	Suppose that $N' \subset N$ contains some $i,j \in N$ such that $\tau(i) \ne \tau(j)$. Then,
    \begin{align*}
       & \min_{i \in N} P\{N_1'(i) = \hat N_1'(i)\} = 1 + O(n\omega_{n,L});\\
       & \min_{i \in N} P\{N_2'(i) = \hat N_2'(i)\} = 1 + O(n\omega_{n,L});\\
       & \min_{i \in N} P\{\overline N_2'(i) = N'(i) \setminus \hat N_1'(i)\} = 1 + O(n\omega_{n,L});\\
       & \min_{i \in N} P\{\overline N_1'(i) = N'(i) \setminus \hat N_2'(i)\} = 1 + O(n\omega_{n,L}).
    \end{align*}
\end{lemma}

\noindent \textbf{Proof: } We show only the first and the third statements. The remaining statements can be proved similarly. Define
\begin{eqnarray*}
	A_L = \left\{\min_{j \in \overline N_1'(i)} \log \hat p_{ij}^- \ge -\frac{r_L}{2} \right\}.
\end{eqnarray*}
Note that
\begin{align*}
	P\left\{N_1'(i) \subset \hat N_1'(i) \right\} &= P \left\{\max_{j \in N_1'(i)} \log \hat p_{ij}^+ - \log \hat p_{ij}^- \le - r_L\right\}\\
	&\ge P \left\{\max_{j \in N_1'(i)} \log \hat p_{ij}^+ - \min_{j \in \overline N_1'(i)} \log \hat p_{ij}^- \le - r_L\right\}\\
	&\ge P \left\{\max_{j \in N_1'(i)} \log \hat p_{ij}^+ \le - \frac{3r_L}{2} \right\} \cap A_L\\
	&\ge P \left\{\max_{j \in N_1'(i)} \log \hat p_{ij}^+ \le - \frac{3r_L}{2} \right\} - P A_L^c\\
	&= 1 - P \left\{\max_{j \in N_1'(i)} \log \hat p_{ij}^+ > - \frac{3r_L}{2} \right\} - P A_L^c
	= 1 + O(n\omega_{n,L}),
\end{align*}
where the last inequality follows by the first and the second statements of Lemma \ref{lemm: tail bounds}(i). Thus we have
\begin{align}
\label{inc1}
    P\left\{N_1'(i) \subset \hat N_1'(i) \right\}  =1 + O(n\omega_{n,L}),
\end{align}
as $n,L\rightarrow \infty$. On the other hand,
\begin{align}
\label{inc2}
	P\left\{\overline N_2'(i) \subset N'(i) \setminus \hat N_1'(i) \right\} &= P \left\{\min_{j \in \overline N_2'(i)} \log \hat p_{ij}^+ - \log \hat p_{ij}^- > - r_L\right\}\\ \notag
	&\ge P \left\{\min_{j \in \overline N_2'(i)} \log \hat p_{ij}^+ > - r_L\right\} = 1 + O(n\omega_{n,L}).
\end{align}
The inequality above follows because $\log \hat p_{ij}^- \le 0$ and the last equality follows by the third statement of Lemma \ref{lemm: tail bounds}(i). Note that the term $O(n\omega_{n,L})$ is uniform over $i \in N$. Since $N_1'(i)$ and $\overline N_2'(i)$ partition $N'(i)$, and $\hat N_1'(i)$ and $N'(i) \setminus \hat N_1'(i)$ also partition $N'(i)$, it follows that
\begin{eqnarray*}
	\min_{i \in N} P\{N_1'(i) = \hat N_1'(i)\} &=& 1 + O(n\omega_{n,L}), \text{ and } \\
	\min_{i \in N} P\{\overline N_2'(i) = N'(i) \setminus \hat N_1'(i)\} &=& 1  + O(n\omega_{n,L}),
\end{eqnarray*}
as $n,L\rightarrow \infty$. The remaining statements can be shown similarly. $\blacksquare$

\begin{definition}
	(i) An ordered partition $(N_1',...,N_s')$ of a subset $N' \subset N$ is said to be a \textbfit{$\tau$-ordered partition}, if for any $r_1 < r_2$, $r_1,r_2 =1,2,...,s$, we have $\tau(i) < \tau(j)$ whenever $i \in N_{r_1}'$ and $j \in N_{r_2}'$.
	
	(ii) Let $\mathscr{N}_\tau(N')$ denote the set of $\tau$-ordered partitions of $N'$.
\end{definition}

When an ordered partition $(N_1',...,N_s')$ is a $\tau$-ordered partition, and $\tau$ partitions $N'$ into $K$ groups (i.e., any two agents, say $i,j$, from two different groups from the $K$ groups have $\tau(i) \ne \tau(j)$), we must have $s \le K$ by the definition of $\tau$-ordered partition. Hence some group in the ordered partition $(N_1',...,N_s')$ can have agents with different types.

\begin{definition}
	An estimated ordered partition $(\hat N_1',...,\hat N_s')$ of a subset $N' \subset N$ is said to be \textbfit{asymptotically $\tau$-compatible at rate $u_{n,L}$}, if 
	\begin{eqnarray*}
		P\{(\hat N_1',...,\hat N_s') \in \mathscr{N}_\tau(N')\} = 1 + O(u_{n,L}), \text{ as } n,L \rightarrow \infty.
	\end{eqnarray*}
\end{definition}

Suppose that we choose $\pi \in \mathscr{N}_\tau(N')$ with $N' \subset N$ such that $\pi = (N_1',...,N_s')$. If we choose a different $\pi \in \mathscr{N}_\tau(N')$, then the group $(N_1',...,N_s')$ changes. Hence we denote $\pi = (N_1'(\pi),...,N_s'(\pi))$ making explicit each group's dependence on $\pi$.

\begin{lemma}
	\label{lemm: conv3}
	For each $\pi = (N_1',...,N_s') \in \mathscr{N}_\tau(N')$, let $R_1(\pi) \subset \{1,2,...,s\}$ be such that for all $r \in R_1(\pi)$, $N_r'$ has some $i,j \in N_r'$ satisfying $\tau(i) \ne \tau(j)$, and let $R_2(\pi) = \{1,2,...,s\} \setminus R_1(\pi)$ so that for all $r \in R_2(\pi)$, and for all $i,j \in N_r'$, we have $\tau(i) = \tau(j)$ . Let $B(\pi), \pi \in \mathscr{N}_\tau(N')$, be disjoint events. Then,
	\begin{eqnarray*}
		\sum_{\pi \in \mathscr{N}_\tau(N')} P\left\{\exists r_1 \in R_1(\pi), \min_{i,j \in N_{r_1}'(\pi): i \ne j} \log \hat p_{ij}^0 > - r_L \right\} \cap B(\pi) &=& O(n^2 \rho_L), \text{ and }\\
		\sum_{\pi \in \mathscr{N}_\tau(N')} P\left\{\exists r_2 \in R_2(\pi), \min_{i,j \in N_{r_2}'(\pi): i \ne j} \log \hat p_{ij}^0 \le -r_L \right\} \cap B(\pi) &=& O(n^2 \omega_{n,L}).
	\end{eqnarray*}
\end{lemma}

\noindent \textbf{Proof: } For each $\pi \in \mathscr{N}_\tau(N')$, we denote $\pi = (N_1'(\pi),...,N_{K(\pi)}'(\pi)) \in \mathscr{N}_\tau(N')$, and let $K(\pi)$ denote the total number of the groups in $\pi$. Then we can write $K(\pi) = |R_1(\pi)| + |R_2(\pi)|$. Let
\begin{eqnarray*}
	B_\tau(N') = \bigcup_{\pi \in \mathscr{N}_\tau(N')} B(\pi).
\end{eqnarray*}
The first statement of the lemma follows because
\begin{eqnarray*}
	&& \sum_{\pi \in \mathscr{N}_\tau(N')} P\left\{\exists r_1 \in R_1(\pi), \min_{i,j \in N_{r_1}'(\pi): i \ne j} \log \hat p_{ij}^0 > - r_L \right\}\cap B(\pi) \\
	&& \le \sum_{\pi \in \mathscr{N}_\tau(N')} \sum_{i,j \in N: \tau(i) \ne \tau(j)} P\left\{ \log \hat p_{ij}^0 > - r_L \right\}\cap B(\pi)\\
	&& = \sum_{i,j \in N: \tau(i) \ne \tau(j)} P\left\{ \log \hat p_{ij}^0 > - r_L \right\}\cap B_\tau(N')\\
	&& \le \sum_{i,j \in N: \tau(i) \ne \tau(j)} P\left\{ \log \hat p_{ij}^0 > - r_L \right\} = O(n^2 \rho_L),
\end{eqnarray*}
by the assumption that $B(\pi)$'s are disjoint. The last equality follows by Lemma \ref{lemm: tail bounds}(ii).

As for the second statement of the lemma, similarly,
\begin{eqnarray*}
    && \sum_{\pi \in \mathscr{N}_\tau(N')} P\left\{\exists r_2 \in R_2(\pi), \min_{i,j \in N_{r_2}'(\pi): i \ne j} \log \hat p_{ij}^0 \le -r_L \right\} \cap B(\pi)\\
	&\le& \sum_{i,j \in N: \tau(i) = \tau(j)}  P\left\{\log \hat p_{ij}^0 \le -r_L \right\}.
\end{eqnarray*}
Again, the last sum is $O(n^2 \omega_{n,L})$ by Lemma \ref{lemm: tail bounds}(ii). $\blacksquare$

\begin{lemma}
	\label{lemm: selection}
	Suppose that an estimated ordered partition $(\hat N_1,...,\hat N_s)$ of $N$ is asymptotically $\tau$-compatible at rate $u_{n,L}$. Then the Selection Step in the Selection-Split Algorithm applied to this ordered partition selects a set, say, $\hat N_{\hat r} \subset N$, with $\hat r = 1,...,s$, such that
	\begin{eqnarray*}
		P\left\{ \exists i,j \in \hat N_{\hat r}, \tau(i) \ne \tau(j) \right\} = 1 + O(n^2 \omega_{n,L} + u_{n,L}), 
	\end{eqnarray*}
	 as $n,L \rightarrow \infty$.
\end{lemma}

\noindent \textbf{Proof:} Let us consider the event that $(\hat N_1,...,\hat N_s)$ coincides with a $\tau$-ordered partition, say, $\pi = (N_1(\pi),...,N_{s}(\pi)) \in \mathscr{N}_\tau(N)$, and denote the event by $A(\pi)$. Note that $A(\pi)$'s are disjoint across $\pi \in \mathscr{N}_\tau(N)$, and
\begin{eqnarray}
\label{asymp compat}
	\sum_{\pi \in \mathscr{N}_\tau(N)} PA(\pi) = 1 + O(u_{n,L}),
\end{eqnarray}
by the assumption that $(\hat N_1,...,\hat N_s)$ of $N$ is asymptotically $\tau$-compatible at rate $u_{n,L}$. Given $\pi \in \mathscr{N}_\tau(\pi)$, let $R_1(\pi)$ and $R_2(\pi)$ be as defined in Lemma \ref{lemm: conv3}. Then,
\begin{eqnarray}
   && \sum_{\pi \in \mathscr{N}_\tau(N)} P \left\{ \hat r \in R_1(\pi) \right\} \cap A(\pi)\\ \notag
   &\ge& \sum_{\pi \in \mathscr{N}_\tau(N)} P\left\{\forall r_1 \in R_1(\pi),\forall r_2 \in R_2(\pi), \min_{i,j \in \hat N_{r_1}} \log \hat p_{ij}^0 < \min_{i,j \in \hat N_{r_2}} \log \hat p_{ij}^0 \right\}\cap A(\pi),
\end{eqnarray}
by the way the Selection Step is defined. Note that
\begin{align}
\label{dev}
	& \sum_{\pi \in \mathscr{N}_\tau(N)} P\left\{\forall r_1 \in R_1(\pi),\forall r_2 \in R_2(\pi), \min_{i,j \in \hat N_{r_1}} \log \hat p_{ij}^0 < \min_{i,j \in \hat N_{r_2}} \log \hat p_{ij}^0 \right\} \cap A(\pi)\\ \notag
	\ge & \sum_{\pi \in \mathscr{N}_\tau(N)}P\left\{\forall r_1 \in R_1(\pi),\forall r_2 \in R_2(\pi), \min_{i,j \in \hat N_{r_1}} \log \hat p_{ij}^0 \le - r_L,  \min_{i,j \in \hat N_{r_2}} \log \hat p_{ij}^0 > - r_L \right\} \cap A(\pi)\\ \notag
	\ge & \sum_{\pi \in \mathscr{N}_\tau(N)}P\left\{\forall r_2 \in R_2(\pi), \min_{i,j \in \hat N_{r_2}} \log \hat p_{ij}^0 > - r_L \right\} \cap A(\pi)\\ \notag
	&- \sum_{\pi \in \mathscr{N}_\tau(N)}P\left\{\exists r_1 \in R_1(\pi), \min_{i,j \in \hat N_{r_1}} \log \hat p_{ij}^0 > - r_L \right\} \cap A(\pi).
\end{align}
By the definition of $A(\pi)$, the second to the last sum in (\ref{dev}) is written as
\begin{eqnarray*}
	&& \sum_{\pi \in \mathscr{N}_\tau(N)} P\left\{ \forall r_2 \in R_2(\pi), \min_{i,j \in N_{r_2}(\pi)} \log \hat p_{ij}^0 > - r_L \right\} \cap A(\pi)\\
	&=&  \sum_{\pi \in \mathscr{N}_\tau(N)} P A(\pi) - \sum_{\pi \in \mathscr{N}_\tau(N)} P\left\{\exists r_2 \in R_2(\pi), \min_{i,j \in N_{r_2}(\pi)} \log \hat p_{ij}^0 \le - r_L \right\} \cap A(\pi).
\end{eqnarray*}
The difference on the right hand side is $1 + O(n^2 \omega_{n,L} + u_{n,L})$ by the second statement of Lemma \ref{lemm: conv3} and (\ref{asymp compat}). On the other hand, the last sum in (\ref{dev}) is equal to
\begin{eqnarray*}
	 \sum_{\pi \in \mathscr{N}_\tau(N)} P\left\{\exists r_1 \in R_1(\pi), \min_{i,j \in N_{r_1}(\pi)} \log \hat p_{ij}^0 > - r_L \right\} \cap A(\pi) = O(n^2 \rho_L),
\end{eqnarray*}
by the first statement of Lemma \ref{lemm: conv3}. Since $\rho_L = O(\omega_{n,L})$, we find that
\begin{eqnarray}
\label{bound4}
	\sum_{\pi \in \mathscr{N}_\tau(N)} P\left\{ \hat r \in R_1(\pi) \right\} \cap A(\pi) = 1 + O(n^2 \omega_{n,L} + u_{n,L}),
\end{eqnarray}
as $n,L\rightarrow \infty$.

Thus, we have
\begin{eqnarray}
\label{dev23}
 && \sum_{\pi \in \mathscr{N}_\tau(N)} P\left\{ \exists i,j \in \hat N_{\hat r}, \tau(i) \ne \tau(j) \right\}\\ \notag
 & \ge & \sum_{\pi \in \mathscr{N}_\tau(N)} P\left\{ \exists i,j \in \hat N_{\hat r}, \tau(i) \ne \tau(j) \right\} \cap A(\pi)\\\notag
  &\ge& \sum_{\pi \in \mathscr{N}_\tau(N)} P\left\{ \exists i,j \in  N_{\hat r}'(\pi),  \tau(i) \ne \tau(j)\right\} \cap A(\pi)\\\notag
  &=& \sum_{\pi \in \mathscr{N}_\tau(N)} P\left\{ \hat r \in R_1(\pi) \right\} \cap A(\pi) = 1 + O(n^2 \omega_{n,L} +u_{n,L}),
\end{eqnarray}
by (\ref{bound4}). Thus we obtain the desired result. $\blacksquare$

\begin{lemma}
	\label{lemm: split}
	For any set $N' \subset N$ which contains $i,j \in N$ such that $\tau(i) \ne \tau(j)$, the ordered partition $(\hat N_1',\hat N_2')$ of $N'$ obtained by the Split Algorithm is asymptotically $\tau$-compatible at rate $n^2 \omega_{n,L}$.
\end{lemma}

\noindent \textbf{Proof:} We use the definitions of $N_1'(i)$, $\overline N_1'(i)$, $N_2'(i)$, and $\overline N_2'(i)$ in (\ref{def N(i)}). By Lemma \ref{lemm: consistency}, for each $i \in N'$, the ordered partitions $\hat T_1(i)=(\hat N_1'(i), N'(i) \setminus \hat N_1'(i))$ and $\hat T_2(i) = (N'(i) \setminus \hat N_2'(i),\hat N_2'(i))$ are such that for $T_1(i) = (N_1'(i),\overline N_2'(i))$ and $T_2(i) = (\overline N_1'(i),N_2'(i))$, we have
\begin{eqnarray*}
	\min_{i \in N} P\{\hat T_1(i) = T_1(i)\} &=& 1 + O(n\omega_{n,L}), \text{ and } \\
	\min_{i \in N} P\{\hat T_2(i) = T_2(i)\} &=& 1 + O(n\omega_{n,L}).
\end{eqnarray*}
 Therefore,
\begin{eqnarray}
\label{conv212}
\max_{i \in N} P\{\hat T_1(i)  \ne T_1(i)\} + P\{\hat T_2(i)  \ne T_2(i)\} = O(n\omega_{n,L}),
\end{eqnarray}
as $n,L \rightarrow \infty$. Define $\hat T = (\hat N_1',\hat N_2')$ and note that
\begin{eqnarray*}
	\hat T &=& (\hat N_1'(i^*), (N'(i^*) \setminus \hat N_1'(i^*)) \cup \{i^*\}), \text{ or }\\
	\hat T &=& ((N'(i^*) \setminus \hat N_2'(i^*)) \cup \{i^*\}, \hat N_2'(i^*)),
\end{eqnarray*}
depending on whether $s_1(i^*) \le s_2(i^*)$ or $s_1(i^*) > s_2(i^*)$. Hence
\begin{align}
\label{decomp}
	P\{\hat T \in \mathscr{N}_\tau(N') \} &= \sum_{i \in N'} P\{\hat T_1(i)  = T_1(i), i^* = i,s_1(i) \le s_2(i)\}\\ \notag
	                                     &+ \sum_{i \in N'} P\{\hat T_2(i)  = T_2(i), i^* = i,s_1(i) > s_2(i)\} + O(n^2\omega_{n,L}),
\end{align}
by (\ref{conv212}). Now, the leading sum on the right hand side is bounded from below by 
\begin{eqnarray*}
	\sum_{i \in N'} (P\{i^* = i,s_1(i) \le s_2(i)\} - P\{\hat T_1(i)  \ne T(i)\}).
\end{eqnarray*}
Applying the same argument to the last sum in (\ref{decomp}), we obtain that
\begin{align*}
   P\{\hat T \in \mathscr{N}_\tau(N') \} \ge 1 - \sum_{i \in N'}\left( P\{\hat T_1(i)  \ne T_1(i)\} + P\{\hat T_2(i)  \ne T_2(i)\} \right) + O(n^2 \omega_{n,L}),
\end{align*}
as $n,L \rightarrow \infty$. Hence by (\ref{conv212}), we conclude that
\begin{eqnarray*}
	P\{\hat T \in \mathscr{N}_\tau(N')\} = 1 + O(n^2 \omega_{n,L}),
\end{eqnarray*}
as $n,L \rightarrow \infty$. $\blacksquare$

\begin{lemma}
	\label{lemm: tau compatibility ext}
	Suppose that for some $s < K_0$, an estimated ordered partition $(\hat N_1,...,\hat N_s)$ of $N$ is asymptotically $\tau$-compatible at rate $u_{n,L}$.
	
	Then the new ordered partition $(\hat N_1',...,\hat N_{s+1}')$ of $N$ obtained by applying the Selection-Split Algorithm to $(\hat N_1,...,\hat N_s)$ is asymptotically $\tau$-compatible at rate $n^2 \omega_{n,L} + u_{n,L}$.
\end{lemma}

\noindent \textbf{Proof:} For each $\pi \in \mathscr{N}_\tau(N)$, let us consider the event that $(\hat N_1,...,\hat N_s)$ coincides with the $\tau$-ordered partition, say, $\pi = (N_1'(\pi),...,N_s'(\pi))$, and denote the event by $A(\pi)$. Let $R_1(\pi)$ be as in the proof of Lemma \ref{lemm: selection}. For each $r \in R_1(\pi)$, let $B_s(r;\pi)$ be the event that the split of $N_r'$ into $\hat N_{r,1}' \cup \hat N_{r,2}'$ (according to the Split Algorithm) coincides with $N_{r,1}' \cup N_{r,2}'$ such that
\begin{eqnarray*}
	(N_1'(\pi),...,N_{r-1}'(\pi),N_{r,1}',N_{r,2}',N_{r+1}'(\pi),...,N_s'(\pi)) \in \mathscr{N}_{\tau}(N).
\end{eqnarray*}
Let $\hat r$ be the chosen group index among $1,...,s$ by the Selection Step. From the proof of Lemma \ref{lemm: selection} (see (\ref{bound4})), we have
\begin{eqnarray}
\label{con4}
	\sum_{\pi \in \mathscr{N}_\tau(N)} P\left\{\hat r \in R_1(\pi) \right\} \cap A(\pi)  = 1 + O(n^2 \omega_{n,L} + u_{n,L}).
\end{eqnarray}
The probability that the new ordered partition $(\hat N_1',...,\hat N_{s+1}')$ of $N$ belongs to $\mathscr{N}_\tau(N)$ is bounded from below by
\begin{eqnarray*}
	\sum_{\pi \in \mathscr{N}_\tau(N)}P(B_s(\hat r;\pi) \cap  A(\pi)) &\ge& \sum_{\pi \in \mathscr{N}_\tau(N)} P\left(\left\{\hat r \in R_1(\pi) \right\} \cap A(\pi) \cap B_s(\hat r;\pi)\right) \\
	&=& \sum_{\pi \in \mathscr{N}_\tau(N)} \sum_{r \in R_1(\pi)} P\left(\left\{\hat r = r \right\} \cap A(\pi) \cap B_s(r;\pi)\right)\\
	&=& \sum_{\pi \in \mathscr{N}_\tau(N)} \sum_{r \in R_1(\pi)} P\left\{\hat r = r \right\} \cap A(\pi) + O(n^2 \omega_{n,L}),
\end{eqnarray*}
by Lemma \ref{lemm: split}. The last double sum is $1 + O(n^2 \omega_{n,L} + u_{n,L})$ by (\ref{con4}). Thus, we conclude that
\begin{eqnarray*}
	P\{(\hat N_1',...,\hat N_{s+1}') \in \mathscr{N}_\tau(N)\} = 1 + O(n^2 \omega_{n,L} + u_{n,L}),
\end{eqnarray*}
as $n,L\rightarrow \infty$. $\blacksquare$\medskip

For each $K \ge 1$, let $\hat T_{K} = (\hat N_1,...,\hat N_{K})$ be the estimated ordered group structure obtained through the Selection-Split algorithm (until the number of the groups reach $K$) and let $T = (N_1,...,N_{K_0})$ be the true ordered group structure. For the remainder of the proof, we assume that the conditions of Theorem \ref{thm: consistency 2 app} hold. 

\begin{lemma}
	\label{lemm: consistency hat T K0}
	Along any sequence of probabilities $P_{n,L} \in \mathcal{P}_{0,\varepsilon}$,
	\begin{eqnarray*}
		P_{n,L}\{ \hat T_{K_0} = T\} = 1 + O(n^2 \omega_{n,L}),
	\end{eqnarray*}
as $n,L\rightarrow \infty$.
\end{lemma}

\noindent \textbf{Proof:} Note that $\hat T_{K_0} = (\hat N_1,...,\hat N_{K_0})$ is asymptotically $\tau$-compatible at rate $n^2 \omega_{n,L}$ by Lemmas \ref{lemm: split} and \ref{lemm: tau compatibility ext}. Since $T$ has $K_0$ groups, this gives the desired result. $\blacksquare$

\begin{lemma} 
	\label{lemm: appendix 2}
    (i) If $K\geq K_{0}$, then $\hat{V}(K)=O_P(1)$, as $n,L \rightarrow \infty .$
	
	(ii) If $K<K_{0}$,  then for any $M>0$, as $n,L\rightarrow \infty$, 
	\begin{equation*}
	P\{\hat{V}(K)>g(L)M\}\rightarrow 1.
	\end{equation*}
\end{lemma}

\noindent \textbf{Proof:} (i) Let $(\hat N_1,...,\hat N_{K_0})$ be the ordered partition obtained by the Selection-Split Algorithm. By Lemma \ref{lemm: consistency hat T K0}, the event that $\tau(i) = \tau(j)$ for all $i,j \in \hat N_k$ has probability approaching one for all $k=1,...,K_0$. Let $(\hat N_1',...,\hat N_K')$ be the ordered partition obtained by the Selection-Split Algorithm with $K \ge K_0$. 

Since $K \ge K_0$, due to the sequential split nature of the algorithm, each of the resulting groups, say, $\hat N_k'$, $k=1,...,K$, is a subset of a group, say, $\hat N_r$, obtained at step $K = K_0$. Therefore, the event that $\tau(i) = \tau(j)$ for all $i,j \in \hat N_k'$ has probability approaching one for each $k=1,...,K$. By Assumption \ref{assump: consistency2b}(i), we have
\begin{equation*}
\hat{V}(K)=\frac{1}{K}\sum_{k=1}^{K}\left\vert \min_{i,j\in \hat N
	_{k}'}\log \hat{p}_{ij}^{0}\right\vert = O_P(1),
\end{equation*}
as $n,L \rightarrow \infty$. Thus (i) follows.\medskip

\noindent (ii) Suppose that $K<K_{0}$, and fix any $M>0$. Let $(\hat N_1,...,\hat N_K)$ be the ordered partition obtained by the Selection-Split Algorithm. Let the event that $\pi \in \mathscr{N}_{\tau}(N)$ coincides with $(\hat N_1,...,\hat N_K)$ be denoted by $A_K(\pi)$. The event is disjoint across $\pi \in \mathscr{N}_{\tau}(N)$. By Lemmas \ref{lemm: split} and \ref{lemm: tau compatibility ext}, the ordered partition $(\hat N_1,...,\hat N_K)$ is asymptotically $\tau$-compatible at rate $n^2 \omega_{n,L}$, that is,
\begin{eqnarray}
\label{AK}
	\sum_{\pi \in \mathscr{N}_{\tau}(N)}P A_K(\pi) = P \left(\bigcup_{\pi \in \mathscr{N}_{\tau}(N)} A_K(\pi)\right) = 1 + O(n^2 \omega_{n,L}).
\end{eqnarray}
Since $K < K_0$, for any $\pi \in \mathscr{N}_{\tau}(N)$ such that $\pi = (\hat N_1,...,\hat N_K)$, there exists $\hat k(\pi) \in \{1,....,K\}$ such that for some $i,j \in \hat N_{\hat k(\pi)}$, $\tau(i) \ne \tau(j)$. Then we have
\begin{eqnarray}
\label{dev2} \quad \quad 
	P\{\hat{V}(K)>g(L)M\} &=& P\left\{\frac{1}{K}\sum_{k=1}^K \min_{i,j \in \hat N_k} \log \hat p_{ij}^0 < - g(L)M \right\}\\ \notag
	&\ge& \sum_{\pi \in \mathscr{N}_\tau(N)} P\left\{\frac{1}{K}\sum_{k=1}^K \min_{i,j \in \hat N_k} \log \hat p_{ij}^0 < - g(L)M \right\} \cap A_K(\pi)\\ \notag
	&\ge & \sum_{\pi \in \mathscr{N}_\tau(N)} P\left\{\min_{i,j \in \hat N_{\hat k(\pi)}} \log \hat p_{ij}^0 < - g(L)K M \right\} \cap A_K(\pi),
\end{eqnarray}
because the log of $p$-values are non-positive. The last sum in (\ref{dev2}) is bounded from below by
\begin{eqnarray*}
	 &&  \sum_{\pi \in \mathscr{N}_\tau(N)} P\left\{ \forall i,j \in N, \text{ s.t. } \tau(i) \ne \tau(j), \log \hat p_{ij}^0 < - g(L)K M \right\} \cap A_K(\pi)\\
	 &\ge& P\left\{\forall i,j \in N, \text{ s.t. } \tau(i) \ne \tau(j), \log \hat p_{ij}^0 < - g(L)K M \right\} + O(n^2 \omega_{n,L}),
\end{eqnarray*}
by (\ref{AK}). By the condition that $g(L)/r_L \rightarrow 0$ as $n,L\rightarrow \infty$, the last probability is bounded from below by (from some large $L$ on)
\begin{eqnarray*}
	&& P\left\{\forall i,j \in N, \text{ s.t. } \tau(i) \ne \tau(j), \log \hat p_{ij}^0 < - r_L \right\} \\
	&=& 1 - P\left\{\exists i,j \in N, \text{ s.t. } \tau(i) \ne \tau(j), \log \hat p_{ij}^0 \ge - r_L \right\}= 1 + O(n^2 \rho_{L}),
\end{eqnarray*}
by Lemma \ref{lemm: tail bounds}(ii). $\blacksquare$

\begin{lemma}
	\label{lemm: consistency hat K}
	 $P\{\hat K = K_0\} \rightarrow 1$ as $n,L\rightarrow \infty$.
\end{lemma}

\noindent \textbf{Proof:} Choose $K$ such that $K_{0}<K$ and write%
\begin{equation*}
\hat{Q}(K_{0})-\hat{Q}(K)=\hat{V}(K_{0})-\hat{V}(K)+(K_{0}-K)g(L).
\end{equation*}
As for the leading term on the left hand side, we have%
\begin{equation*}
\hat{V}(K_{0})-\hat{V}(K)= O_P(1),
\end{equation*}%
by Lemma \ref{lemm: appendix 2}(i). Since $g(L) \rightarrow \infty$, we find that whenever $K>K_{0}$, we have 
\begin{equation*}
P\left\{ \hat{Q}(K_{0})<\hat{Q}(K)\right\} \rightarrow 1.
\end{equation*}

And for all $K<K_{0},$ we have by Lemma \ref{lemm: appendix 2}(ii), for any $M>0,$ 
\begin{equation}
\label{conv}
P\left\{ \hat{V}(K)>g(L)M\right\} \rightarrow 1,
\end{equation}%
whereas $\hat{V}(K_{0})=O_P(1)$. Therefore, choose $\varepsilon>0$ and take any $M_\varepsilon' > 0$ such that 
\begin{eqnarray*}
	P\{\hat V(K_0) \ge M_\varepsilon' \} \le \varepsilon.
\end{eqnarray*}
We take large $M >K_0 - K$ such that $0< M_\varepsilon' \le (K - K_0 + M)g(L)$. We find that 
\begin{eqnarray*}
P\left\{ \hat{Q}(K_{0})<\hat{Q}(K)\right\} &=& P\left\{\hat V(K_0) < \hat V(K) + (K - K_0) g(L) \right\}\\
&\ge& P\left\{\hat V(K_0) < (K - K_0 + M) g(L) \right\} + o(1), (\text{by (\ref{conv})})\\
&\ge& P\left\{\hat V(K_0) < M_\varepsilon' \right\} + o(1) \ge 1 - \varepsilon + o(1),
\end{eqnarray*}
as $n,L \rightarrow \infty$. By sending $\varepsilon$ down to zero, we conclude that $P\{\hat{K}=K_{0}\}\rightarrow 1$, as $n,L\rightarrow \infty$. $\blacksquare$\medskip

\noindent \textbf{Proof of Theorem \ref{thm: consistency 2 app}:} The desired result follows from Lemmas \ref{lemm: consistency hat T K0} and \ref{lemm: consistency hat K}. $\blacksquare$

\section{Confidence Sets for the Group Structure}
The web appendix of \citet*{KrasnokutskayaSongTang2020JPE} proposes a method to construct a confidence set for each group of agents having the same type. Here for the sake of readers' convenience, we reproduce the procedure here using the notation of this paper. Let us consider a set-up where we have $K_{0}$ groups and the set $N$ of agents. Let $\hat K$ be the consistent estimator of $K_0$ as proposed in \citet*{KrasnokutskayaSongTang2020}. As for confidence sets, we construct a confidence set for each group of agents who have the same type. First, we fix $k=1,...,\hat K$ and construct a confidence set for the $k$-th type group $N_k$. In other words, we construct a random set $\hat{\mathcal{C}}_{k}\subset N$
such that
\begin{equation*}
\text{liminf}_{L\rightarrow \infty }P\{N_{k}\subset \hat{\mathcal{C}}
_{k}\}\geq 1-\alpha,
\end{equation*}
For this, we need to devise a way to approximate the finite sample probabilities like $%
P\{N_{k}\subset \hat{\mathcal{C}}_{k}\}$. Since we do not know the
cross-sectional dependence structure among the agents, we use a bootstrap
procedure that preserves the dependence structure from the original sample. The remaining issue is to determine the space in which the random set $\hat{\mathcal{C}}_{k}\subset N$ can take values in. It is computationally infeasible to consider all possible such sets. Instead, we proceed as follows. First we estimate $\hat{N}_{k}$ as prescribed in the paper and also obtain $\hat{\delta}_{ij}^{0}$, the test statistic defined in the main text. Given the estimate $\hat{N}_{k}$, we construct a sequence of sets as follows:\medskip 

\noindent \textbf{Step 1:} Find $i_{1}\in N\backslash \hat{N}_{k}$\ that minimizes min$_{j\in 
	\hat{N}_{k}}\hat{\delta}_{i_{1}j}^{0}$, and construct $\hat{\mathcal{C}}%
_{k}(1)=\hat{N}_{k}\cup \{i_{1}\}.$

\noindent \textbf{Step 2:} Find $i_{2}\in N\backslash \hat{\mathcal{C}}_{k}(1)\ $that
minimizes min$_{j\in \hat{C}_{k}(1)}\hat{\delta}_{i_{2}j}^{0}$, and construct 
$\hat{\mathcal{C}}_{k}(2)=\hat{C}_{k}(1)\cup \{i_{2}\}.$

\noindent \textbf{Step $m$:} Find $i_{m}\in N\backslash \hat{\mathcal{C}}_{k}(m-1)$\ that minimizes min$_{j\in \hat{C}_{k}(m-1)}\hat{\delta}_{i_{m}j}^{0}$ and
construct $\hat{\mathcal{C}}_{k}(m)=\hat{\mathcal{C}}_{k}(m-1)\cup
\{i_{m}\}.$
\medskip

Repeat Step $m$ up to $n=|N|$. \medskip

Now, for each bootstrap iteration $s=1,...,B$, we construct
the sets $\hat{N}_{k,s}^{\ast }$ and $\{\hat{\mathcal{C}}_{k,s}^{\ast
}(m)\}$ following the steps described above but using the bootstrap sample.
(Note that this bootstrap sample should be drawn independently of the bootstrap sample used
to construct bootstrap $p$-values $\hat p_{ij}$ in the classification.)

Then, we compute the following:%
\begin{equation*}
\hat{\pi}^{k}(m)\equiv \frac{1}{B}\sum_{s=1}^{B}1\left\{ \hat{N}_{k}\subset 
\hat{\mathcal{C}}_{k,s}^{\ast }(m)\right\} .
\end{equation*}%
Note that the sequence of sets $\hat{\mathcal{C}}_{k,s}^*(m)$
increases in $m$. Hence the number $\hat{\pi}^{k}(m)$ should also increase
in $m$. An $(1-\alpha )$\%-level confidence set is given by $\hat{\mathcal{C}}_{k}^*(m)$ with $1 \le m \le n$ such that 
\begin{equation*}
\hat{\pi}^{k}(m-1)<1-\alpha \leq \hat{\pi}^{k}(m).
\end{equation*}
Note that such $m$ always exists, because $\hat{\mathcal{C}}_{k,s}^*(n) = N$. 

\section{Further Simulation Results}

Tables E.1 and E.2 summarize the distribution of estimation errors in our group classification algorithm from 500 simulated data sets, when the number of groups is $ K_0 = 2 $ and assumed known to the econometrician. The column $ D_{\mu} $ shows the difference between the group means chosen in the simulation. 

When $ K_0 = 2 $, the results show that the estimation error, as measured by the expected average discrepancy (EAD), decreases with the distance between group means. Such a reduction in EAD is most substantial when the number of agents is larger ($n = 40$) and the size of the data is small ($L = 100$). Given group difference, EAD decreases as sample size increases moderately from $L = 100$ to $400$. This pattern is most obvious when $ D_{\mu} = 0.2 $.

The other measure of estimation errors, HAD(p), also shows encouraging results. HAD(p) is zero for most of the cells in both panels (a) and (b), which shows that the empirical distribution of proportion of mis-classified bidders is reasonably skewed to the right. Besides, the reduction in HAD(p) as the sample size increases is most pronounced with closer group means, regardless of the number of bidders in the population.

When $ K_0 = 4 $, the results demonstrate very similar patterns. Most remarkably, both measures of mis-classification errors only increase very marginally relative to the case with $ K_0 = 2$. 

Tables E.3 and E.4 report results from the full, feasible classification procedure when the number of groups is estimated through the penalization scheme proposed in the text. For most of the specifications used in these two tables, the estimates for the number of groups $ \hat{K}_0 $ are tightly clustered around the correct $ K_0 $. Compared with the results for infeasible classification under known $ K_0 $, EAD and HAD(p) increase in most cases. Nonetheless such an increase is quite moderate,  suggesting that our feasible classification algorithm performs reasonably well relative to its infeasible counterpart.

In Tables E.3 - E.4, we report the analysis of computation time for the classification procedure. In Table 5.3, we give a decomposition of the time that it took for the classification procedure. The table clearly shows that the major computation time spent is when we construct bootstrap p-values. Once the p-values are constructed, the classification algorithm itself runs fairly fast.

In Table E.4 , the computation time is shown to vary depending on the number of the agents ($n$), the number of the true groups ($K_0$), and the number of the markets ($L$). The results show that the most computation time increase arises when the number of the bidders increases rather than when the number of the markets or the number of the groups increases. Our simulation studies are based on our MatLab code. The program was executed using a computer with the following specifications: Intel(R) Xeon (R) CPU X5690 @3.47 GHz 3.46 GHz.


\normalsize

\begin{table}[h!]
	\begin{center}
		\small
		Table E.1 : Performance of the Classification Estimator with Two Groups:
		\par
		($K_{0}=2$ and known)\medskip
		\par
		\begin{tabular}{cccc|c|cccccc}
			\hline\hline
			\small
			&  $n$	&	$L$	&	$D_\mu$ &	EAD	&	HAD(.10)	&	HAD(.25)	&	HAD(.50)	&	HAD(.75)	&	HAD(.90)&	\\ \cline{2-11} 
			\multicolumn{11}{c}{}\\ 
			&	12	&	400	&	0.6	&	0.012	&	0.012	&	0	    &	0	    &	0	&	0	&\\
			&	12	&	400	&	0.4	&	0.014	&	0.014	&	0	    &	0	    &	0	&	0	&\\
			&	12	&	400	&	0.2	&	0.004	&	0.004	&	0	    &	0	    &	0	&	0	&\\
			&\\\cline{2-11}
			\multicolumn{11}{c}{}\\ 
			&	12	&	200	&	0.6	&	0.004	&	0.004	&	0	    &	0	    &	0	&	0	&\\
			&	12	&	200	&	0.4	&	0.006	&	0.006	&	0	    &	0	    &	0	&	0	&\\
			&	12	&	200	&	0.2	&	1.118	&	0.560	&	0.252	&	0.158	&	0	&	0	&\\
			&\\\cline{2-11}
			\multicolumn{11}{c}{}\\ 
			&	12	&	100	&	0.6	&	0.006	&	0.006	&	0	    &	0	    &	0	&	0	&\\
			&	12	&	100	&	0.4	&	0.084	&	0.078	&	0.006	&	0	    &	0	&	0	&\\
			&	12	&	100	&	0.2	&	1.794	&	0.682	&	0.478	&	0.284	&	0	&	0	&\\
			\cline{2-11} 
			\multicolumn{11}{c}{}\\ 
			&	40	&	400	&	0.6	&	0.018	&	0	    &	0	    &	0	    &	0	    &	0	    &\\
			&	40	&	400	&	0.4	&	0.022	&	0	    &	0	    &	0	    &	0	    &	0	    &\\
			&	40	&	400	&	0.2	&	1.170	&	0.178	&	0.014	&	0	    &	0	    &	0	    &\\
			&\\\cline{2-11}
			\multicolumn{11}{c}{}\\ 
			&	40	&	200	&	0.6	&	0.018	&	0	    &	0	    &	0	    &	0	    &	0	    &\\
			&	40	&	200	&	0.4	&	0.020	&	0	    &	0	    &	0	    &	0	    &	0	    &\\
			&	40	&	200	&	0.2	&	2.726	&	0.404	&	0.210	&	0.122	&	0.021	&	0.001	&\\
			&\\\cline{2-11}
			\multicolumn{11}{c}{}\\ 
			&	40	&	100	&	0.6	&	0.020	&	0	    &	0	    &	0	    &	0	    &	0    &\\
			&	40	&	100	&	0.4	&	0.452	&	0.010	&	0	    &	0	    &	0	    &	0	    &\\
			&	40	&	100	&	0.2	&	3.720	&	0.902	&	0.578	&	0.234	&	0.132	&	0.043	&\\
			\hline\hline
		\end{tabular}%
	\end{center}
	\par
	\medskip 
	
	\begin{flushleft}
		\noindent {\footnotesize Note: This table summarizes the distribution of estimation errors in our classification algorithm from 500 Monte Carlo replications when $ K_0 = 4$ and known. Here $n$ represents the number of the individual agents, $L$ the number of the observed games in the data, $ D_\mu $ the distance between population means, EAD the expected average discrepancy, and HAD(p) the hazard rate of average discrepancies at $p$.}
	\end{flushleft}
\end{table}

\normalsize
\begin{table}[h!]
	\small
	\begin{center}
		Table E.2 : Performance of the Classification Estimator with Four Groups:
		\par
		($K_{0}=4$ and known)\medskip
		\par
		\begin{tabular}{cccc|c|cccccc}
			\hline\hline
			\small
			&  $n$	&	$L$	&	$D_\mu$ &	EAD	&	HAD(.10)	&	HAD(.25)	&	HAD(.50)	&	HAD(.75)	&	HAD(.90)&	\\ \cline{2-11}
			\multicolumn{11}{c}{}\\ 
			&	12	&	400	&	0.6	&	0.011	&	0.014	&	0.004	&	0	&	0	&	0	&\\
			&	12	&	400	&	0.4	&	0.018	&	0.016	&	0.010	&	0	&	0	&	0	&\\
			&	12	&	400	&	0.2	&	0.017	&	0.022	&	0.006	&	0	&	0	&	0	&\\
			\cline{2-11}
			\multicolumn{11}{c}{}\\ 
			&	12	&	200	&	0.6	&	0.013	&	0.018	&	0.004	&	0	    &	0	    &	0	&\\
			&	12	&	200	&	0.4	&	0.004	&	0.008	&	0	    &	0	    &	0	    &	0	&\\
			&	12	&	200	&	0.2	&	1.112	&	0.764	&	0.188	&	0.024	&	0.008	&	0	&\\\cline{2-11}
			\multicolumn{11}{c}{}\\ 
			&	12	&	100	&	0.6	&	0.003	&	0.006	&	0	    &	0	    &	0	    &	0	&\\
			&	12	&	100	&	0.4	&	0.044	&	0.040	&	0.024	&	0.002	&	0	    &	0	&\\
			&	12	&	100	&	0.2	&	1.504	&	0.868	&	0.342	&	0.106	&	0.04	&	0	&\\
			\cline{2-11} 
			\multicolumn{11}{c}{}\\ 
			&	40	&	400	&	0.6	&	0.115	&	0.020	&	0.020	&	0	&	0	&	0	&\\
			&	40	&	400	&	0.4	&	0.121	&	0.020	&	0.020	&	0	&	0	&	0	&\\
			&	40	&	400	&	0.2	&	2.450	&	0.680	&	0.368	&	0.018	&	0.018	&	0	&\\
			&\\\cline{2-11}
			\multicolumn{11}{c}{}\\ 
			&	40	&	200	&	0.6	&	0.109	&	0.018	&	0.018	&	0	&	0	&	0	&\\
			&	40	&	200	&	0.4	&	0.140	&	0.026	&	0.026	&	0	&	0	&	0	&\\
			&	40	&	200	&	0.2	&	3.172	&	0.810	&	0.366	&	0.246	&	0.026	&	0	&\\
			\cline{2-11}
			\multicolumn{11}{c}{}\\ 
			&	40	&	100	&	0.6	&	0.141	&	0.024	&	0.024	&	0	&	0	&	0	&\\
			&	40	&	100	&	0.4	&	1.003	&	0.176	&	0.176	&	0.006	&	0	&	0	&\\
			&	40	&	100	&	0.2	&	4.557	&	0.904	&	0.652	&	0.526	&	0.202	&	0.053	&\\
			\hline\hline
		\end{tabular}%
	\end{center}
	\par
	\medskip 
	
	\begin{flushleft}
		\noindent {\footnotesize Note: This table summarizes the distribution of estimation errors in our classification algorithm from 500 Monte Carlo replications when $ K_0 = 4$ and known. Here $n$ represents the number of the individual agents, $L$ the number of the observed games in the data, $ D_\mu $ the distance between population means, EAD the expected average discrepancy, and HAD(p) the hazard rate of average discrepancies at $p$.}
	\end{flushleft}
\end{table}	
\clearpage

\begin{table}[t]
	\small
	\begin{center}
		Table E.3 : Computational Time for Various Steps of the Procedure
		\par
		($n=60$, $K_{0}=2$, unknown, time measured in \textbf{seconds})\medskip
		\par
		\begin{tabular}{cc|l|r|r|rr}
			\hline\hline
			\small
			&       &                                 &         &           &         &\\
			&	Step	&	Description	                    &	$L$=100	  &	$L$=200	    &	$L$=400	  &\\\cline{2-6}
			&       &                                 &         &           &         &\\
			&	1	    &	generating pairwise indexes from the data	&	0.2987	&	0.3543	  &	0.4607	&\\
			&	2	    &	constructing bootstrap pairwise indexes	                &	81.2178	&	81.4871	  &	82.0807	&\\
			&	3	    &	computing bootstrap p-values	            &	0.0012	&	0.0014	  &	0.0014	&\\
			&	4	    &	division of a group into two	  &	0.0008	&	0.0008	  &	0.0008	&\\
			&	n+4	  &	number of groups selection	    &	0.0002	&	0.0002	  &	0.0002	&\\\cline{2-6}
			&       &                                 &         &           &         &\\
			&		    &	Total Time	                    &	81.528	&	81.852	  &	82.552	&\\
			\hline\hline
		\end{tabular}%
	\end{center}
	\par
	\medskip 
	
	\begin{flushleft}
		\noindent {\footnotesize Note: The table shows a decomposition of a total time it has taken for the classification procedure. The table shows that the major portion of the time comes from constructing the bootstrap pairwise indexes. Once the bootstrap p-values are constructed, the classification algorithm runs quite fast.}
	\end{flushleft}
\end{table}

\begin{table}[t]
	\small
	\begin{center}
		Table E.4 : Total Computational Time: across $n$, $L$, and $K_0$
		\par
		($K_0$ unknown, time measured in \textbf{seconds})\medskip
		\par
		\begin{tabular}{cc|r|r|r|r|rr}
			\hline\hline
			\small
			&                  &         &         &         &         &        &\\
			&    &	$L$=100	 &	$L$=200	 &	$L$=400	 &	$L$=200	 &	$L$=200	&\\
			&    &	$K_0$=2	   &	$K_0$=2	   &	$K_0$=2	   &	$K_0$=4	   &	$K_0$=6	  &\\\cline{2-7}
			&                  &         &         &         &         &        &\\
			&	$n=$ 12	&	3.246	  &	3.224	  &	3.239	  &	3.216	  &	3.219	  &\\
			&	$n=$ 24	&	13.057	&	13.177	&	13.259	&	13.185	&	13.189	&\\
			&	$n=$ 48	&	51.987	&	52.272	&	52.700	&	52.281	&	52.291	&\\
			&	$n=$ 60	&	81.528	&	81.852	&	82.552	&	81.862	&	82.874	&\\
			&	$n=$ 72	&	116.949	&	117.213	&	117.577	&	116.912	&	117.328 &\\
			&	$n=$ 96	&	209.426	&	209.971	&	209.834	&	209.884	&	210.058 &\\
			\hline\hline
		\end{tabular}%
	\end{center}
	\par
	\medskip 
	\begin{flushleft}
		\noindent {\footnotesize Note: The table shows the change in the computation time as one changes the number of the groups ($K$), the number of the markets and the number of the agents (i.e., bidders) ($(n)$). The major increase in the computation time arises when the number of the bidders increases rather than when the number of the markets or the groups increases.}
	\end{flushleft}
\end{table} 

\clearpage

\section{Additional Materials for the Empirical Application}

Table F.1 reports summary statistics
for this set of projects. The table indicates that the projects are worth
\$523,000 and last for around three months on average; 38\% of these projects are partially supported through federal funds. There are 25 firms
that participate regularly in this market. All other firms in the data are
treated as fringe bidders. An average auction attracts six regular potential
bidders and eight fringe bidders. Since only a fraction of potential bidders
submits bids, an entry decision plays an important role in this market.
Finally, the distance to the company location varies quite a bit and is
around 28 miles on average for regular potential bidders.

\begin{table}[!h]
	\small
	\begin{center}
		Table F.1 : Summary Statistics for California Procurement Market\medskip
		\par
		\begin{tabular}{ll|ccc}
			\hline\hline
			& Variable                          & Mean & Std.\ Dev         &  \\ \hline
			& Engineer's estimate (mln)         & 0.523& 0.261             &  \\ 
			& Duration, large projects (months) & 3.01 & 1.56              &  \\ 
			& Federal Aid                       & 0.384&                   &\\\hline
			\multicolumn{2}{l|}{Number of Potential Bidders:} & 14.1 & 8.4 &  \\ 
			& Fringe Bidders & 8.2 & 4.8 &  \\ 
			& Regular Bidders & 5.5 & 3.3 &  \\ \hline
			\multicolumn{2}{l|}{Number of Entrants:} & 5.4 & 2.8 &  \\ 
			& Fringe Bidders & 3.5 & 2.7 &  \\ 
			& Regular Bidders & 1.9 & 1.8 &  \\ \hline
			\multicolumn{2}{l|}{Distance (miles):} & 18.72 & 6.33 &  \\ 
			& Fringe Bidders & 11.21 & 5.42 &  \\ 
			& Regular Bidders & 28.34 & 11.73 &  \\ \hline\hline
		\end{tabular}%
	\end{center}
	\par
	\medskip 
	\begin{flushleft}
		\noindent {\footnotesize Note: This table reports summary statistics for
			the set of medium size bridge work and paving projects auctioned in the
			California highway procurement market between years of 2002 and 2012. It
			consists of 1,054 projects. The distance variable is measured in miles. It reflects the
			driving time between the project site and the nearest company plant. The ``Federal Aid'' variable is equal to one if the project receives federal aid and zero otherwise. }
	\end{flushleft}
\end{table}

In Table F.2, we present an extended version of Table 6 in Section 6 of the main paper. This table includes the estimates of the group-specific fixed effects.

\begin{table}[!t]
	\begin{center}
		\small
		Table F.2 : Parameter Estimates (Extended Version of Table 6.) \medskip
		\par
		\begin{tabular}{cc|cc|cc|cc}
			\hline\hline
&                 & Estimate                   & Std. Error & Estimate                   & Std. Error & P-value&\\ \hline
\multicolumn{7}{l}{The Distribution of Project Costs}  \\ 
&	 Constant ($\bar{q}_0$) 	 &	 0.127$^{\ast \ast \ast }$ 	 &	(0.0129)	  &	 0.113$^{\ast \ast \ast }$    &	(0.0119)	  & 0.216 &	\\ 
&	 Eng. Estimate 	           &	-0.0004$^{\ast \ast \ast }$  &	(0.0002)	  &	-0.0005$^{\ast \ast \ast }$  &	(0.0002)	  &	0.392&\\ 
&	 Duration 	               &	 0.00026$^*$ 	               &	(0.00036)	  &	0.00022$^{\ast }$            &	(0.00027)  &	0.212&\\ 
&	 Distance 	               &	 0.0012$^{\ast \ast \ast }$  &	(0.00022)	  &	0.00086$^{\ast \ast \ast }$   &	(0.00019)	  &	0.041&\\ 
&  Bridge                    & -0.0092$^{\ast \ast \ast }$  &  (0.0018)    & -0.012$^{\ast \ast \ast }$   &  (0.0011)    & 0.074&\\
&	 Federal Aid	             &	-0.043$^{\ast \ast \ast }$   &	(0.0103)	  &	-0.078$^{\ast \ast \ast }$   &	(0.009)	    & 0.012 &\\
&  Regular Bidder&                               &              & -0.035$^{\ast \ast \ast }$   &  (0.003)     &  &\\
&	 $\bar{q}_1-\bar{q}_0$     &	 -0.051$^{\ast \ast \ast }$  &	(0.008)	    &		                           &		       &&	\\ 
&	 $\bar{q}_2-\bar{q}_0$     &	 -0.012$^{\ast \ast \ast }$   &	(0.005)	    &		                           &		       &&\\ 
&	 $\bar{q}_3-\bar{q}_0$     &	 -0.032$^{\ast \ast \ast }$  &	(0.009)	    &		                           &		       &&	\\ 
&	 $\bar{q}_4-\bar{q}_0$     &	 -0.058$^{\ast \ast \ast }$  &	(0.008)	    &		                           &		       &&	\\ 
&	 $\bar{q}_5-\bar{q}_0$     &	 -0.014$^{\ast \ast \ast }$  &	(0.007)	    &		                           &		       &&	\\ 
&	 $\bar{q}_6-\bar{q}_0$     &	 -0.008$^{\ast \ast \ast }$  &	(0.006)	    &		                           &		       &&	\\ 
&	 $\bar{q}_7-\bar{q}_0$     &	 -0.009$^{\ast \ast \ast }$  &	(0.007)	    &		                           &		       &&	\\ 
&	 $\bar{q}_8-\bar{q}_0$     &	 -0.050$^{\ast \ast \ast }$  &	(0.006)	    &		                           &		       &&	\\ 
&	 $\sigma _{C}$ 	&	 0.087$^{\ast \ast \ast }$ 	 &	(0.032)	    &	0.112$^{\ast \ast \ast }$    &	(0.022)	    &	0.087&\\ 
&	 $\sigma _{U}$ 	&	 0.021$^{\ast \ast \ast }$ 	 &	(0.009)	    &	0.0207$^{\ast \ast \ast }$   &	(0.008)	    &	0.452&\\\hline 
\multicolumn{7}{l}{The Distribution of Entry Costs}  \\ 
&	 Constant ($\tilde{q}_0$)  &	 -0.0114$^{\ast }$ 	       &	(0.0078)	&	-0.0161$^{\ast }$	&	(0.0091)	   &	0.212 &  \\ 
&	 Eng. Estimate             &	 0.0055$^{\ast \ast \ast }$&	(0.0016)	&	 0.0051$^{\ast \ast \ast }$	   &	(0.0012)	&	0.333 &  \\ 
&	 Number of Items           &	 0.0018$^{\ast }$	         &	(0.0011)	&	 0.0011$^{\ast \ast \ast }$	   &	(0.0005)	&	0.082 &  \\ 
&  Regular Bidder            &                             &            & -0.022 $^{\ast \ast \ast }$    &   (0.004)   &   \\
&	 $\tilde{q}_1-\tilde{q}_0$ &	-0.019$^{\ast \ast \ast }$ &	(0.005)	  &		      &		      &	&  \\ 
&	 $\tilde{q}_2-\tilde{q}_0$ &	-0.018$^{\ast \ast \ast }$ &	(0.007)	  &		      &		      &	&  \\ 
&	 $\tilde{q}_3-\tilde{q}_0$ &	-0.016$^{\ast \ast \ast }$ &	(0.007)	  &		      &		      &	&  \\ 
&	 $\tilde{q}_4-\tilde{q}_0$ &	-0.024$^{\ast \ast \ast }$ &	(0.006)	  &		      &		      &	&  \\ 
&	 $\tilde{q}_5-\tilde{q}_0$ &	-0.022$^{\ast \ast \ast }$ &	(0.008)	  &		      &		      &	& \\ 
&	 $\tilde{q}_6-\tilde{q}_0$ &	-0.018$^{\ast \ast \ast }$ &	(0.006)	  &		      &		      &	&  \\ 
&	 $\tilde{q}_7-\tilde{q}_0$ &	-0.017$^{\ast \ast \ast }$ &	(0.008)	  &		      &		      &	&  \\ 
&	 $\tilde{q}_8-\tilde{q}_0$ &	-0.019$^{\ast \ast \ast }$ &	(0.008)	  &		      &		      &	&  \\ 
			\hline\hline
		\end{tabular}%
	\end{center}
	\par
	\medskip 
	\begin{flushleft}
		\noindent {\footnotesize Note: In the results above the distance is measured in miles. The fringe bidders are the reference group. The first two columns correspond to the specification which allows for the unobserved bidder heterogeneity; the next two columns correspond to the specification without unobserved bidder heterogeneity. 
The last column reports the p-value of the bootstrap-based test of the equality of coefficients estimated under the specifications with and without unobserved bidder heterogeneity. Details of the test are explained below. The results are based on the data for 1,054 medium-sized projects that involve paving and bridge work.}
\end{flushleft}
\end{table}
\newpage

The bootstrap $p$ values in the last column of Table 6 are obtained by the following procedure. First, let $\hat \theta_R$ and $\hat \theta_U$ be estimators of a scalar parameter $\theta_0$, where $\hat \theta_R$ is obtained under the bidder homogeneity restriction and $\hat \theta_U$ is obtained under the group heterogeneity of bidder types. We define
\begin{eqnarray*}
	\mathcal{T} = \sqrt{L}|\hat \theta_R - \hat \theta_U| 
\end{eqnarray*}
For critical values, for each boostrap sample (indexed by $b=1,...,B$), we construct both $\hat \theta_{R,b}^*$ and $\hat \theta_{U,b}^*$ using the same bootstrap sample. Then we construct a bootstrap version of $\mathcal{T}$ as follows:
\begin{eqnarray*}
	\mathcal{T}_b^* = \sqrt{L}|\hat \theta_{R,b}^* - \hat \theta_{U,b}^* - (\hat \theta_R - \hat \theta_U)|. 
\end{eqnarray*}
The bootstap $p$ value for the equality of the two parameters, one identified under homogeneity and the other identified under group heterogeneity, is given by
\begin{eqnarray*}
	\frac{1}{B}\sum_{b=1}^B 1\{\mathcal{T}_b^* > \mathcal{T}\}.
\end{eqnarray*}
We compute this $p$ value for each parameter in Table 6 that is defined for both specifications and report it in the last column of the table.

\putbib[refs_comp]
\end{bibunit}
\end{document}